\documentclass[prd,preprint,eqsecnum,nofootinbib,amsmath,amssymb,
               tightenlines,dvips,superscriptaddress,floatfix]{revtex4}
\usepackage{graphicx}
\usepackage{bm}
\usepackage{amsfonts}
\usepackage{amssymb}
\usepackage{slashed}

\usepackage{dsfont}
\def\openone{\mathds{1}}

\def\b{{\bm b}}

\def\p{{\bm p}}
\def\q{{\bm q}}

\def\B{{\bm B}}
\def\C{{\bm C}}
\def\P{{\bm P}}

\def\bcalB{{\bm{\mathcal B}}}
\def\bcalP{{\bm{\mathcal P}}}

\def\uOmega{\underline{\Omega}}
\def\Ybar{\overline{Y}}
\def\eps{\epsilon}

\def\Nc{N_{\rm c}}
\def\Nf{N_{\rm f}}
\def\CA{C_{\rm A}}
\def\dA{d_{\rm A}}

\def\alphas{\alpha_{\rm s}}
\def\alphaqed{\alpha_{\scriptscriptstyle\rm EM}}
\def\Re{\operatorname{Re}}

\def\tr{\operatorname{tr}}
\def\sgn{\operatorname{sgn}}

\def\grad{{\bm\nabla}}

\def\ix{{\rm i}}
\def\fx{{\rm f}}
\def\xx{{\rm x}}
\def\xbx{{\bar{\rm x}}}
\def\yx{{\rm y}}
\def\ybx{{\bar{\rm y}}}
\def\zx{{\rm z}}

\def\yfrake{{\mathfrak y}_e}

\def\Time{\mathbb{T}}

\def\seq{{\rm seq}}
\def\seqNf{{\seq}}
\def\new{{\rm new}}

\def\calX{{\cal X}}

\def\gammaE{\gamma_{\rm\scriptscriptstyle E}}

\def\Beta{\operatorname{B}}

\def\four{{(4)}}

\def\xe{x_e}
\def\ye{y_e}
\def\ze{z_e}

\def\bbI{\mathbb{I}}
\def\bbA{\mathbb{A}}
\def\MSbar{\overline{\mbox{MS}}}
\def\barOmega{\bar\Omega}

\newcounter{savefootnote}
\newcommand{\astfootnote}[1]{%
  \setcounter{savefootnote}{\value{footnote}}%
  \setcounter{footnote}{0}%
  \let\oldthefootnote=\thefootnote
  \renewcommand{\thefootnote}{\fnsymbol{footnote}}%
  \footnote{#1}%
  \let\thefootnote=\oldthefootnote%
  \setcounter{footnote}{\value{savefootnote}}%
}

\begin {document}



\title
    {
      In-medium loop corrections and longitudinally polarized gauge
      bosons in high-energy showers
    }

\author{Peter Arnold}
\affiliation
    {%
    Department of Physics,
    University of Virginia,
    Charlottesville, Virginia 22904-4714, USA
    \medskip
    }%
\author{Shahin Iqbal}
\affiliation
    {%
    National Centre for Physics, \\
    Quaid-i-Azam University Campus,
    Islamabad, 45320 Pakistan
    \medskip
    }%

\date {\today}

\begin {abstract}%
{%
The splitting processes of bremsstrahlung and pair production in a medium
are coherent over large distances in the very high energy limit,
which leads to a suppression known as the Landau-Pomeranchuk-Migdal
(LPM) effect.  We continue study of the case when the coherence
lengths of two consecutive splitting processes overlap (which is
important for understanding corrections to standard treatments
of the LPM effect in QCD), avoiding soft-emission approximations.
In this particular paper, we show (i) how the ``instantaneous''
interactions of Light-Cone Perturbation Theory must be included in
the calculation to account for effects of
longitudinally-polarized gauge bosons in intermediate states,
and (ii) how to compute virtual corrections to LPM emission rates,
which will be necessary in order to make infrared-safe calculations
of the characteristics of in-medium QCD showering of high-energy partons.
In order to develop these topics in as simple a context as possible,
we will focus in the current paper not on QCD but on
large-$\Nf$ QED, where $\Nf$ is the number of electron flavors.
}%
\end {abstract}

\maketitle
\thispagestyle {empty}

{\def\boldmath{}\tableofcontents}
\newpage


\begin{center}
  \textbf{Note Added}
\end {center}

\bigskip

This version of the preprint incorporates the published erratum
JHEP \textbf{12}, 098 (2023) to the original publication
JHEP \textbf{12}, 120 (2018).
This preprint version also corrects eq.\ (A29c) by adding
absolute value signs that are needed for the same reason given in
the earlier erratum for
corrections to the original publication's eqs.\ (A30) and (E27).

\newpage


\section{Introduction}
\label{sec:intro}

\subsection{Overview}

When passing through matter, high energy particles lose energy by
showering, via the splitting processes of hard bremsstrahlung and pair
production.  At very high energy, the quantum mechanical duration of
each splitting process, known as the formation time, exceeds the mean
free time for collisions with the medium, leading to a significant
reduction in the splitting rate known as the Landau-Pomeranchuk-Migdal
(LPM) effect \cite{LP,Migdal}.%
\footnote{
  The papers of Landau and Pomeranchuk \cite{LP} are also available in
  English translation \cite{LPenglish}.
}
A long-standing problem in field theory has
been to understand how to implement this effect in cases where
the formation times of two consecutive splittings overlap.

\begin {figure}[t]
\begin {center}
  \includegraphics[scale=0.5]{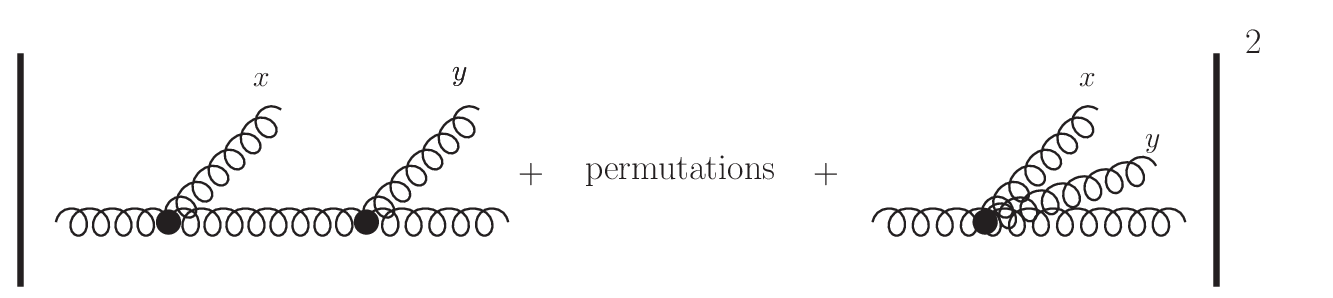}
  \caption{
     \label{fig:gTOggg}
     Real double gluon bremsstrahlung processes $g \to ggg$
     considered in the analysis of overlapping formation time
     effects at high energy in refs.\ \cite{2brem,seq,dimreg,4point}.
     Only the high energy particles are shown: their many interactions
     with the medium are not shown explicitly.  $x$ and $y$ represent the
     momentum fractions (relative to the initial particle) of two of
     the daughters of this splitting process.
  }
\end {center}
\end {figure}

Let $x$ and $y$ be the longitudinal
momentum fractions of two consecutive bremsstrahlung gauge bosons.
In the limit $y \ll x \ll 1$, the problem of overlapping formation
times has been analyzed at leading logarithm order in
refs.\ \cite{Blaizot,Iancu,Wu}
in the context of
energy loss of high-momentum partons traversing
a QCD medium (such as a quark-gluon plasma).
Together with Chang, we
subsequently developed and implemented field theory
formalism needed for the more general case where $x$ and $y$ are
arbitrary \cite{2brem,seq,dimreg,4point}.
We used the formalism to calculate the LPM interference effects
on real double gluon bremsstrahlung $g \to ggg$ in medium,
given by the processes depicted in fig.\ \ref{fig:gTOggg},
in the high-energy limit.  [For the sake of simplicity, that specific
calculation also made other
simplifying approximations by taking the multiple scattering
($\hat q$) limit and the large-$\Nc$ limit.]
But that calculation was incomplete for two reasons.

\begin {figure}[t]
\begin {center}
  \includegraphics[scale=0.5]{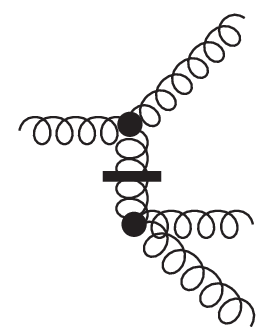}
  \caption{
     \label{fig:gTOgggINST}
     The contribution to $g \to ggg$ due to longitudinal polarization
     of the intermediate gluon line (represented by a bar across the
     line).  This process must be added
     to the $g \to ggg$ amplitude of fig.\ \ref{fig:gTOggg} if the
     intermediate lines in fig.\ \ref{fig:gTOggg} are summed only
     over transverse polarizations as in
     refs.\ \cite{2brem,seq,dimreg,4point}.
  }
\end {center}
\end {figure}

First, the calculations of refs. \cite{2brem,seq,dimreg,4point}
only included gluons that were transversely polarized.
But the polarization of the intermediate-state gluon
in the first diagram of fig.\ \ref{fig:gTOggg} does not
have to be transverse: a full calculation should include
the effects of the longitudinal polarization as well.%
\footnote{
  Some historical explanation:
  Longitudinal polarizations were left out of our earlier analysis
  of refs.\ \cite{2brem,seq,dimreg,4point} because of an unstated
  and incorrect assumption.  Consider the two consecutive vertices
  appearing in
  the first diagram of fig.\ \ref{fig:gTOggg}.
  Our interest is in the case where the formation times associated
  with these vertices overlap, in which case the characteristic scale
  of the time separation of those vertices (and so also their
  spatial separation in the direction of motion $z$) is of order the formation
  time.  That formation time is parametrically large in the large-energy
  limit ($t_{\rm form} \sim \sqrt{E/\hat q}$, ignoring dependence on
  daughter momentum fractions).  The effects of longitudinal polarizations
  do not extend over large distances, and so ref.\ \cite{2brem}
  implicitly ignored them.  But that's not a valid argument that
  the effects of nearly-coincident
  gluon emissions can be ignored,
  as was made clear in our own calculations by the subsequent
  calculation in ref.\ \cite{4point} of the
  4-point vertex contributions represented by the last diagram of
  fig.\ \ref{fig:gTOggg}.
  It was not until we started computing virtual corrections
  (discussed in a moment)
  that we clearly realized our mistake in ignoring intermediate longitudinal
  polarizations.
}
It has long been known how to integrate out the effects of
longitudinal polarizations and find the effective theory
of the transverse polarizations, which is the program of
Light Cone Perturbation Theory (LCPT) \cite{LB,BL,BPP}.%
\footnote{
  For readers not familiar with time-ordered
  LCPT who would like
  the simplest possible example of how it reassuringly
  reproduces the results of
  ordinary Feynman diagram calculations,
  we recommend section
  1.4.1 of Kovchegov and Levin's monograph \cite{KL}.
  For some less-simple examples, see also ref.\ \cite{LCPTcheck}.
}
In LCPT, where light-cone gauge $A^+ = 0$ is used,
the longitudinal polarization gives rise to an
interaction that is instantaneous in light-cone time
$x^+ \equiv x^0+x^z$ (similar to how in Coulomb gauge
the non-transverse polarization $A^0$ of the gauge field
responds to sources instantaneously in normal time $x^0$,
even though physical quantities ultimately do not).
In LCPT, the longitudinal polarization can be integrated out
to yield a Hamiltonian that is supplemented by
just an extra 4-point interaction among the fields.
The result is to add one extra diagram (and its permutations),
fig.\ \ref{fig:gTOgggINST},
to the sum of diagrams depicting the
amplitude in fig.\ \ref{fig:gTOggg}.
In fig.\ \ref{fig:gTOgggINST}, the bar across the intermediate gluon
line indicates that it is an (integrated-out) longitudinally
polarized gluon.  That gluon line is drawn vertically to emphasize that
it represents an instantaneous interaction in the context of
time($x^+$)-ordered perturbation theory, with $x^+$ running from
left to right in the figure.  One goal of this paper is to include
this type of process into our earlier calculations of the effect of
overlapping formation times on sequential bremsstrahlung.

\begin {figure}[t]
\begin {center}
  \includegraphics[scale=0.5]{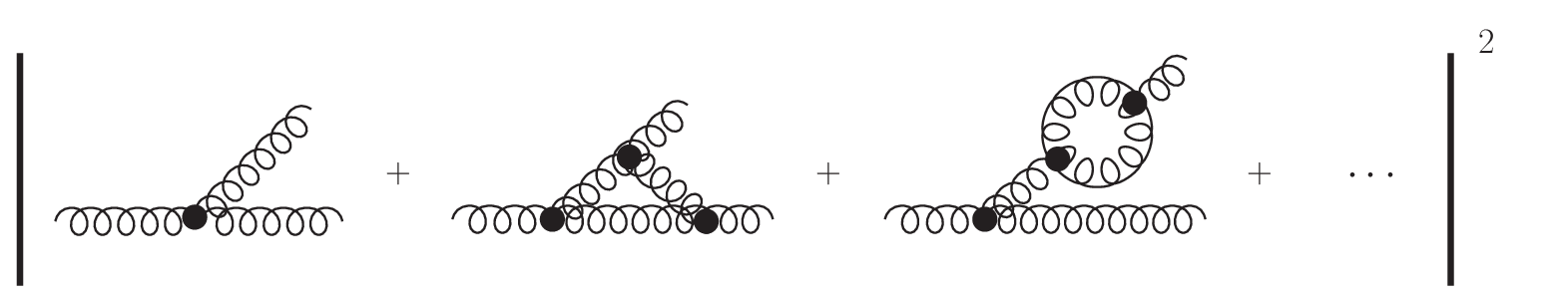}
  \caption{
     \label{fig:gTOggVIRT}
     Examples of one-loop virtual corrections to single bremsstrahlung
     $g \to gg$.
  }
\end {center}
\end {figure}

A calculation of in-medium real double bremsstrahlung $g \to ggg$ is not
enough, however, to study the effects on in-medium energy loss and the
characteristics of in-medium shower development.
Crudely analogous to what happens for vacuum bremsstrahlung in QED, there
are infrared divergences whose treatment at this order
requires additionally
computing virtual corrections to single bremsstrahlung $g \to gg$,
such as depicted by the diagrams in fig.\ \ref{fig:gTOggVIRT}.
The second goal of this paper is to show how to calculate such
loop corrections (without soft gluon approximations)
in the context of finding the effects of
overlapping formation times on energy loss and shower development.
A new technical feature, compared to our calculations of $g \to ggg$,
is that loop calculations require UV regularization and
renormalization.
In particular, one of the effects of loops (besides mitigating
certain infrared divergences of the $g{\to}ggg$ calculation)
will be to replace the
coupling $\alphas$ in the leading-order single-bremsstrahlung
process $g \to gg$ by the running coupling $\alphas(Q_\perp)$ evaluated
at a characteristic transverse momentum scale of the
LPM-modified bremsstrahlung process.

In this particular paper, we work out all of these issues
for a warm-up theory: QED in the large-$\Nf$ limit, where $\Nf$ is
the number of electron flavors.  
The large-$\Nf$ limit is taken just to reduce the complexity of
the calculations and so streamline the presentation of
the important developments.
The method itself should be straightforwardly
generalizable to the QCD analysis of the LPM effect in
figs.\ \ref{fig:gTOggg}--\ref{fig:gTOggVIRT}
(at least in the large-$\Nc$ limit of
refs.\ \cite{2brem,seq,dimreg,4point}), but we leave that
QCD calculation for the future.

There are several aspects of this paper that are related to
recent work by Beuf \cite{Beuf1,Beuf2} and H\"anninen, Lappi, and
Paatelainen \cite{LaP,HLaP} on next-to-leading-order
deep inelastic scattering (NLO DIS).
Here, making use of LCPT,
we will study the combination of
$1{\to}3$ splitting $e \to e \bar e e$ and UV-renormalized
one-loop corrections to
$1{\to}2$ splitting ($e \to \gamma e$ or $\gamma \to e\bar e$) in
the presence of a thick medium (though the formalism we use can
in principle handle more general situations).
In NLO DIS,
Beuf and H\"anninen et al.\ instead used LCPT to study the combination of
$1{\to}3$ splitting $\gamma^* \to q \bar q g$
and UV-renormalized one-loop corrections to
$1{\to}2$ splitting ($\gamma^* \to q \bar q$) in
the presence of an extremely {\it thin} medium.
We briefly comment further on the similarities and
dissimilarities in appendix \ref{app:smallx}.

The goal of this paper is to develop calculational methods.
We leave discussion of application of the results to computing
IR-safe characteristics of energy loss and shower development to
a later paper \cite{qedNfstop}.


\subsection{Overlapping formation times
  in large-\boldmath$\Nf$ QED}

\begin {figure}[t]
\begin {center}
  \begin{picture}(400,120)(0,0)
  \put(0,20){\includegraphics[scale=0.5]{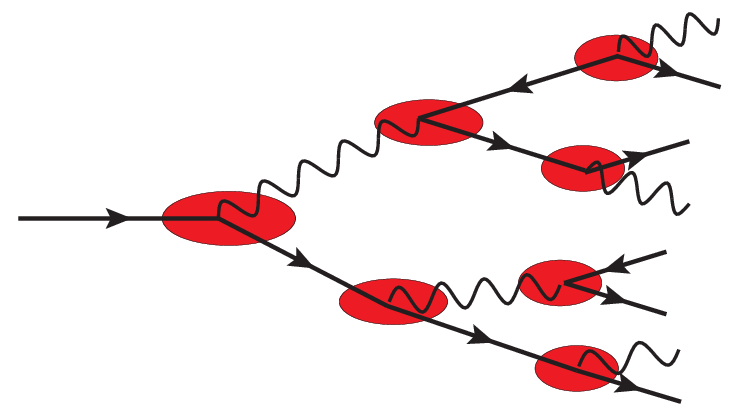}}
  \put(75,5){(a)}
  \put(230,15){\includegraphics[scale=0.5]{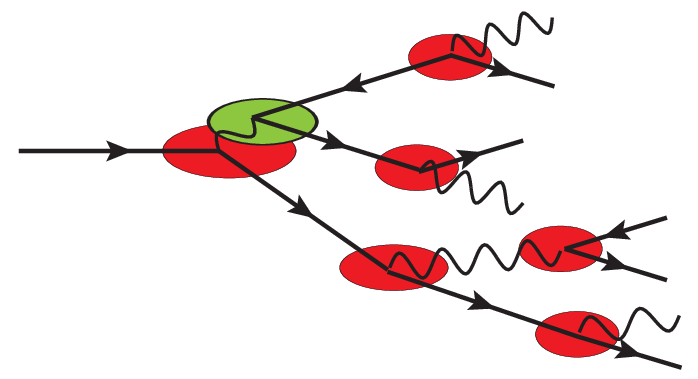}}
  \put(305,5){(b)}
  \end{picture}
  \caption{
     \label{fig:qedshower}
     Schematic picture of in-medium QED showers of an initial high energy
     electron.  The ovals represent LPM formation lengths for each splitting.
     In (a), splittings are independent and do not interfere with
     each other.  In (b), two splittings have overlapping
     formation lengths and so may not be treated as independent.
     Showering is extremely collinear in the high energy limit, but
     the transverse directions in these schematic drawings have been
     drastically magnified for the purpose of illustration.
  }
\end {center}
\end {figure}

Fig.\ \ref{fig:qedshower} depicts the showering of a high-energy
particle in a medium.  For simplicity of discussion, let's focus for
the moment on roughly-democratic splitting processes, meaning that
neither daughter of any splitting is soft (i.e.\ neither
has very small energy compared to the parent).
Let's also first discuss the case of QED with only one electron
flavor ($\Nf=1$).
If the coupling associated with the splitting vertices is small,
then there is a hierarchy of scale in fig.\ \ref{fig:qedshower} between
(i) the formation length $l_{\rm form}$ associated with each splitting
(denoted by the length of the ovals) and (ii) the mean free path
$l_{\rm rad}$ between consecutive roughly-democratic splittings.
Parametrically,
\begin {equation}
   l_{\rm rad} \sim \frac{l_{\rm form}}{\alpha} \,.
\label {eq:lrad}
\end {equation}
A rough mnemonic for this result is that each formation length of media
traversed offers one opportunity for splitting, with probability of
order $\alpha$.  On average it then takes $\sim 1/\alpha$ such opportunities
to radiate.

Because of this scale hierarchy, the probability for two consecutive
splittings to have overlapping formation times, as depicted in
fig.\ \ref{fig:qedshower}b, is suppressed by a factor of $\alpha$.
To leading order in $\alpha$, one may treat the splittings in showers
as non-overlapping, as depicted in fig.\ \ref{fig:qedshower}a,
and so may treat the
probabilities of each splitting in the shower as independent.
The goal of the current program is to understand how to calculate
(beyond soft-emission approximations) the $\alpha$-suppressed effect of
overlapping formation times, such as the overlapping pair of
splittings in fig.\ \ref{fig:qedshower}b.

\begin {figure}[tp]
\begin {center}
  \includegraphics[scale=0.6]{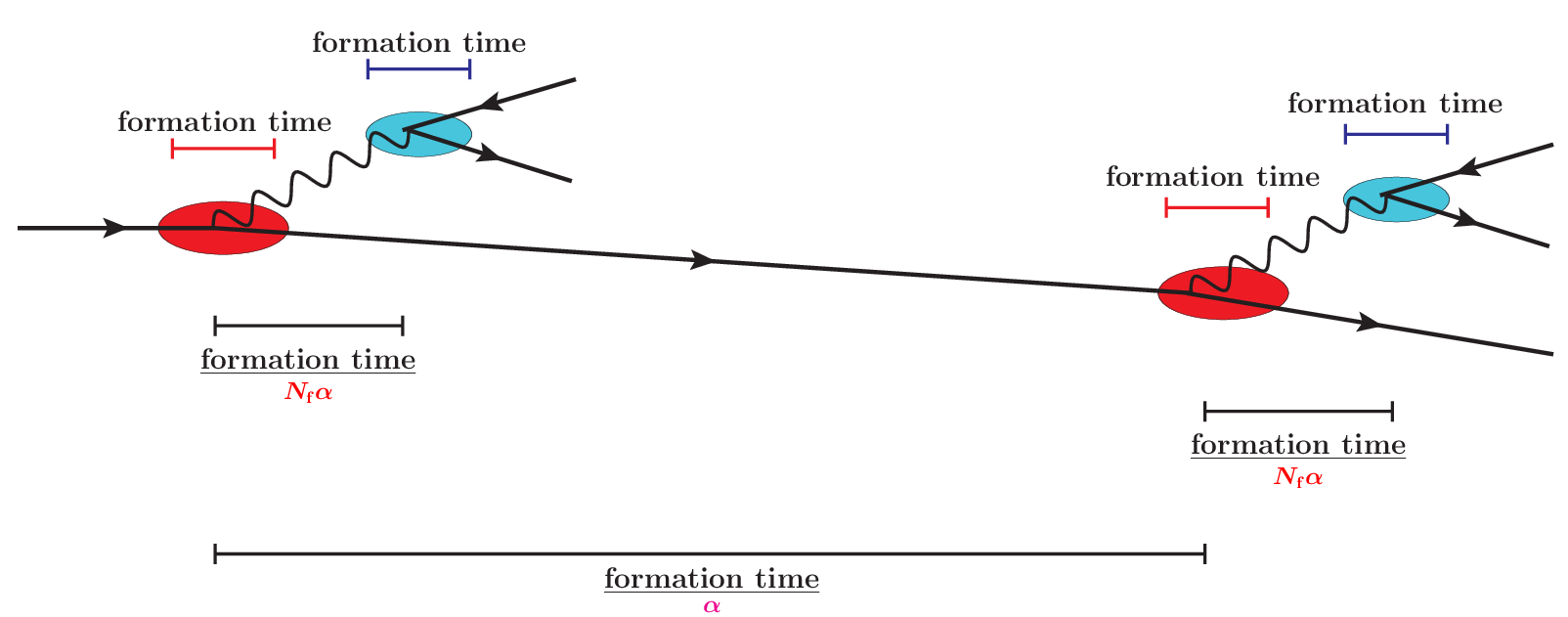}
  \caption{
     \label{fig:typicalNf}
     Parametric summary of relative size of typical formation lengths
     and distances between splittings for large-$\Nf$ QED when
     $\alpha \ll \Nf\alpha \ll 1$.  In this picture, splittings
     are assumed to be roughly democratic, i.e.\ we are not depicting
     the case of soft daughters.  We have not bothered to show
     the part of the shower representing the subsequent evolution of
     pair-produced electrons and positrons,
     but they behave similarly to the evolution of the original
     electron.
  }
\end {center}
\end {figure}

\begin {figure}[tp]
\begin {center}
  \includegraphics[scale=0.7]{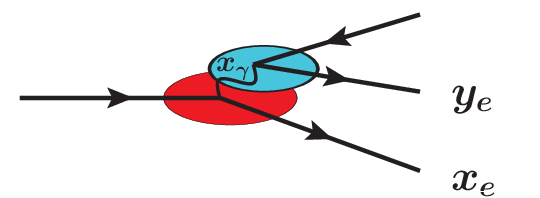}
  \caption{
     \label{fig:overlapNf}
     The dominant type of overlap correction when
     $\alpha \ll \Nf\alpha \ll 1$.  (Also shown here is the
     convention we will use later to label
     the longitudinal momentum fractions $\xe$ and $\ye$
     of the two final-state electrons relative to the initial
     electron.  The corresponding fraction of the intermediate photon is
     $x_\gamma = 1{-}\xe$.)
  }
\end {center}
\end {figure}

So far, the qualitative discussion has been the same as the
discussion for QCD in the introduction of ref.\ \cite{2brem}.
But now consider the large-$\Nf$ limit, which will reduce the
number of calculations needed in this paper.
If the properties of the medium are held constant, then
$\Nf$ does not affect the rate for bremsstrahlung $e \to \gamma e$.
Formation times are also not affected.
But the rate for pair production $\gamma \to e\bar e$ is proportional
to $\Nf$, and so the mean free path for pair production is smaller
than for bremsstrahlung by a factor of $\Nf$.
The hierarchy of scales relevant to typical showering is then summarized
by fig.\ \ref{fig:typicalNf},
assuming that $\Nf$ is large but $\Nf\alpha$ is still small:%
\footnote{
  We will not attempt it here, but it may well be possible to sum
  all orders of $\Nf\alpha$ for $\alpha \ll 1$, similar in spirit
  to the
  analysis of transport and pressure in refs.\ \cite{LargeNfTransport}
  and \cite{LargeNfPressure}.
}
\begin {equation}
   \alpha \ll \Nf\alpha \ll 1 .
\label {eq:sizes}
\end {equation}
The probability that two of the closer splittings in the figure
($e \to \gamma e \to e \bar e e$) might overlap will be of
order $\Nf\alpha$.  That's overlap of (i) bremsstrahlung with (ii) the
subsequent pair production later initiated by the bremsstrahlung photon.
The chance of any other type of overlap, such as two consecutive
bremsstrahlung processes $e \to \gamma e \to \gamma\gamma e$, is
only of order $\alpha$ and so is parametrically less likely in the
large-$\Nf$ limit.  So, to get the dominant correction due to
overlapping formation lengths in this problem under the formal
assumption (\ref{eq:sizes}), the only type of
overlap we need to compute is the type shown in
fig.\ \ref{fig:overlapNf} (plus related virtual processes).

\begin {figure}[tp]
\begin {center}
  \includegraphics[scale=0.43]{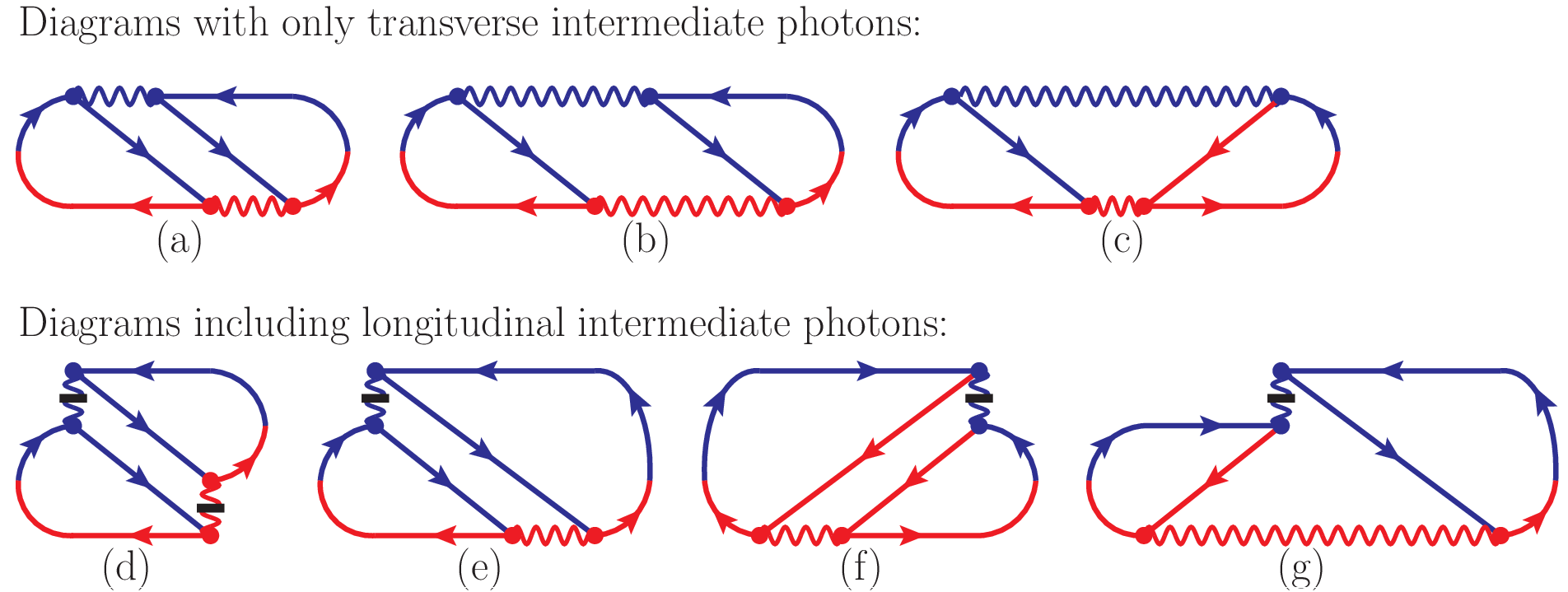}
  \caption{
     \label{fig:diags}
     Time-ordered interference diagrams for
     $e \to e \bar e e$ in large-$\Nf$ QED.
     As in refs.\ \cite{2brem,seq},
     blue represents a contribution to
     the amplitude and red represents a contribution to the conjugate
     amplitude.  As in the other figures of this paper, repeated
     interactions with the medium are present but not explicitly shown.
     The complex conjugates of the above interference diagrams should also be
     included by taking $2\Re[\cdots]$ of the above.
  }
\end {center}
\end {figure}

\begin {figure}[tp]
\begin {center}
  \includegraphics[scale=0.43]{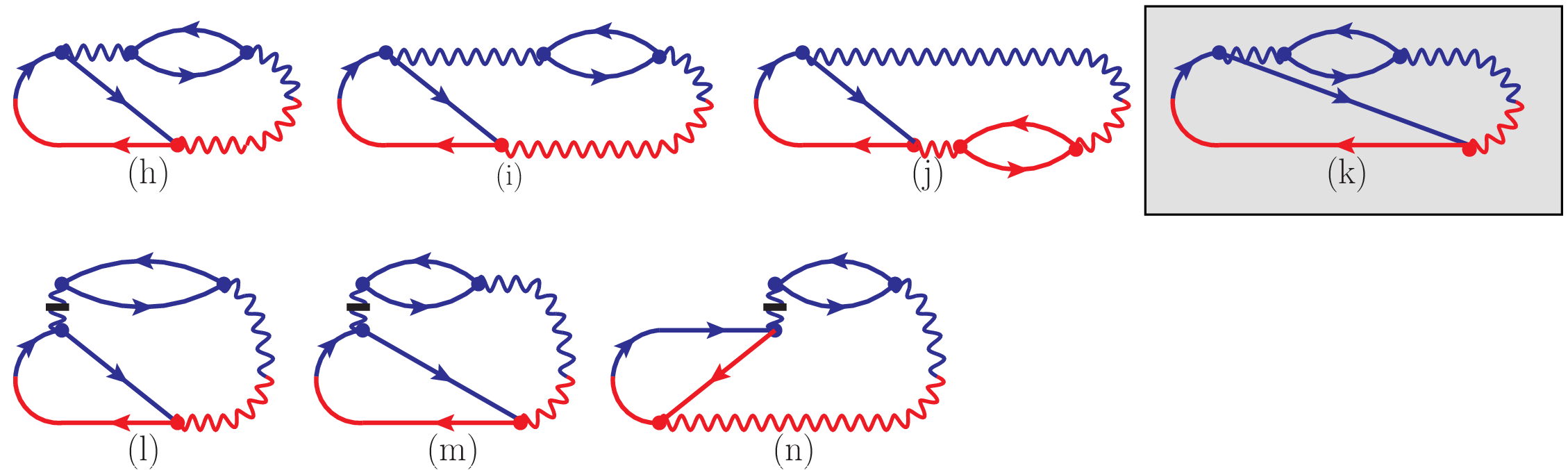}
  \caption{
     \label{fig:diagsVIRT}
     Time-ordered interference diagrams for the virtual correction to
     $e \to \gamma e$ in large-$\Nf$ QED.
     The boxed diagram is the only one whose result
     cannot be simply related
     to one of the $e \to e\bar e e$ diagrams of
     fig.\ \ref{fig:diags}.
     Again, complex conjugates should be included by taking $2\Re[\cdots]$.
  }
\end {center}
\end {figure}

\begin {figure}[tp]
\begin {center}
  \includegraphics[scale=0.43]{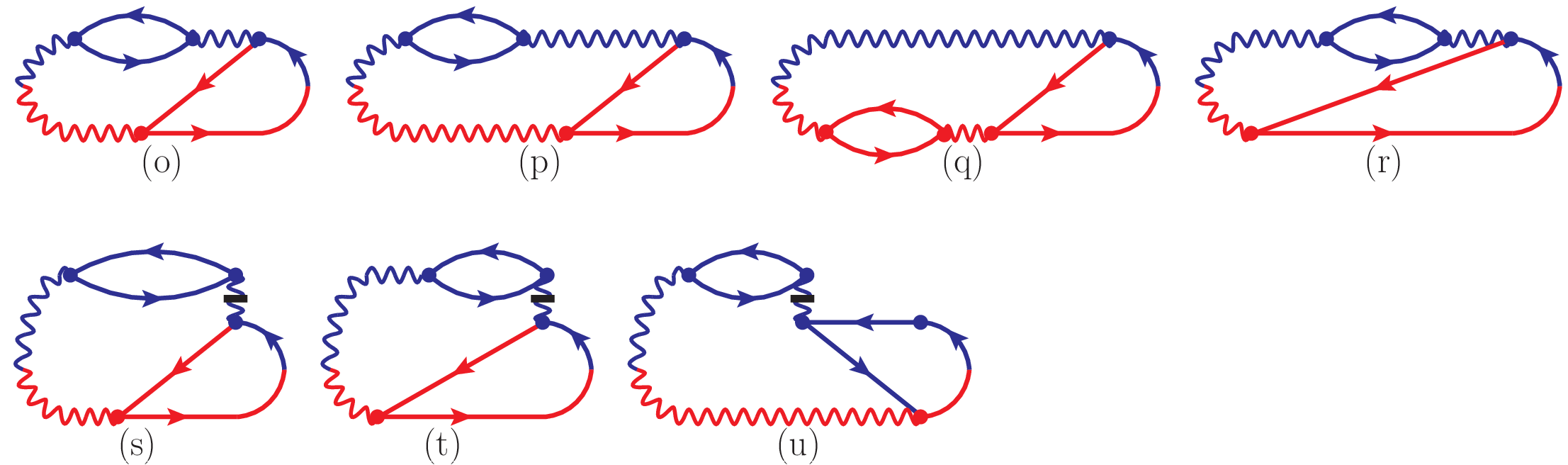}
  \caption{
     \label{fig:diagsVIRT2}
     Time-ordered interference diagrams for the virtual correction to
     $\gamma \to e\bar e$ in large-$\Nf$ QED.
     All of these diagrams can be related to one of the diagrams of
     figs.\ \ref{fig:diags} or \ref{fig:diagsVIRT}.
  }
\end {center}
\end {figure}

\begin {figure}[tp]
\begin {center}
  \includegraphics[scale=0.43]{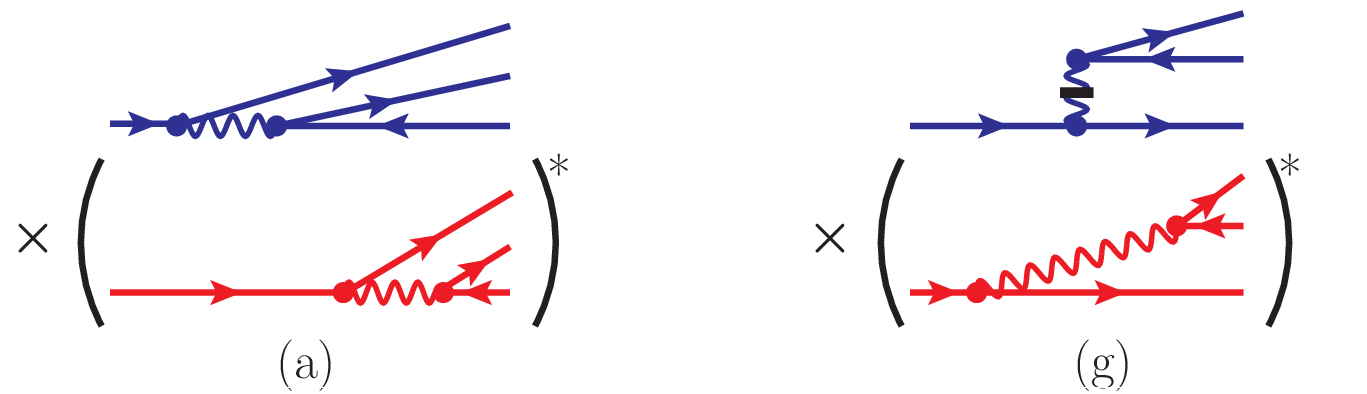}
  \caption{
     \label{fig:interference}
     The interferences
     represented by figs.\ \ref{fig:diags}a and \ref{fig:diags}g.
     These interferences are still time-ordered.  For example,
     in (a), the vertex times in the conjugate amplitude are
     restricted to be larger than those in the amplitude, as depicted.
  }
\end {center}
\end {figure}

The interference diagrams needed to calculate the effect of overlapping
formation times in large-$\Nf$ QED are shown in
figs.\ \ref{fig:diags}--\ref{fig:diagsVIRT2}.
These diagrams are drawn using the conventions of
refs.\ \cite{2brem,seq} and represent contributions to the
{\it rate} for the process.
Each diagram shows the product of
a term in the amplitude for the process (colored blue in the diagram)
times a term in the conjugate amplitude (colored red).
The diagrams are time-ordered (more precisely in this paper,
light-cone time ordered), with time running from left to right.
We will only consider rates which are integrated over
the transverse momenta of the final particles, which allows one to
ignore the evolution of any final-state particle after it has
been emitted in {\it both}\/ the amplitude and conjugate amplitude.%
\footnote{
  See the discussion in section IV.A of ref. \cite{2brem} and
  appendix F of ref.\ \cite{seq}.
}
So, for instance, the diagrams of fig.\ \ref{fig:diags}a and
\ref{fig:diags}g represent
the ($\p_\perp$-integrated) interferences of fig.\ \ref{fig:interference}.

We will see later that the results for most of the virtual diagrams
can be related to results for $e \to e \bar e e$ (fig.\ \ref{fig:diags}),
or to each other.  There will only be one quintessential virtual diagram
that we will have to compute from scratch, which will turn out to
be the boxed diagram shown in fig.\ \ref{fig:diagsVIRT}k.
That diagram, and the related diagram of fig.\ \ref{fig:diagsVIRT2}r,
contain the only true UV divergence in these large-$\Nf$ calculations%
\footnote{
  One simplification of the large-$\Nf$ limit is that we do not have
  one-loop fermion self energies nor one-loop vertex corrections,
  and so we need not also compute the divergent diagrams
  associated with those.
}
and will be the diagrams responsible for the usual
renormalization of the QED coupling $\alpha$.

\begin {figure}[tp]
\begin {center}
  \includegraphics[scale=0.6]{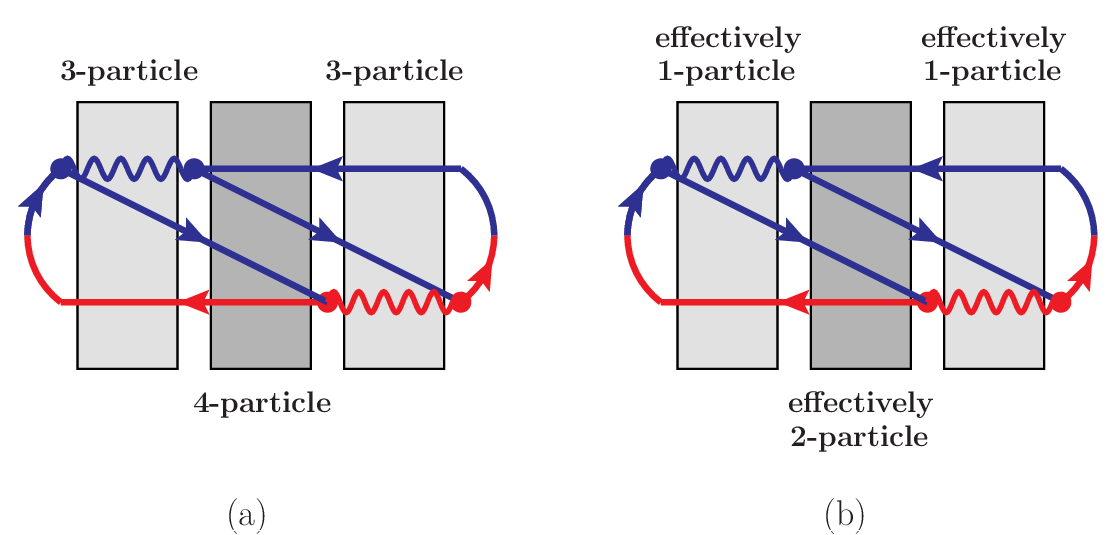}
  \caption{
     \label{fig:interp}
     (a) Interpretation of the interference diagram
     as simultaneous evolution in time of the particles in the amplitude
     and conjugate amplitude (all interacting with the medium, and
     averaged over the randomness of the medium).  The labels
     denote how many high-energy particles need to be evolved
     between splitting vertices.
     (b) Figure (a) relabeled according to the effective number of
     particles needed after symmetries are used to reduce the problem
     \cite{2brem}.
  }
\end {center}
\end {figure}

In the approach of our earlier work \cite{2brem,seq,dimreg,4point}, inspired by
Zakharov's treatment of the LPM effect \cite{Zakharov}, we
interpret the interference diagram of fig.\ \ref{fig:diags}a, for
instance, as the evolution of an initial $e \bar e$ pair, where the $e$
represents the initial electron in the amplitude and the $\bar e$
represents the same particle in the conjugate amplitude.
In this language, the time evolution of fig.\ \ref{fig:diags}a is
interpreted, as shown in fig.\ \ref{fig:interp}a,
as a phase of 3-particle evolution ($\gamma e$ in the amplitude
and $\bar e$ in the conjugate amplitude), followed by 4-particle
evolution ($e \bar e e$ in the amplitude and $\bar e$ in the conjugate
amplitude), followed by 3-particle evolution ($e\bar e$ in the amplitude
and $\gamma$ in the conjugate amplitude).  We then used symmetries
of the problem to replace each medium-averaged $N$-particle
evolution problem by an effectively $(N{-}2)$-particle evolution problem,%
\footnote{
  For a full discussion of this reduction,
  see section III of ref.\ \cite{2brem}.
  Let $z$ be the direction of motion of the initial particle.
  In the high-energy limit,
  the propagation of the $N$ particles can be formulated as a
  two-dimensional non-relativistic Schr\"odinger problem
  in the transverse ($xy$) plane, with longitudinal momenta $p_{zi}$
  playing the role of the ``masses'' in the Schr\"odinger equation. 
  One of the symmetries is transverse translation invariance
  (over the small transverse distance
  scales probed by the splitting), which means that one can
  eliminate one degree of freedom by taking out the uninteresting
  ``center of mass'' motion, reducing the problem to effectively
  $N{-}1$ (two-dimensional) degrees of freedom.
  The other symmetry is that of small 3-dimensional
  rotations which change the direction of the $z$ axis by a
  very tiny amount (and so preserve $|\p_\perp| \ll p_z$ for all
  the particles).  For our particular problem, this symmetry
  allows one to remove one {\it additional}\/ degree of freedom.
  This is not obvious; refer to ref.\ \cite{2brem}
  for details.  The specific case of reducing 3 particles to 1 particle
  was implicit to the analysis of BDMPS \cite{BDMPS12,BDMPS3} and
  Zakharov \cite{Zakharov}, though presented in different language.
}
as summarized in fig.\ \ref{fig:interp}b.
One of our tasks in the current paper will be to translate the
LCPT diagrammatic rules for longitudinal photon interactions into
corresponding rules within our framework.

Before going forward, we should mention that formally there are some
additional virtual diagrams in Light Cone
Perturbation Theory, which will be shown and discussed later 
(fig.\ \ref{fig:diagsEXTRA}) but which are negligible in the
high-energy limit provided we use dimensional regularization.


\subsection{Assumptions and Approximations}

As with the earlier work in refs.\ \cite{2brem,seq,dimreg,4point}, the
formalism we will present is quite general, but for now we will implement
it in simple cases that make the actual calculations much easier.

First, we will assume that the medium is large, static, and uniform.
More precisely, we approximate the medium as (statistically)
uniform and unchanging over the scale
of formation lengths/times.

Second, we will make the ``multiple scattering approximation'' that
interactions with the medium can be characterized by the
parameter $\hat q$ that is often used in discussions of quark-gluon
plasmas, and which is the proportionality constant in the relationship
$\langle Q_\perp^2 \rangle = \hat q \, \Delta z$, where $Q_\perp$ is the
transverse momentum that a high-energy charged particle picks up
passing through length $\Delta z$ of the medium.
For a variety of reasons, the effective value of
$\hat q$ can have logarithmic dependence on
energy at fixed order in the coupling $\alpha$ that controls
high-energy splittings.  There are important qualitative
differences between QCD and QED, which we briefly review in
appendix \ref{app:qhat}, but it is off the main topic of this
paper.  Here, we will make the approximation that $\hat q$ is a
constant and assume that one is using a value of $\hat q$ appropriate
for the overall energy scale of the initial particle.


\subsection{Some Qualitative Results}
\label {sec:Results}

This paper further develops methods for
next-to-leading order calculations of the LPM effect and gives
analytic results for the case of large-$\Nf$ QED.
The details of the full analytic results are complicated enough
that we will not present them here in the introduction.
And we are mostly leaving numerical analysis and application
of those results to a later paper.  But there are two features
of our results which we will present here.

The first feature regards the appropriate choice of
renormalization scale for the factor of $\alpha$ that controls
the cost of high-energy splitting and determines the overall
importance of overlapping formation times in showering---that is,
the scale of the $\alpha$ associated with each
high-energy splitting vertex.
In earlier work \cite{2brem}, we asserted based on qualitative
physical reasoning that this coupling
should be taken to be $\alpha(Q_\perp)$, where $Q_\perp$ is the
characteristic transverse momentum scale of the LPM-modified
splitting process.  Our explicit next-to-leading-order LPM results
bear this out.  For roughly-democratic splitting processes,
we find that the logarithmic terms in our NLO results for single
splitting are%
\footnote{
   Specifically, see eq.\ (\ref{eq:xyyxRen2}) later in this paper,
   specializing here to the roughly-democratic case
   where neither $\xe$ nor $1{-}\xe$ are small.
   We have not drawn from our results
   any conclusions about the best renormalization
   scale in limiting cases such as $1{-}\xe \ll 1$, because then
   the logarithmic corrections due to choice of renormalization scale are
   overwhelmed by other, power-law corrections which dominate.
   Those power-law corrections are the same parametric size as
   (\ref{eq:yintParametric}).
}
\begin {equation}
   \left[ \frac{d\Gamma}{dx} \right]_{e\to\gamma e}
   =
   \left[ \frac{d\Gamma}{dx} \right]_{e\to\gamma e}^{\rm LO}
   \left[
     1 + \beta_0 \alpha \ln\Bigl( \frac{|E\Omega|^{1/2}}{\mu} \Bigr)
     + \cdots
   \right] ,
\end {equation}
where $\mu$ is the renormalization scale for $\alpha(\mu)$,
$\beta_0$ is the first coefficient of the renormalization group
$\beta$-function for $\alpha$, and $|\Omega|$ is a frequency that
is of order the inverse formation time.  To avoid
poor convergence of the perturbative expansion due to
large logarithms, one should therefore choose
$\mu \sim |E \Omega|^{1/2}$ above.
As we review in section \ref{sec:QEDvQCD}, this is indeed equivalent
(for roughly-democratic splittings) to $\mu \sim Q_\perp$.

\begin {figure}[t]
\begin {center}
  \includegraphics[scale=0.5]{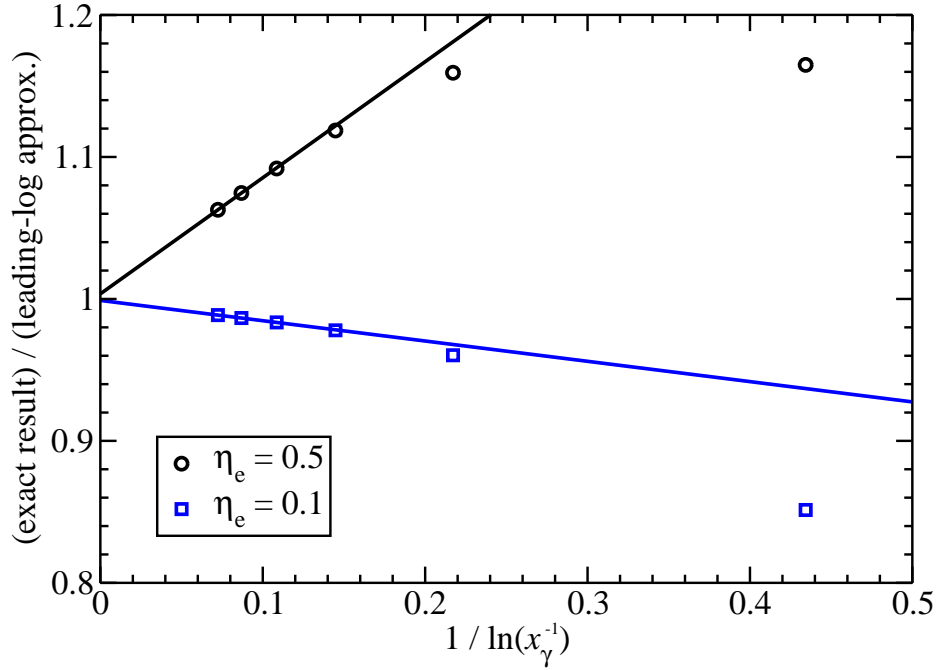}
  \caption{
     \label{fig:softN}
     A plot demonstrating the success of the leading-log approximation
     (\ref{eq:soft}) as $x_\gamma{\to}0$.  The vertical axis is the
     ratio of full numerical results to the leading-log approximation,
     and the horizontal axis is the inverse logarithm
     $1/\ln(x_\gamma^{-1})$.  The data points
     come from numerical calculations of our general $e \to e\bar e e$
     results, given in appendix \ref{app:real}.  The extrapolation to
     $x_\gamma{\to}0$ is taken by a straight-line fit to the leftmost two
     data points of each set.
     Extrapolations are shown for two different values of $\yfrake$.
  }
\end {center}
\end {figure}

The other feature regards our result for the
effect of overlapping formation times on the
real double splitting process $e \to \gamma e \to e\bar e e$.
Though our analytic result is complicated for the general case,
there is a simple leading-log formula in the limit
that the intermediate photon is soft.  That formula can be derived
with a relatively simple analysis based on (i) the leading-order
LPM formula for pair production $\gamma \to e\bar e$ combined
with (ii) Dokshitzer-Gribov-Lipatov-Altarelli-Parisi
(DGLAP) evolution of parton distributions to get the
probability of seeing the photon $\gamma$ inside the initial
electron $e$.  We explain this analysis in section \ref{sec:QEDvQCD},
which yields the formula
\begin {equation}
   \left[ \Delta \, \frac{d\Gamma}{dx_\gamma\,d\yfrake} \right]_{e \to e\bar e e}
   \approx
   \frac{\Nf \alpha^2 [\yfrake^2+(1{-}\yfrake)^2]}
        {2\pi^2 [\yfrake(1{-}\yfrake)]^{1/2} x_\gamma^{3/2}}
   \ln \Bigl( \frac{1}{x_\gamma [\yfrake(1{-}\yfrake)]^{1/2}} \Bigr)
   \sqrt{ \frac{\hat q}{E} }
   \qquad
   (x_\gamma \ll 1) ,
\label {eq:soft}
\end {equation}
where $x_\gamma$ is the longitudinal momentum fraction of the
intermediate photon relative to the initial electron in
the bremsstrahlung $e \to \gamma e$, and
$\yfrake$ is the longitudinal momentum fraction of the
final electron relative to the photon in the subsequent
pair production $\gamma \to e\bar e$.
The $\approx$ sign in (\ref{eq:soft}) is our notation for indicating
that this formula is only valid at leading-log order.
The $\Delta$ in front of $d\Gamma/dx_\gamma\,d\yfrake$ is our notation
\cite{seq} to denote the effect of overlapping formation times on
rates for double splitting.  That is, $\Delta [d\Gamma/dx_\gamma\,d\yfrake]$
is the difference between (i) $d\Gamma/dx_\gamma\,d\yfrake$ and
(ii) what one would
get by always treating the two consecutive medium-induced
splittings $e \to \gamma e$ and $\gamma \to e\bar e$ as quantum-mechanically
independent.  Fig.\ \ref{fig:softN} verifies the leading-log
approximation by
showing that the ratio of our full numerical results for $e \to e\bar e e$
divided by the approximation (\ref{eq:soft}) extrapolates to 1 as
$x_\gamma{\to}0$.  The convergence is slow because logarithms
grow slowly.

As we will discuss in section \ref{sec:QEDvQCD}, there are two
interesting qualitative differences between QED $e\to e\bar e e$ and
QCD $g \to ggg$: (i) the $x_\gamma{\ll}1$ behavior (\ref{eq:soft})
will not give rise to any infrared (IR) divergences in energy-loss calculations,
and so needs no IR cancellation between real and virtual diagrams,
and (ii) the in-medium collinear logarithmic enhancement factor
in (\ref{eq:soft}) does not cancel between real double-splitting
diagrams as it did for $g \to ggg$ \cite{seq} via a
Gunion-Bertsch cancellation.%
\footnote{
  Specifically, see appendix B of ref.\ \cite{seq}.
}
This reflects a difference between soft pair production and soft
bremsstrahlung.

Even though QED is qualitatively different from QCD in
these respects, it nonetheless provides a good training ground for
working out calculational methods for overlap effects in QCD.
And overlap effects in QED are interesting in their own right,
even if so far only treated here in the large-$\Nf$ limit.


\subsection{Outline}

We have tried to organize this paper so that the main text gives
an introduction and overview of the techniques we use while the
fine details of the calculation are left to an extensive set of
appendices.
In section \ref{sec:QEDvQCD}, we summarize qualitative differences
between QCD and (large-$\Nf$) QED regarding the LPM effect in
both single-splitting and overlapping double-splitting processes.
A quantitative review of single-splitting formulas is left
to appendix \ref{app:LO}, and appendix \ref{app:qhat}
contains a translation of
modern notation using $\hat q$ to the QED
results originally presented by Migdal \cite{Migdal}.
Section \ref{sec:LCPT} introduces the elements needed for
organizing calculations
of overlapping formation-time effects in terms of
light-cone perturbation theory (LCPT), with the full list of
details left to appendix \ref{app:VertexRules}.
The use of those rules to calculate overlap effects in the
real double-splitting process $e{\to}e\bar e e$
of fig.\ \ref{fig:diags} is
left to appendix \ref{app:real}, where much of the calculation
is adapted from our previous work on $g{\to}ggg$ in QCD.
In section \ref{sec:virt}, we turn to the virtual corrections
of figs.\ \ref{fig:diagsVIRT} and \ref{fig:diagsVIRT2}.
We first discuss the techniques needed, which we call
back- and front-end transformations, to
easily relate almost all virtual diagrams to non-virtual diagrams.
We then turn to the one remaining virtual diagram
(the boxed diagram of fig.\ \ref{fig:diagsVIRT}k)
and outline its computation and renormalization
but leave details for appendices \ref{app:Fund1}
and \ref{app:bbI}.%
\footnote{
  If, like us, you think it's really fun to figure out how to do a
  new type of dimensionally-regularized integral, then appendix
  \ref{app:bbI} is the appendix for you.
}
Finally, we end with a brief conclusion in section \ref{sec:conclusion}.
A complete summary of the next-to-leading order LPM rate formulas derived
in this paper is given in Appendix \ref{app:summary}.


\subsection{Reference Acronyms}

When discussing detailed formulas in appendices and sometime footnotes,
we will often need to refer to particular sections or equations of
our earlier work \cite{2brem,seq,dimreg,4point}.
To streamline such references, we will often refer
to our earlier work in such cases by the author acronyms and numbers
AI1 \cite{2brem}, ACI2 \cite{seq}, ACI3 \cite{dimreg}, and ACI4 \cite{4point}.
So, for instance, ``ACI3 (3.4)'' will be shorthand for eq.\ (3.4) of
ref.\ \cite{dimreg}, and ``AI1 section II.A'' will be shorthand for
section II.A of ref.\ \cite{2brem}.


\section{Qualitative differences between QCD and QED}
\label {sec:QEDvQCD}

\subsection {Single splitting}

The crucial difference to keep in mind between the LPM effect in QCD
and QED is that the photon is neutral but the gluon has color charge.
In terms of parametric behavior, this makes a huge difference in
how LPM suppression behaves when a bremsstrahlung photon or gluon is soft.
The LPM effect relies on the near-collinearity of high-energy
splitting processes, and the formation time $l_{\rm form}$ (and
so the amount of LPM rate suppression) is smaller when the splitting
is less collinear.  Since it is easier to deflect a low-momentum
particle than a high-momentum particle, the collinearity of
QCD gluon bremsstrahlung is controlled by how soft the
bremsstrahlung gluon is: high-energy
gluon bremsstrahlung is {\it less} LPM-suppressed the softer
the bremsstrahlung gluon is.
A soft photon, however, does not scatter in first approximation,
and so the collinearity of QED photon bremsstrahlung is insensitive
to this.  But a softer photon means a longer-wavelength photon,
which means a photon with less resolving power, which is why
the QED case has the opposite behavior: high-energy
photon bremsstrahlung is {\it more} LPM-suppressed the softer
the photon is.

\begin {figure}[tp]
\begin {center}
  \includegraphics[scale=0.6]{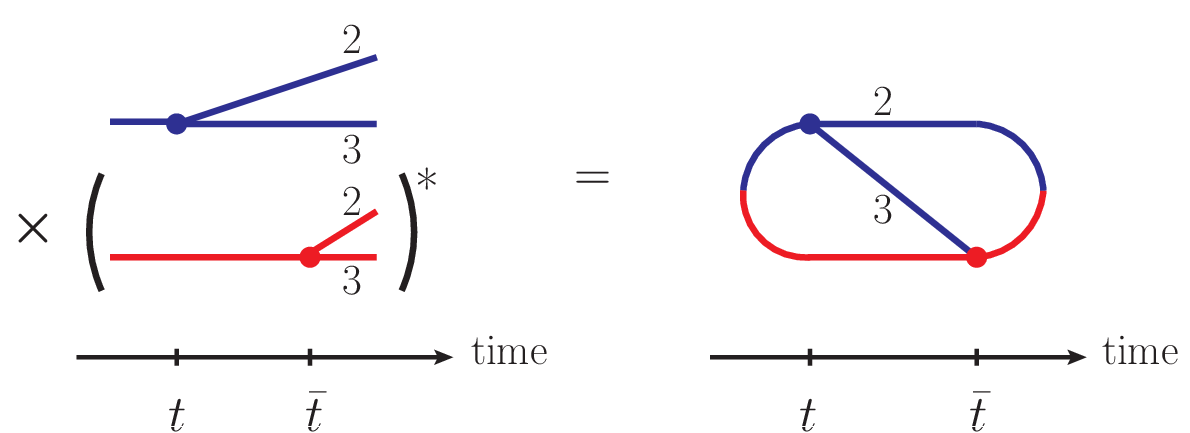}
  \caption{
     \label{fig:LOschematic}
     Schematic (time-ordered) interference diagram for a leading-order
     single-splitting process  $E \to x_2E$ and $x_3 E$, where here the
     lines can be any particle type.
  }
\end {center}
\end {figure}


\subsubsection {Formation Times}

A little more concretely, let's review, in a general way,
some standard parametric estimates of
formation times.  Consider a single splitting process of
a high-energy particle with energy $E$ into daughters of energy
$E_2{=}x_2E$ and $E_3{=}x_3 E$ and ask for the formation time.  That's the
time scale $|t-\bar t|$ over which it's possible for splitting in the
amplitude at time $t$ to interfere with splitting in the conjugate
amplitude at time $\bar t$, as depicted by the interference
diagram in fig.\ \ref{fig:LOschematic}.  That time scale
$t_{\rm form}$ is determined by the difference $\delta E$
of the energies of (i) the
two-particle state $|2,3\rangle$ after splitting in the amplitude and
(ii) the one-particle state before splitting in the conjugate amplitude:
\begin {equation}
   \delta E = (E_{\p_2} + E_{\p_3}) - E_{\p_1}
   \simeq
   - \frac{\p_1^2 + m_1^2}{2 E}
   + \frac{\p_2^2 + m_2^2}{2 E_2}
   + \frac{\p_3^2 + m_2^2}{2 E_2} ,
\label {eq:deltaE}
\end {equation}
where $\p_i$ are the transverse momenta, which we take to be small
compared the energies $E_i$.
In this paper, we will assume throughout that energies
are high enough that we can ignore the effective masses $m_i$ of
the high-energy particles in the medium.  One can then make a rough
parametric estimate $t \sim 1/|\delta E|$ of the formation time by using
the definition of $\hat q$ to take
$\p_i^2 \sim \hat q_i t$ above, where $\hat q_i$ is the $\hat q$ appropriate
for particles of type $i$.  Then, from (\ref{eq:deltaE}),
\begin {equation}
  t \sim \frac{1}{|\delta E|} \sim
  \left[ - \frac{\hat q_1 t}{2E} + \frac{\hat q_2 t}{2x_2 E}
         + \frac{\hat q_3 t}{2x_3 E} \right]^{-1}
\end {equation}
and so
\begin {equation}
  t_{\rm form}
  \sim \sqrt{ \frac{E}{-\hat q_1 + (\hat q_2/x_2) + (\hat q_3/x_3)} }
  \,.
\label {eq:tformGeneral}
\end {equation}
For photon bremsstrahlung $e \to \gamma e$, we have $\hat q_\gamma{=}0$,
and (\ref{eq:tformGeneral}) gives
\begin {equation}
  t_{\rm form}^{e\to\gamma e}
  \sim
  \sqrt{ \frac{E}{-\hat q + (\hat q/\xe)} }
  = \sqrt{ \frac{\xe E}{(1{-}\xe)\hat q} }
  = \sqrt{ \frac{(1{-}x_\gamma) E}{x_\gamma\hat q} }
  \,,
\label {eq:tQED}
\end {equation}
where $\hat q \equiv \hat q_e$.
Above, $\xe$ and $x_\gamma$ (which in this case is $x_\gamma = 1{-}\xe$)
are respectively the longitudinal momentum fractions
of the electron and photon daughters of the splitting.
In contrast, for gluon bremsstrahlung $q \to qg$ or $g \to gg$,
we have $\hat q_q \sim \hat q_g$ and
\begin {equation}
  t_{\rm form}^{X \to g X}
  \sim
  \sqrt{ \frac{E}{-\hat q_X + (\hat q_g/x_g) + (\hat q_X/(1{-}x_g)))} }
  \sim \sqrt{ \frac{x_g(1{-}x_g)E}{\hat q_{\rm QCD}} }
  \,.
\label {eq:tQCD}
\end {equation}
The parametric estimates (\ref{eq:tQED}) and (\ref{eq:tQCD})
show that the photon formation time becomes long (more LPM suppression)
as $x_\gamma \to 0$ whereas the gluon formation time becomes short
(less LPM suppression) as $x_g \to 0$.

We've used the very general formula (\ref{eq:tformGeneral}) to
derive standard parametric estimates of LPM formation times
because it is then easy in the same breath to obtain the
standard parametric result for QED pair production
$\gamma \to e \bar e$, which is qualitatively different from
photon bremsstrahlung.  (\ref{eq:tformGeneral}) gives
\begin {equation}
  t_{\rm form}^{\gamma\to e\bar e}
  \sim
  \sqrt{ \frac{E}{\bigl(\hat q/(1-\xe)\bigr) + \bigl(\hat q/\xe\bigr)} }
  = \sqrt{ \frac{\xe (1{-}\xe) E}{\hat q} }
  \,.
\label {eq:tpair}
\end {equation}
The QCD pair production formation time ($g \to q\bar q$)
is parametrically similar
to this, as well as to
the QCD bremsstrahlung formation time
(\ref{eq:tQCD}).


\subsubsection {Splitting Rates}

Below, we will need the differential rates $d\Gamma/dx$ for single
splitting.  For nearly-democratic splittings, the relation given
by (\ref{eq:lrad}) is that $\Gamma \sim \alpha/t_{\rm form}$.
But we will also be interested in non-democratic splittings (one daughter
soft) and will want to discuss the differential rate
$d\Gamma/dx$, where $x$ is one of the daughter momentum fractions.
The differential rate is sensitive to the DGLAP splitting function
for the process:
\begin {equation}
   \frac{d\Gamma^{1\to 23}}{dx} \sim
   \frac{\alpha P_{1\to23}(x)}{t_{\rm form}^{1\to23}} \,,
\end {equation}
which gives
\begin {equation}
   \left[ \frac{d\Gamma}{dx} \right]_{\rm brem}
      \sim \frac{\alpha}{x_{\gamma,g} t_{\rm form}}
\label {eq:dGestBrem}
\end {equation}
for bremsstrahlung ($e \to \gamma e$ or $X \to g X$, with
$x_g$ being the softest gluon in the case $g\to gg$) and
\begin {equation}
   \left[ \frac{d\Gamma}{dx} \right]_{\rm pair}
      \sim \frac{\Nf \alpha}{t_{\rm form}}
\label {eq:dGestPair}
\end {equation}
for pair production ($\gamma\to e\bar e$ or $g\to q\bar q$).

For a review of precise, quantitative formulas for single splitting
rates (in the high energy limit where the $\hat q$ approximation can be
used), see appendix \ref{app:LO}.
Appendix \ref{app:qhat} discusses exactly how,
in the QED case, these formulas match up to the original results presented
by Migdal \cite{Migdal} in the case where the medium is an atomic
gas.


\subsubsection {An aside on nearly-democratic splitting}

We mention in passing that QED and QCD are parametrically
similar in the case of nearly-democratic splitting
(neither daughter soft compared to the parent).
All the formation times discussed above then have the same order of
magnitude:
\begin {equation}
  t_{\rm form} \sim \sqrt{ \frac{E}{\hat q} }
  \qquad
  (\mbox{nearly-democratic splitting}) .
\end {equation}
(The splitting rates are then also all parametrically the same
except for factors
of $\Nf$.)
The amount of transverse momentum $Q_\perp$ transferred from the medium
in one formation time is of order
$Q_\perp^2 \sim \hat q t_{\rm form}$, which can be written in a number
of equivalent ways:
\begin {equation}
  Q_\perp \sim (\hat q t_{\rm form})^{1/2}
  \sim (\hat q E)^{1/4}
  \sim \Bigl( \frac{E}{t_{\rm form}} \Bigr)^{1/2}
  \qquad
  (\mbox{nearly-democratic splitting})
  .
\label {eq:QperpEst}
\end {equation}
We used the last form to identify $\mu \sim |E\Omega|^{1/2}$
with $Q_\perp$ in the discussion of renormalization scale in
section \ref{sec:Results}.

One could of course also write down case-by-case parametric estimates
of $Q_\perp$ for {\it non}-democratic splittings, but we do not have
need of them.


\subsection {Overlapping double splitting}

In ref.\ \cite{seq} (ACI2),%
\footnote{
  Specifically ACI2 section I.D.
}
we discussed how to parametrically
estimate the size (though not the sign)
of QCD overlapping formation-time effects for real double splitting
$g \to ggg$
by using formation times to estimate rates.  Here we will first review
that QCD estimate and then see what changes when we switch to
(large-$\Nf$) QED.


\subsubsection {Review of QCD estimate for $g \to ggg$}

Let $yE$ and $xE$ be the energies of the softest and next-softest of the three
final-state gluons, so that $y \lesssim x \lesssim 1{-}x{-}y$.
Since this is QCD, the formation time $t_{{\rm form},y}$
for radiating the $y$ gluon will be shorter than the formation time
$t_{{\rm form},x}$ for radiating the $x$ gluon.  The probability that
a $y$ emission happens to take place {\it during} the $x$ emission
is then just $t_{{\rm form,x}}$ times the rate of $y$ emission.
Overall, that means that the joint differential rate for overlapping
$x$ and $y$ emissions is
\begin {equation}
  \left[ \frac{d\Gamma}{dx\,dy} \right]_{\rm overlap}
  \sim
  \frac{d\Gamma_x}{dx} \times \frac{d\Gamma_y}{dy} \, t_{{\rm form},x} .
\label {eq:est1}
\end {equation}
Using (\ref{eq:dGestBrem}), that's
\begin {equation}
  \left[ \frac{d\Gamma}{dx\,dy} \right]_{\rm overlap}
  \sim
  \frac{\alphas^2}{xy t_{{\rm form},y}}
\end {equation}
and thence, from (\ref{eq:tQCD}),
\begin {equation}
  \left[ \frac{d\Gamma}{dx\,dy} \right]_{\rm overlap}
  \sim
  \frac{\alphas^2}{xy^{3/2}} \sqrt{\frac{\hat q}{E}} \,.
\label {eq:QCDest}
\end {equation}
The fact that two emissions overlap does not {\it a priori} mean
they will  influence each other, and so the qualitative argument for
(\ref{eq:QCDest}) only provides an estimate for how large
overlap effects {\it might} be.
However, detailed calculations \cite{seq} of overlap effects for
$g \to ggg$ confirm it.

Note that if one multiples (\ref{eq:QCDest})
by the energy $(x+y)E$ lost by the
leading parton and integrates over $x$ and $y$, one would find a
power-law divergent contribution from overlap effects to energy loss.
No such power-law divergence appears in the soft-gluon bremsstrahlung
calculations of ref.\ \cite{Blaizot,Iancu,Wu}, but those calculations
inextricably combine the effects of soft virtual emission with those
of soft real emission, for which there are cancellations in QCD.
Here we have only estimated the size of soft real emission alone.


\subsubsection {QED estimate for $e \to e\bar e e$}

For large-$\Nf$ QED we've already identified that the dominant overlap
correction comes from the
$e \to e\bar e e$ process of fig.\ \ref{fig:overlapNf}.
Let $\xe E$ and $\ye E$ be the energies of the two final-state electrons.
As shown in the figure,
our convention throughout this paper will be that $\xe$ is the
daughter whose electron line is connected to the initial-state electron,
and $\ye$ is the electron in the $e\bar e$ pair produced by the
intermediate photon.  This is a distinction made possible by
the large-$\Nf$ limit, in which the chance that the electron in the
$\gamma \to e\bar e$
pair has the same flavor as the initial electron (which would
allow interference terms involving exchange of the two electrons)
is $1/\Nf$ suppressed.
In this section, we will focus on the case $\xe \sim 1$, which includes
the case $x_\gamma \to 0$ but not $\xe \to 0$.  (Our later explicit
calculations of diagrams make no such assumption.)

The formation time for the initial bremsstrahlung $e \to \gamma e$ can be
taken from (\ref{eq:tQED}):
\begin {equation}
  t_{{\rm form},x} \sim \sqrt{ \frac{E}{x_\gamma\hat q} }
  \qquad (\xe\sim 1) .
\label {eq:txQED}
\end {equation}
To get the formation time for the subsequent pair production process
$\gamma \to e\bar e$, we have to be careful applying the
single splitting estimate (\ref{eq:tpair}) because we are considering
the case where the photon is already soft.
The $E$ in (\ref{eq:tpair}) is the photon energy, which we are
now calling $x_\gamma E$.  The $\xe$ in (\ref{eq:tpair}) is the
longitudinal momentum fraction of the pair's electron relative to the photon,
which in the application here (see fig.\ \ref{fig:overlapNf}) is
\begin {equation}
   \yfrake \equiv \frac{\ye E}{x_\gamma E} = \frac{\ye}{1-\xe} \,.
\label {eq:yfrake}
\end {equation}
With these substitutions, (\ref{eq:tpair}) gives
\begin {equation}
   t_{{\rm form},y} \sim
   \sqrt{ \frac{\yfrake (1{-}\yfrake) x_\gamma E}{\hat q} }
\label {eq:typair}
\end {equation}
in the present context.  As in the QCD estimate, this is $\lesssim$
$t_{{\rm form},x}$, and so we can estimate the probability of
overlap the same way, using (\ref{eq:est1}), which can also be written
\begin {equation}
  \left[ \frac{d\Gamma}{dx\,d\yfrake} \right]_{\rm overlap}
  \sim
  \frac{d\Gamma_x}{dx} \times \frac{d\Gamma_y}{d\yfrake} \, t_{{\rm form},x} .
\end {equation}
Here we need to
make sure to use the {\it pair}-production formula (\ref{eq:dGestPair})
to get $d\Gamma/d\yfrake \sim \Nf\alpha/t_{{\rm form},y}$.
So, using (\ref{eq:typair}) above,
\begin {equation}
  \left[ \frac{d\Gamma}{d\xe\,d\yfrake} \right]_{\rm overlap}
  \sim
  \frac{\Nf\alpha^2}{x_\gamma t_{{\rm form},y}}
  \sim
  \frac{\Nf\alpha^2}{x_\gamma^{3/2} [\min(\yfrake,1{-}\yfrake)]^{1/2}}
    \sqrt{\frac{\hat q}{E}}
  \qquad (\xe\sim1) .
\end {equation}
Equivalently, using (\ref{eq:yfrake}),
\begin {equation}
  \left[ \frac{d\Gamma}{d\xe\,d\ye} \right]_{\rm overlap}
  \sim
  \frac{\Nf\alpha^2}{x_\gamma^{3/2} [\min(\ye,x_\gamma{-}\ye)]^{1/2}}
    \sqrt{\frac{\hat q}{E}} \,.
\end {equation}

Unlike the QCD case, the integral over $\ye$ does not diverge:
\begin {equation}
  \left[ \frac{d\Gamma}{d\xe} \right]_{\rm overlap}^{e \to e\bar e e}
  \sim
  \frac{\Nf\alpha^2}{x_\gamma^{3/2}}
  \int_0^1 \frac{d\yfrake}{[\min(\yfrake,1{-}\yfrake)]^{1/2}}
  \sim
  \frac{\Nf\alpha^2}{x_\gamma^{3/2}}
  \qquad (\xe\sim1) .
\label {eq:yintParametric}
\end {equation}
Furthermore, were we to use this formula to calculate the overlap effects
on energy loss, we would get finite results, even without accounting
(as one still should) for virtual corrections.
QED has much better infrared behavior than QCD for these calculation because
the LPM suppression of photon radiation increases as the photon becomes
softer.



\subsubsection {Leading-log formula for $e \to e\bar e e$}

Before moving on to the details of our complete calculation
of overlap effects in the general case, we first
briefly discuss a relatively simple argument for
the leading-log formula (\ref{eq:soft}) for $x_\gamma \ll 1$.
As mentioned earlier, the existence of such a logarithm is
another difference between $e \to e\bar e e$ and $g \to ggg$.

\begin {figure}[t]
\begin {center}
  \includegraphics[scale=0.7]{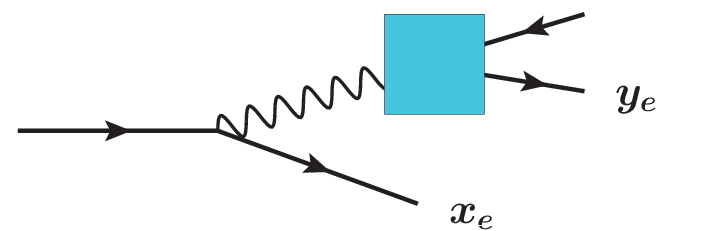}
  \caption{
     \label{fig:soft}
     The double splitting process $e \to \gamma e \to e\bar e e$ viewed
     in an approximation (relevant to overlap effects in the limit
     $x_\gamma \ll 1$) treating it as fundamentally
     pair production $\gamma \to e\bar e$ with the $\gamma$ 
     interpreted
     as a parton inside the original electron, described by DGLAP
     evolution.
  }
\end {center}
\end {figure}

Consider the $e \to e\bar e e$ process of
fig.\ \ref{fig:overlapNf} and re-interpret it as
the Feynman diagram shown in fig.\ \ref{fig:soft}.  In the latter figure,
we have not explicitly drawn
the (still overlapping) ovals corresponding to formation times.
Instead, the box here denotes the leading-order pair-production
process $\gamma \to e\bar e$ via interaction with the medium,
and we will consider that photon $\gamma$ as a parton contained
inside the initial electron.  In a related context, ref.\ \cite{seq}
(ACI2) argues that the corresponding parton-model-like (i.e.\ DGLAP)
approximation to such a process has the generic form%
\footnote{
  See specifically ACI2 eqs.\ (B5--B6) in ACI2 appendix B.1 \cite{seq}.
}
\begin {equation}
   \frac{d\Gamma}{dx\, dy}
   \approx
   \frac{\alpha}{2\pi} \, P(1{-}x) \,
   \ln\Bigl( \frac{Q^2}{Q_0^2} \Bigr) \,
   \frac{d\Gamma}{d y}
   \approx
   \frac{\alpha}{2\pi} \, P(1{-}x) \,
   \ln\Bigl( \frac{t_{{\rm form},x}}{t_{{\rm form},y}} \Bigr) \,
   \frac{d\Gamma}{d y} \,,
\label {eq:soft1}
\end {equation}
where $d\Gamma/dy$ is the differential rate for the
fundamental parton-level process (represented here by the
box in fig.\ \ref{fig:soft}) and here the rest of (\ref{eq:soft1})
represents the probability of finding the
photon inside the initial electron.  The analysis of that
probability is different from what it would be in vacuum
because scattering from the medium cuts off collinear logarithms.
[See ref.\ \cite{seq} for the argument that the
DGLAP logarithm $\ln(Q^2/Q_0^2)$ 
of the virtuality ratio translates to the logarithm
$\ln(t_{{\rm form},x}/t_{{\rm form},y})$ of the formation-time ratio.]
The translation of (\ref{eq:soft1}) to fig.\ \ref{fig:soft} is then
\begin {equation}
   \frac{d\Gamma}{d\xe\, d\ye}
   \approx
   \frac{\alpha}{2\pi} \, P_{e\to e}(\xe) \,
   \ln\Bigl( \frac{t_{{\rm form},x}}{t_{{\rm form},y}} \Bigr) \,
   \left[ \frac{d\Gamma}{d\ye} \right]_{\gamma \to e\bar e}^{\rm LO}
\end {equation}
or equivalently
\begin {equation}
   \frac{d\Gamma}{dx_\gamma\, d\yfrake}
   \approx
   \frac{\alpha}{2\pi} \, P_{e\to\gamma}(x_\gamma) \,
   \ln\Bigl( \frac{t_{{\rm form},x}}{t_{{\rm form},y}} \Bigr) \,
   \left[ \frac{d\Gamma}{d\yfrake} \right]_{\gamma \to e\bar e}^{\rm LO}
   .
\label {eq:soft2}
\end {equation}
The parametric estimate of $d\Gamma/d\yfrake$ was given earlier as
$\sim \Nf\alpha/t_{{\rm form},y}
 \sim \Nf\alpha \sqrt{\hat q/\yfrake(1{-}\yfrake)E}$,
but here we want the exact leading-order result.
That's reviewed in appendix \ref{app:LO} and (using
the same translations as in our parametric analysis) is
\begin {equation}
   \left[ \frac{d\Gamma}{d\yfrake} \right]_{\gamma \to e\bar e}^{\rm LO}
   =
   \frac{\Nf\alpha}{\pi} P_{\gamma\to e}(\yfrake)
      \, \Re(i \Omega_0^{\gamma\to e\bar e})
\end {equation}
here, with
\begin {equation}
   \Omega_0^{\gamma\to e\bar e}
    = \sqrt{ \frac{-i \hat q}{2\yfrake(1{-}\yfrake) E_\gamma} }
    = \sqrt{ \frac{-i \hat q}{2\yfrake(1{-}\yfrake) x_\gamma E} } \,.
\label {eq:Omega0pair}
\end {equation}
Combining (\ref{eq:soft2}--\ref{eq:Omega0pair}) with the
formation times (\ref{eq:txQED}) and (\ref{eq:typair}), using the
explicit formulas for the DGLAP splitting functions
[see (\ref{eq:Pee}) and (\ref{eq:Pge})], and taking the limit
$x_\gamma \ll 1$ (and hence $\xe \to 1$) yields the
leading-log approximation (\ref{eq:soft}) quoted earlier
and verified from our full calculation of overlap effects.


\section{Light Cone Perturbation Theory for Overlapping LPM}
\label {sec:LCPT}

\subsection {Diagrammatic Rules and their translation}
\label {sec:vertex}

\begin {figure}[tp]
\begin {center}
  \includegraphics[scale=0.43]{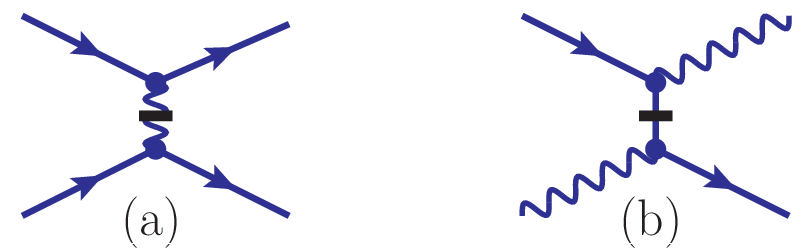}
  \caption{
     \label{fig:LCPT}
     (a) Depiction of the interactions that arise in QED after integrating out
     longitudinal-polarized photons in light-cone gauge.
     (b) Similar interaction arising in LCPT from integrating out
     unphysical fermion states.
  }
\end {center}
\end {figure}

\begin {figure}[tp]
\begin {center}
  \includegraphics[scale=0.43]{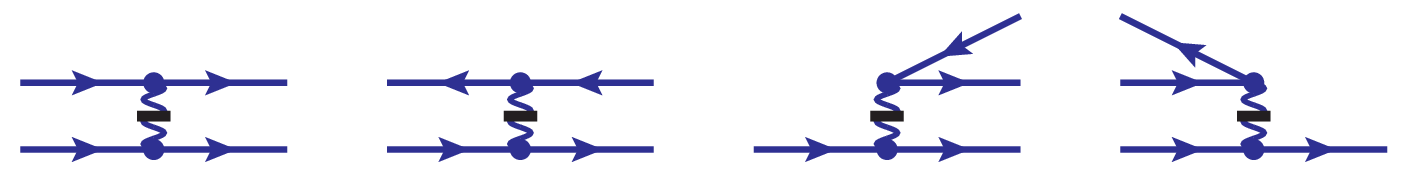}
  \caption{
     \label{fig:LCPTh}
     Various light-cone time ordered versions of
     fig.\ \ref{fig:LCPT}a, appropriate for Hamiltonian formalism.
     Light-cone time runs from left to right.
  }
\end {center}
\end {figure}

Our earlier work did not include the effects of
longitudinally-polarized gauge bosons in intermediate states.  One may
retain a description solely in terms of transverse polarizations by
first integrating out the longitudinal polarizations in light-cone
gauge, as in Light Cone Perturbation Theory.  This gives rise to
4-fermion interactions such as shown in fig.\ \ref{fig:LCPT}a,
which are  instantaneous in light-cone time $x^+$, local in transverse
position ${\bm x}^\perp$, and non-local in $x^-$.
In Hamiltonian formalism
(as in LCPT), the legs can be incoming or outgoing, such as shown in
fig.\ \ref{fig:LCPTh} for fig.\ \ref{fig:LCPT}a.
In LCPT, a similar 4-particle interaction arises from fermion exchange,
corresponding to fig.\ \ref{fig:LCPT}b.
Loosely speaking, this interaction term
accounts for the difference between using (i) the actual
off-shell exchanged fermion and (ii) treating that fermion as though
it had on-shell polarization $u_p$ or $v_p$.
An important feature of LCPT is that fig.\ \ref{fig:LCPT} for QED
(and similar 4-field diagrams for QCD) is all that is needed to
account for the effects of longitudinal gauge bosons or
off-shell fermion polarizations:
there are no $n$-field interaction terms in the light-cone
Hamiltonian with $n>4$.

\begin {figure}[t]
\begin {center}
  \begin{picture}(468,85)(0,0)
  \put(0,55){
    \begin{picture}(220,30)(0,0)
    \put(0,0){\includegraphics[scale=0.5]{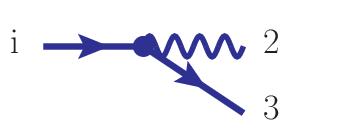}}
    \put(100,12){$
      \langle 2,3 | \delta H | \ix \rangle_{\rm rel} =
      -g \bar u_3 \slashed{\varepsilon}_2^* u_\ix
    $}
    \put(-1,-1){\framebox(222,32){}}
    \end{picture}
  }
  \put(0,0){
    \begin{picture}(470,45)(0,0)
    \put(-2,0){\includegraphics[scale=0.5]{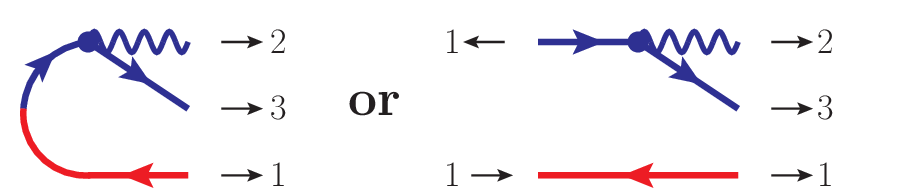}}
    \put(216,30){$
      \langle \P | {-}i\,\delta H | \rangle =
      -i \frac{\rm above}
              {\sqrt{|2 E_1| |2 E_2| |2 E_3|}}
      = \frac{i g}{2 E^{3/2}} \, \bcalP_{e\to\gamma e}\cdot\P_{31}
    $}
    \put(216,5){$
      \langle \B | {-}i\,\delta H | \rangle
      = \frac{g}{2 E^{3/2}} \, \bcalP_{e\to\gamma e}\cdot
        \grad \delta^{(2)}(\B_{31})
    $}
    \end{picture}
  }
  \end{picture}
  \caption{
     \label{fig:VertexEtoGE}
     The box shows the Hamiltonian matrix element for $e \to \gamma e$,
     as in LCPT,
     normalized with relativistic normalization as in \cite{KL}.
     The next line shows the corresponding formula, in the normalization
     and conventions of AI1 \cite{2brem}, for an initial splitting
     $e \to \gamma e$ in the amplitude (as opposed to conjugate amplitude),
     like the left-most (earliest time) vertex in fig.\ \ref{fig:interp}b
     or in figs.\ \ref{fig:diags}(a--c) and \ref{fig:diagsVIRT}(h-k).
     As written, the latter formulas only apply to $2{\to}3$ particle
     transitions in the language of fig.\ \ref{fig:interp}a, which is
     equivalent to effectively $0{\to}1$ particle transitions in
     the language of fig.\ \ref{fig:interp}b.  Note that these formulas
     show the matrix element of $-i\,\delta H$ rather than $\delta H$,
     according to the convention of AI1 \cite{2brem}.  [The origin of
     the $-i$ factor is the $-i$ in the evolution operator
     $e^{-i H t}$ for amplitudes.]
  }
\end {center}
\end {figure}

\begin {figure}[t]
\begin {center}
  \begin{picture}(468,130)(0,0)
  \put(0,85){
    \begin{picture}(220,30)(0,0)
    \put(0,0){\includegraphics[scale=0.5]{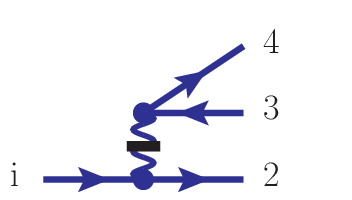}}
    \put(100,19){$
      \langle 2,3,4 | \delta H | \ix \rangle_{\rm rel} =
      \frac{g^2
        (\bar u_2 \gamma^+ u_\ix)
        (\bar u_4 \gamma^+ v_3)
      }
      {(p_3^+ + p_4^+)^2}
    $}
    \put(-1,-1){\framebox(270,45){}}
    \end{picture}
  }
  \put(0,0){
    \begin{picture}(470,70)(0,0)
    \put(-2,0){\includegraphics[scale=0.5]{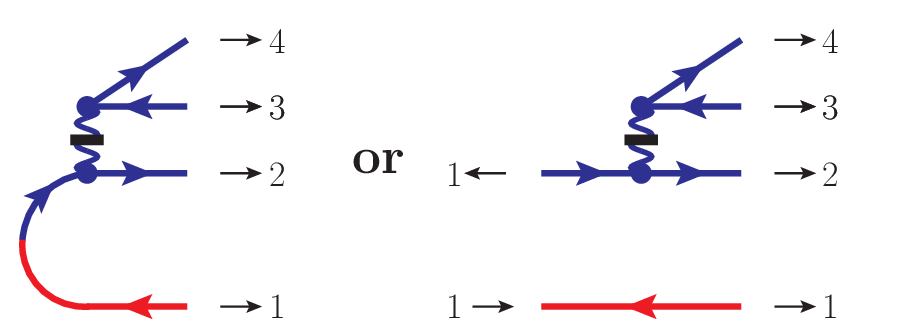}}
    \put(216,58){$
      \langle \P_{34},\P_{12} | {-}i\,\delta H | \rangle =
      -i
      \frac{\rm above}{\sqrt{|2 E_1| |2 E_2| |2 E_3| |2 E_4|}}
      \,
      |x_3{+}x_4|^{-1}
    $}
    \put(350,35){$
      = \frac{-ig^2}{|x_3+x_4|^3 E^2}
    $}
    \put(216,7){$
      \langle \B_{34},\B_{12} | {-}i\,\delta H | \rangle
      = \frac{-ig^2}{|x_3+x_4|^3 E^2} \,
        \delta^{(2)}(\B_{34}) \, \delta^{(2)}(\B_{12})
    $}
    \end{picture}
  }
  \end{picture}
  \caption{
     \label{fig:VertexEtoEEE}
     Like fig.\ \ref{fig:VertexEtoGE} but for an example of an
     ``instantaneous'' LCPT interaction, shown in the box.
     Here, in the language of fig.\ \ref{fig:interp}, the formulas
     outside of the box give an example of a $2{\to}4$ (effectively
     $0{\to}2$) transition.  Chirality is conserved following each
     fermion line through the vertex.
  }
\end {center}
\end {figure}

Figs.\ \ref{fig:VertexEtoGE} and \ref{fig:VertexEtoEEE} show two
examples of standard LCPT rules for vertices (shown inside the boxes)
and examples of their translation to corresponding vertices of
interference diagrams in the reduced-particle description used in our
earlier work \cite{2brem,4point}, such as depicted by
fig.\ \ref{fig:interp}b.  Other basic vertices are covered
in appendix \ref{app:VertexRules}.
In the language of fig.\ \ref{fig:interp}a,
the interference diagram elements, shown outside the boxes in
figs.\ \ref{fig:VertexEtoGE} and \ref{fig:VertexEtoEEE}, represent
vertices for $2{\to}3$ and $2{\to 4}$ particle transitions
respectively.  In the reduced-particle description of
fig.\ \ref{fig:interp}b, these become effectively $0{\to}1$ and
$0{\to}2$ transitions, with the states described by one
transverse momentum or position variable ($\P$ or $\B$) per effective
particle.

The first expression for $\langle \P|{-}i\delta H|\rangle$ in fig.\
\ref{fig:VertexEtoGE} shows the general rule for converting ordinary
matrix elements (the boxed formula) into the conventions of
AI1 \cite{2brem} for effectively $0{\to}1$ particle transitions.
The only change here to the $\delta H$ matrix element is that
the normalization conventions of AI1 \cite{2brem} include a
factor of $|2 E_n|^{-1/2}$ for each individual particle state.
Our convention is that $E_n = x_n E$,
where $E$ is the energy of the original high-energy parent at the
very start of the processes depicted in
figs.\ \ref{fig:diags}--\ref{fig:diagsVIRT2}, and where $x_n$ is
the longitudinal momentum fraction of particle $n$ with the convention
that $x_n$ is negative for particles in the conjugate amplitude,
so that $\sum_n x_n$ is always zero.

The final expression for $\langle \P|{-}i\delta H|\rangle$ in fig.\
\ref{fig:VertexEtoGE} shows what you get if you express
$\bar u_3 \slashed{\varepsilon}_2^* u_\ix$ in the same notation used in our
previous work \cite{2brem}.  Here
\begin {equation}
   \P_{ij} \equiv x_j \p_i - x_i \p_j ,
\label {eq:Pdef}
\end {equation}
where the $\p$'s represent transverse momenta.  For 3-particle
(effectively 1-particle) states, the $\P_{ij}$ are related by
momentum conservation:%
\footnote{
  For a discussion of (\ref{eq:Pdef}),
  (\ref{eq:Pcycle}) and (\ref{eq:Bdef}) in the
  context of our application and notation here, see specifically
  AI1 sections II.E and III \cite{2brem}.
}
\begin {equation}
  \P_{12} = \P_{23} = \P_{31} .
\label {eq:Pcycle}
\end {equation}
The $\bcalP$ are related to the square roots of helicity-dependent
DGLAP splitting functions.
Here,
\begin {equation}
  \bcalP_{\ix\to 2,3}(x_\ix{\to}x_2,x_3)
  =
  \frac{ ({\bm e}_x{\pm}i{\bm e}_y) }{ |x_\ix x_2 x_3| }
  \sqrt{ P_{\ix\to 23}(x_\ix{\to}x_2,x_3) }
  =
  \frac{ ({\bm e}_x{\pm}i{\bm e}_y) }{ |x_\ix x_2 x_3| }
  \sqrt{ |x_\ix| \, P_{\ix\to 23}(1{\to}z_2,z_3) }
  ,
\end {equation}
where $z_n \equiv x_n/x_\ix$ are the momentum fractions of the daughters
relative to their immediate parent,
$P_{\ix\to 23}(1{\to}z_2,z_3)$ is the helicity-dependent DGLAP splitting
function appropriate to the particle types and
helicities of the initial and final particles in the
splitting process, and the $\pm$ in the circular basis
vector ${\bm e}_x {\pm} i{\bm e}_y$ is chosen accordingly based
on those helicities.%
\footnote{
  See AI1 section IV.E and
  AI1 appendix C \cite{2brem}
  for more details of our conventions.
}

The Fourier conjugate of $\P_{ij}$ is, in our notation,
\begin {equation}  
  \B_{ij} \equiv \frac{\b_i - \b_j}{x_i+x_j} \,.
\label {eq:Bdef}
\end {equation}
The $\langle \B | {-}i\,\delta H |\rangle$ formula in fig.\
\ref{fig:VertexEtoGE} is simply the Fourier transform of the
$\langle \P | {-}i\,\delta H |\rangle$ formula.

Now turn to fig.\ \ref{fig:VertexEtoEEE}.
The first expression for $\langle \P_{34},\P_{12}|{-}i\,\delta H|\rangle$
there shows the general rule for converting ordinary
matrix elements (the boxed formula) into the conventions of
refs.\ \cite{2brem,4point} (AI1,ACI3) for instantaneous,
effectively $0{\to}2$ particle transitions.
In addition to the $|2 E_n|^{-1/2}$ factors, there is an additional
factor of $|x_3+x_4|^{-1}$ that is associated with the normalization
of the state $|\P_{34},\P_{12}\rangle$ that one obtains when reducing
from a 4-particle description, of the state just after the interaction,
to an effective 2-particle description.%
\footnote{
  See specifically AI1 section IV.D and AI1 appendix B \cite{2brem}.
  It is the same factor as in the bottom diagram of
  AI1 fig.\ 16.
}
There is nothing special here
about the pairing of the indices $1234$ in this
normalization factor; one could just as well have written
\begin {equation}
      \langle \P_{23},\P_{41} | {-}i\,\delta H | \rangle =
      -i
      \frac{\rm above}{\sqrt{|2 E_1| |2 E_2| |2 E_3| |2 E_4|}}
      \,
      |x_1+x_4|^{-1}
\end {equation}
in fig.\ \ref{fig:VertexEtoEEE}.
(Note that $|x_3{+}x_4|^{-1} = |x_1{+}x_2|^{-1}$ and
$|x_1{+}x_4|^{-1} = |x_2{+}x_3|^{-1}$ since
$\sum_n x_n = 0$.)
The only advantage to choosing $\langle\P_{34},\P_{12}|$
in fig.\ \ref{fig:VertexEtoEEE} is that then the $|x_3{+}x_4|^{-1}$ from
the normalization factor there
neatly combines with the $(x_3{+}x_4)^{-2}$ coming from the
denominator $(p_3^+{+}p_4^+)^{-2}$
in the boxed LCPT rule.
Finally, the $\langle \B_{34},\B_{12}|{-}i\,\delta H|\rangle$ formula
in fig.\ \ref{fig:VertexEtoEEE} is
just the Fourier transform of the
$\langle \P_{34},\P_{12}|{-}i\,\delta H|\rangle$ formula.

We need not state whether our convention for light-cone components
is $v^\pm = v^0 \pm v^3$ or $v^\pm = (v^0 \pm v^3)/\sqrt2$ because it
does not matter.
The formulas for matrix elements, such as in
fig.\ \ref{fig:VertexEtoEEE}, give the same result either way.

\begin {figure}[t]
\begin {center}
  \begin{picture}(468,290)(0,0)
  \put(0,235){
    \begin{picture}(470,53)(0,-8)
    \put(-2,-4){\includegraphics[scale=0.5]{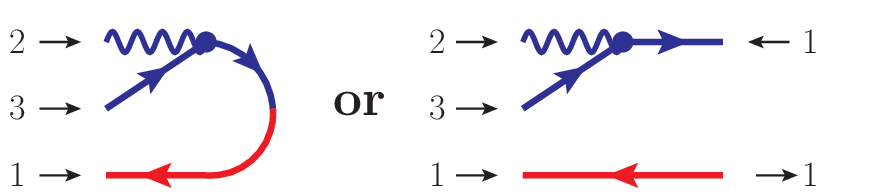}}
    \put(215,30){$
      \langle | {-}i\,\delta H | \P \rangle =
      -\langle \P | {-}i\,\delta H | \rangle^*
      = \frac{i g}{2 E^{3/2}} \, \bcalP_{e\to\gamma e}^*\cdot\P_{31}
    $}
    \put(215,9){$
      \langle | {-}i\,\delta H | \B \rangle
      = -\langle \B | {-}i\,\delta H | \rangle^*
    $}
    \put(250,-4){$
      = -\frac{g}{2 E^{3/2}} \, \bcalP_{e\to\gamma e}^*\cdot
        \grad \delta^{(2)}(\B_{31})
    $}
    \end{picture}
  }
  \put(0,160){
    \begin{picture}(470,53)(0,-8)
    \put(-2,-4){\includegraphics[scale=0.5]{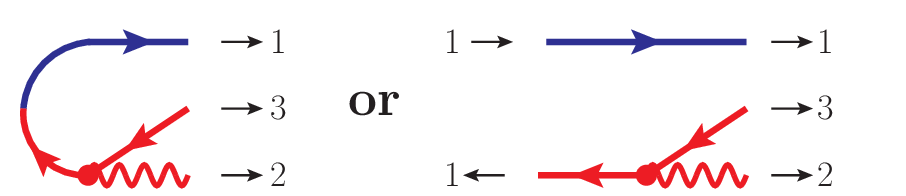}}
    \put(215,37){$
      \langle \P | {+}i\,\overline{\delta H} | \rangle =
      \langle \P | {-}i\,\delta H | \rangle^*_{(x,\p)\to(-x,-\p)}
    $}
    \put(250,22){$
      = -\frac{i g}{2 E^{3/2}} \, \bcalP_{e\to\gamma e}^*\cdot\P_{31}
    $}
    \put(215,3){$
      \langle \B | {+}i\,\overline{\delta H} | \rangle
      = \langle \B | {-}i\,\delta H | \rangle^*_{(x,\b)\to(-x,\b)}
    $}
    \put(250,-12){$
      = -\frac{g}{2 E^{3/2}} \, \bcalP_{e\to\gamma e}^*\cdot
        \grad \delta^{(2)}(\B_{31})
    $}
    \end{picture}
  }
  \put(0,85){
    \begin{picture}(470,53)(0,-8)
    \put(-2,-4){\includegraphics[scale=0.5]{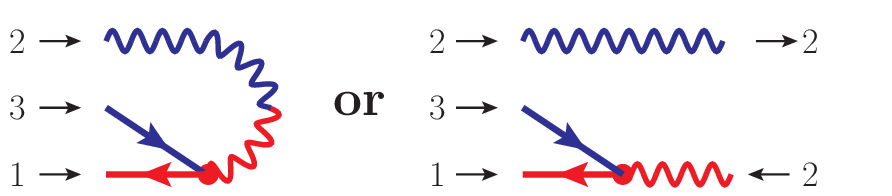}}
    \put(215,37){$
      \langle | {+}i\,\overline{\delta H} | \P \rangle =
      \langle \P | {-}i\,\delta H | \rangle^*
    $}
    \put(250,22){$
      = -\frac{i g}{2 E^{3/2}} \, \bcalP_{e\to\gamma e}^*\cdot\P_{31}
    $}
    \put(215,3){$
      \langle | {+}i\,\overline{\delta H} | \B \rangle
      = \langle \B | {-}i\,\delta H | \rangle^*
    $}
    \put(250,-12){$
      = \frac{g}{2 E^{3/2}} \, \bcalP_{e\to\gamma e}^*\cdot
        \grad \delta^{(2)}(\B_{31})
    $}
    \end{picture}
  }
  \put(0,0){
    \begin{picture}(470,53)(0,-18)
    \put(98,-14){\includegraphics[scale=0.5]{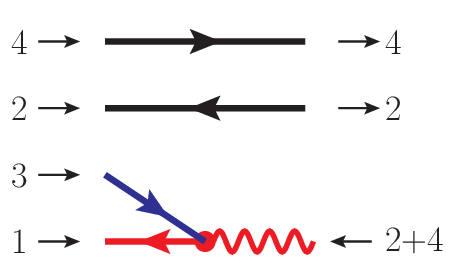}}
    \put(215,37){$
      \langle \P' | {+}i\,\overline{\delta H} | \P_{31},\P_{42} \rangle
    $}
    \put(250,22){$
      = {\rm previous} \times
        |x_1+x_3|^{-1} (2\pi)^2 \delta^{(2)}(\P'_{42}{-}\P_{42})
    $}
    \put(215,3){$
      \langle \B' | {+}i\,\overline{\delta H} | \B_{13},\B_{24} \rangle
    $}
    \put(250,-12){$
      = {\rm previous} \times |x_1+x_3|^{-1} \delta^{(2)}(\B'_{42}{-}\B_{42})
    $}
    \end{picture}
  }
  \end{picture}
  \caption{
     \label{fig:VertexEtoGEmore}
     Variations on fig.\ \ref{fig:VertexEtoGE} that are related by
     complex conjugation and sometimes, depending on the directions drawn for
     momentum flow, by reversal $(x_n,\p_n) \to (-x_n,-\p_n)$ of momentum
     variables (see Appendix \ref{app:VertexRules} for more detail).
     The overall factor of $|x_1{+}x_3|^{-1} = |x_2{+}x_4|^{-1}$ in the last
     diagram arises from
     the same effectively-2-particle state normalization
     factor for $|\P_{31},\P_{42}\rangle$ that was discussed earlier in
     the text for the $\langle\P_{34},\P_{12}|$ of fig.\ \ref{fig:VertexEtoEEE}.
     The two spectator lines in this case are colored black here to
     indicate that their color (red or blue) does not matter.
     [Note: The above vertex diagrams are best viewed in color.]
  }
\end {center}
\end {figure}

From formulas such as
figs.\ \ref{fig:VertexEtoGE} and \ref{fig:VertexEtoEEE}
for basic interactions,
many variations follow from substitutions and complex conjugation.
As an example,
the interference-diagram rule in fig.\ \ref{fig:VertexEtoGE}
also gives the additional interference-diagram rules shown in
fig.\ \ref{fig:VertexEtoGEmore}.  This procedure has the advantage
of definiteness, but one may also formulate a more generic rule that
covers all the different variations---see appendix
\ref{app:VertexRules}.

\begin {figure}[tp]
\begin {center}
  \includegraphics[scale=0.43]{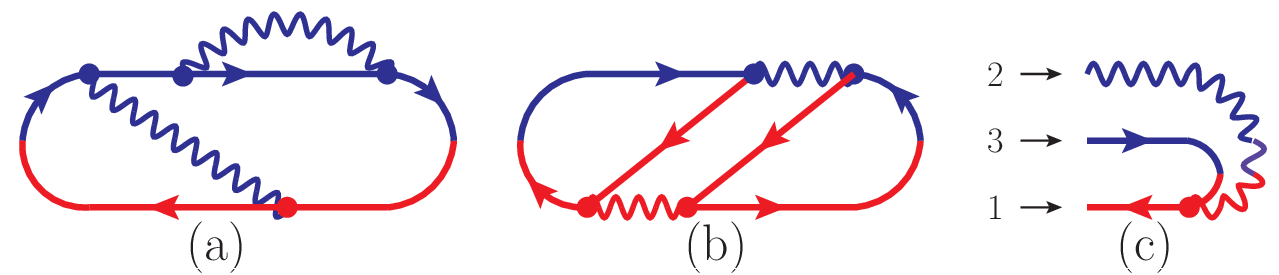}
  \caption{
     \label{fig:Vertex1cdeUse}
     (a) A virtual correction not considered in this paper because it
     is sub-leading in $1/\Nf$.
     (b) The complex conjugate of fig.\ \ref{fig:diags}a.
     (c) An alternative way that we could have drawn the
     vertex in the third line of diagrams in fig.\ \ref{fig:VertexEtoGEmore},
     which would have made it clearer that the vertex correspond to
     $e \to \gamma e$ in the conjugate amplitude.
  }
\end {center}
\end {figure}

To illustrate the possible relations, we've shown more variations
in fig.\ \ref{fig:VertexEtoGEmore} than we will actually need.
The first line of vertex diagrams
is not relevant for large $\Nf$, but at sub-leading order in $1/\Nf$
would appear in the virtual-correction diagram of
fig.\ \ref{fig:Vertex1cdeUse}a.  The second line of vertex diagrams would
appear in the diagram for the complex conjugate of
fig.\ \ref{fig:diags}a, shown in
fig.\ \ref{fig:Vertex1cdeUse}b.  But this diagram need not be
evaluated separately since we will be taking $2\Re[\cdots]$ of the diagrams
in fig.\ \ref{fig:diags}.  The third and fourth lines of vertex diagrams
in fig.\ \ref{fig:VertexEtoGEmore} appear in
figs.\ \ref{fig:diagsVIRT}(i,k,m) and figs.\ \ref{fig:diagsVIRT}(h,l),
respectively.  Note that these vertices correspond to
$(e \to \gamma e)^*$ in the conjugate amplitude and {\it not} to
$e \bar e \to \gamma$.  They are drawn as they are just because
the diagrams for the amplitude and conjugate amplitude have been
sewn together.  Depending on your taste, it might have been clearer to
draw the third line of fig.\ \ref{fig:VertexEtoGEmore}
as fig.\ \ref{fig:Vertex1cdeUse}c,
but we have chosen to keep with the
same style of drawings as in ref.\ \cite{2brem} (AI1).

Using these rules, and the additional vertex rules of appendix
\ref{app:VertexRules}, the calculation of the real double splitting
diagrams of fig.\ \ref{fig:diags} proceeds almost the same as
our gluon-splitting calculations in our earlier work
\cite{2brem,seq,dimreg,4point}.  In particular, the calculation of
the second line of diagrams in fig.\ \ref{fig:diags}, which
involve the instantaneous 4-fermion vertex of LCPT, is very similar to
the calculation of QCD diagrams involving the 4-gluon vertex
\cite{4point}.  Because these calculations are so close to previous
work, we leave the details and analytic results to appendix
\ref{app:real}.

We'll mention just one qualitative detail here: During effectively 2-particle
evolution (e.g.\ the middle shaded region of fig.\ \ref{fig:interp}b),
the system
evolves in medium
like a coupled pair of 2-dimensional non-Hermitian
harmonic oscillators, with
two complex eigenfrequencies.%
\footnote{
  See AI1 section V.B \cite{2brem}.  Two complex frequencies $\Omega$
  also appear in the earlier work of
  refs.\ \cite{Blaizot,Iancu,Wu} on double, small-$x$
  gluon bremsstrahlung, where each $\Omega$ is determined by
  one of the two bremsstrahlung gluon energies.
}
In (large-$\Nc$) QCD double bremsstrahlung, both
eigenfrequencies are non-zero.  In QED,
however, one vanishes.
See appendix \ref{app:real} for details,
but this difference does not have any particular impact on the method
of calculation.


\subsection {An implicit approximation}

\begin {figure}[tp]
\begin {center}
  \includegraphics[scale=0.5]{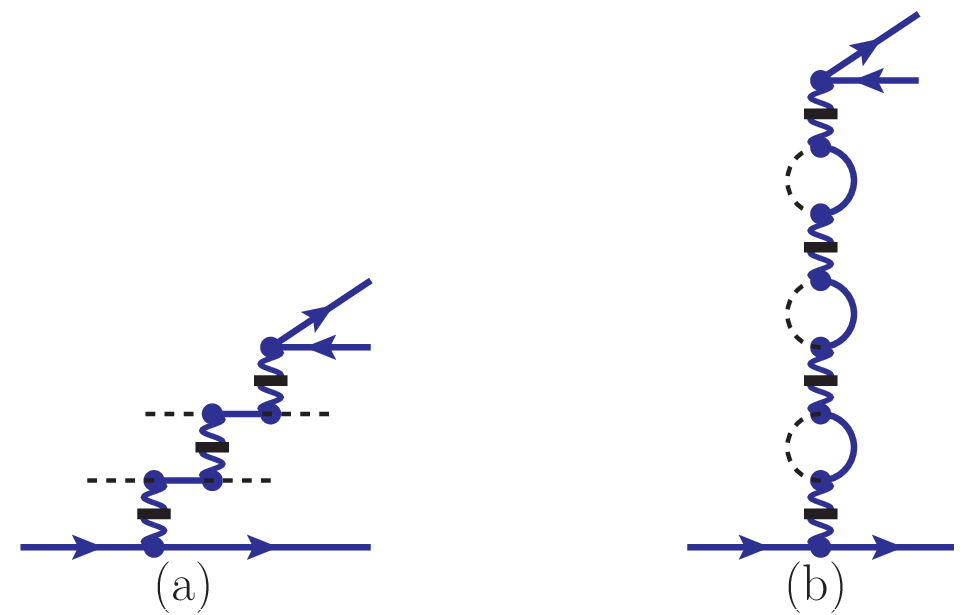}
  \caption{
     \label{fig:dress}
     (a) Possible soft medium corrections to the instantaneous longitudinal
     photon interaction of LCPT. Dashed lines represent soft electrons or
     positrons;
     intermediate solid lines represent intermediate high-energy particles.
     (b) How such corrections might alternatively
     be drawn in the language of thermal or finite-density loops.
     [(b) is expressed here
     as a Feynman diagram rather than a time-ordered diagram.]
  }
\end {center}
\end {figure}

There was an implicit approximation made in the above treatment of
instantaneous LCPT interactions.  Remember that the high-energy particles
shown in our diagrams, such as fig.\ \ref{fig:diags}, are implicitly
interacting many times with the medium.  We should take a minute
to think about what these diagrams would look like if we explicitly
drew all the interactions with the soft particles in the medium.
In particular, let's think about what happens to the longitudinal
intermediate photon lines in the second row of diagrams in
fig.\ \ref{fig:diags}.  In principle, these could be dressed by
soft interactions as in fig.\ \ref{fig:dress}a.  (Note that we've
been careful not to add any new {\it high}-energy particles to the final
state, because otherwise the figures we have drawn previously
would not have been complete drawings of all the high-energy particles.)
In the language of thermal field theory Feynman diagrams,
the contribution from fig.\ \ref{fig:dress}a would be captured
by dressing the photon propagator with thermal loops as in
fig.\ \ref{fig:dress}b.  These soft interactions have not been
included in the vertex rule we have given in
fig.\ \ref{fig:VertexEtoEEE}: the formulas there
were based on the {\it vacuum} LCPT rule given inside the box.

Fortunately, medium corrections to the vertex rule of
fig.\ \ref{fig:VertexEtoEEE} will be suppressed by some
power of the high energy scale $E$.  The typical transverse
separation $b$ of the high-energy particles during an LPM formation
time are
of order
\begin {equation}
   b \sim \frac{1}{Q_\perp}
   \sim \sqrt{\frac{t_{\rm form}}{E}}
   \sim \frac{1}{(\hat q E)^{1/4}}
\end {equation}
(suppressing dependence on longitudinal momentum fractions, and
working in the thick-media limit being considered in this paper).
In the limit of large $E$, this is parametrically
small compared to any distance scale that characterizes the medium,
such as the typical distance between medium particles.
So, the chance that there is a medium interaction during the instant
of the instantaneous interaction is parametrically small and
can be ignored.%
\footnote{
  In contrast, the effect of the medium on the evolution between
  vertex times in our diagrams (e.g.\ the shaded regions of
  fig.\ \ref{fig:interp})
  cannot be ignored because the times between these vertices are
  of order formation times, which are parametrically large in the
  high energy limit.
}

\begin {figure}[tp]
\begin {center}
  \includegraphics[scale=0.5]{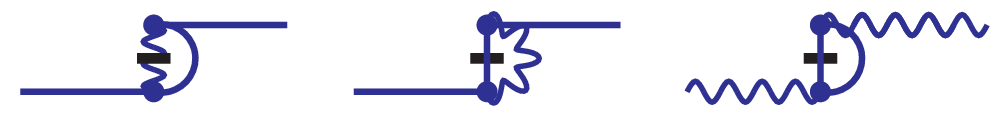}
  \caption{
     \label{fig:LCPT2point}
     Additional 2-point LCPT interactions arising from normal ordering
     the instantaneous interactions of fig.\ \ref{fig:LCPT}.
  }
\end {center}
\end {figure}

\begin {figure}[tp]
\begin {center}
  \includegraphics[scale=0.5]{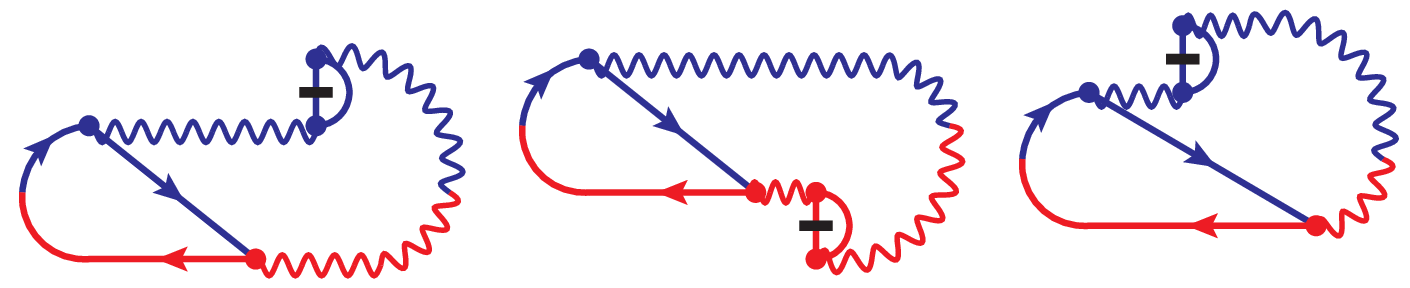}
  \caption{
     \label{fig:diagsEXTRA}
     Additional virtual diagrams for an (in-medium) LCPT calculation of
     $e \to \gamma e$, which should in principle be
     added to fig.\ \ref{fig:diagsVIRT} but which are ignorable in
     our calculation.
  }
\end {center}
\end {figure}

On a similar note, there are some other types of interactions in
time-ordered LCPT calculations that we have also ignored.  The LCPT
rules that we have quoted assume that the LCPT Hamiltonian has
been normal ordered.  When normal ordering the instantaneous
interactions, such as fig.\ \ref{fig:LCPT},
there are contractions that produce additional
2-point interactions in the normal-ordered Hamiltonian, which are
often depicted diagrammatically \cite{BPP}
as in fig.\ \ref{fig:LCPT2point}.
In our large-$\Nf$ QED calculation, these 
can contribute to
the virtual correction to $e \to \gamma e$
as in fig.\ \ref{fig:diagsEXTRA}, which should be added to the diagrams
we listed earlier in fig.\ \ref{fig:diagsVIRT2}.
Fortunately, we can ignore these additional diagrams for reasons somewhat
similar to those for the LCPT analysis
of NLO deep inelastic scattering in refs.\ \cite{Beuf1,LaP}.
Because we are in the high-energy limit, we have ignored the masses
of our high-energy particles.  We will be using dimensional regularization,
and normal-ordering
contractions
such as fig.\ \ref{fig:LCPT2point} vanish in dimensional regularization
for massless particles in vacuum, which is ultimately a consequence of
dimensional analysis.
Unlike refs.\ \cite{Beuf1,LaP}, however, our
loops are {\it not}\/ in vacuum, and the medium introduces a scale that
in principle invalidates the argument that these loops will vanish.
But because these particular loops involve an instantaneous
interaction, and because transverse separations are suppressed by powers
of the energy $E$, the effect of the medium on the loops of
fig.\ \ref{fig:LCPT2point} are negligible in the
high-energy limit.  So we may ignore the diagrams of
fig.\ \ref{fig:diagsEXTRA}.

\section{Virtual Corrections to Single Splitting}
\label {sec:virt}

Consider the diagrams of fig.\ \ref{fig:diagsVIRT}, which represent
next-to-leading order corrections to {\it single}-splitting.
Each of these diagrams contain a virtual $e\bar e$ loop.
We shall now see that almost all of these diagrams are related in
a relatively simple way to the double-splitting diagrams of
fig.\ \ref{fig:diags}.  As mentioned earlier, there is one exception:
the boxed diagram of fig.\ \ref{fig:diagsVIRT}k, which will require
a new (and quite non-trivial) calculation.
But let's start with the simpler cases.


\subsection{Back-end transformation}

\begin {figure}[tp]
\begin {center}
  \includegraphics[scale=0.5]{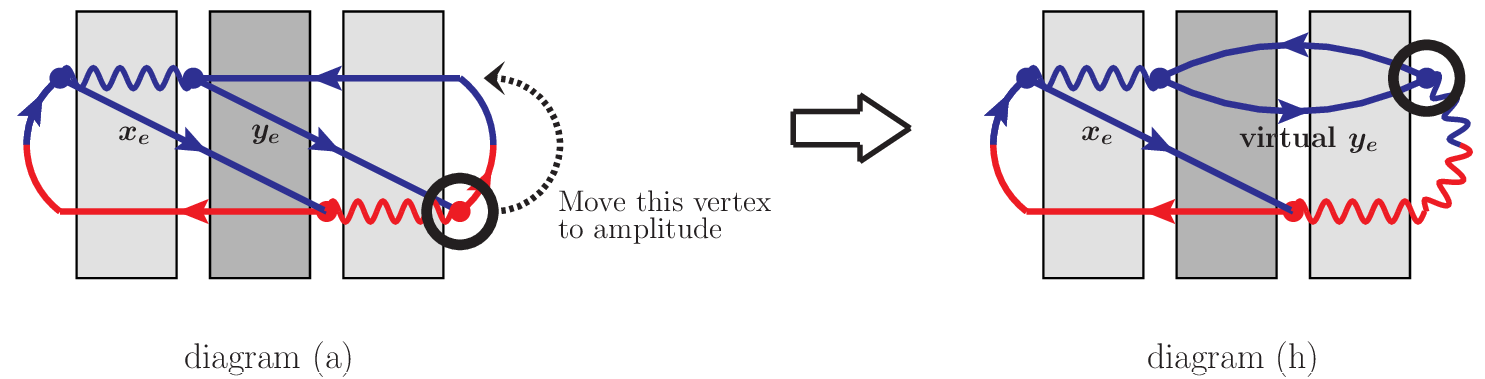}
  \caption{
     \label{fig:backend}
     Example of two diagrams related by a back-end transformation,
     in which the final vertex in one diagram is slid around the
     right-hand side (back end) of the diagram to become the final
     vertex of the other diagram.  We've shaded the evolution between
     vertices as in fig.\ \ref{fig:interp}, just to emphasize that
     the general pattern of these evolutions does not change.
  }
\end {center}
\end {figure}

Most of the diagrams of fig.\ \ref{fig:diagsVIRT} (namely h,i,j,l,n)
are related to various double emission diagrams of fig.\ \ref{fig:diags}
(a,b,c,e,g respectively) by a simple diagrammatic procedure:
Take the latest-time vertex in the diagram, and slide it around the
right end of the diagram from being a vertex in the amplitude to being
a vertex in the conjugate amplitude, or vice versa.  For concreteness,
a specific example is shown in fig.\ \ref{fig:backend}.  We will call
this a ``back-end'' transformation.  Provided that the vertex being
moved is the latest-time vertex in {\it both}\/ diagrams, there is an
extremely simple relationship between the values of the diagrams: Before
integrating over any longitudinal momentum fractions,
the two diagrams differ only by an overall minus
sign.  Heuristically, this minus sign can be roughly understood from
the relation of virtual loops, through the
optical theorem, to the probability of something {\it not}\/
happening.  So, for example, figs.\ \ref{fig:diags}(b,c) are,
roughly speaking, related to the probability $P_2$ of one splitting
$e \to \gamma e$ later being followed by another, $\gamma \to e\bar e$.
Figs.\ \ref{fig:diagsVIRT}(i,j), on the other hand, are roughly related
to the probability $P_1$ of one splitting $e \to \gamma e$ {\it not}\/ being
later followed by another.%
\footnote{
  This characterization makes sense for (1) a formal calculation
  of probabilities in the limit where the medium has some finite (but
  very large) size $L$, formally expanding to second-order
  in perturbation theory in hard splittings.  That is, the limit
  $\alpha \to 0$ for fixed $L$ and fixed particle energies
  (with $L$ large compared to formation
  times and $\alpha$ here referring to the $\alpha$ at the scale characterizing
  the high-energy splitting).  More precisely, this is the limit where
  the mean free time between splittings is large compared to $L$.
  But one may be interested in
  the opposite order of limits
  (2) $L \to \infty$ with fixed $\alpha \ll 1$ and fixed particle
  energies---that is, the case where the mean free time between splittings
  is small compared to $L$. 
  As discussed in detail in ACI2 \cite{seq}, the
  formal calculation (1) can be used to figure out a correction to
  splitting probabilities that allows for the calculation (2),
  and so it is in the formal context of (1) that we always discuss
  our diagrams, even for those cases where our ultimate interest may be (2).
}
Any increase in $P_2$ should be accompanied
by a decrease in $P_1$ by conservation of probability, and so one may
expect these diagrams to be the same size but with opposite signs.
(In principle, the total probability that's actually conserved at this
order is $P_0+P_1+P_2$, where $P_0$ is the probability that
{\it neither} emission occurs.  But the cases of the
diagrams that we are relating by back-end transformations are cases where
changes to $P_0$ do not come into play.)

By writing expressions for diagrams that are related by
a back-end transformation, using the vertex rules of section
\ref{sec:vertex} and appendix \ref{app:VertexRules}, and then
tying them together with $n$-particle evolution in the medium
as in fig.\ \ref{fig:interp} and refs.\ \cite{2brem,seq} (AI1,ACI2),
one may verify that back-end transformations really are that simple:
there is just an overall sign difference.
Note in particular that the longitudinal
momentum fractions of the particles are the same in the two diagrams
of fig.\ \ref{fig:backend},
for each of the three shaded regions of in-medium evolution.
So those evolution factors match up identically between the two diagrams.

Virtual loops should be integrated over the longitudinal momenta
of the particles in the loop.  So the final relationship between
the pairs of back-end related diagrams discussed above can be written
as
\begin {equation}
   \left[ \frac{d\Gamma}{d\xe} \right]_{\rm (h,i,j,l,n)}
   = - \int_0^{1-\xe} d\ye \>
   \left[ \frac{d\Gamma}{d\xe\,d\ye} \right]_{\rm (a,b,c,e,g)} ,
\label {eq:backend}
\end {equation}
where $\Gamma$ is the rate,
$\xe$ is the momentum of the daughter whose electron line
is connected to the original electron, and $\ye$
is the other electron in the final state of real double splitting
processes $e \to e\bar e e$, as in fig.\ \ref{fig:backend}.
The integration limits come from the facts that (i) the longitudinal
momentum fractions in the produced $e\bar e$ pair are $\ye$ and
$1{-}\xe{-}\ye$ in these diagrams (relative to the initial electron),
and (ii)
longitudinal momentum fractions are always positive in LCPT.

Thinking about the UV divergences of loop diagrams vs.\ tree diagrams,
one might be uneasy with the idea that diagrams (a) and (h) in
fig.\ \ref{fig:backend} could possibly be related so simply.
In particular, diagram (h) involves a photon self-energy, and
photon self-energies are UV divergent.%
\footnote{
  The back-end transformation says that diagrams (a) and (h) differ
  by an overall sign {\it before} integrating (h) over $\ye$.
  In LCPT, the range of integration in (\ref{eq:backend})
  for $\ye$ is finite ($\ye$ is not integrated to infinity)
  because both virtual particles in the loop are forced to
  have positive longitudinal momentum $p^+$.
  So any UV divergence will already be present even
  before integrating over $\ye$.
}
In contrast, diagram (a) represents an interference term that
contributes to the calculation of the magnitude-square of a tree-level
amplitude, and tree-level rates are not UV divergent.
Nonetheless, diagram (a) by itself {\it is} UV divergent, even though
those divergences cancel in the sum $2\Re(\mbox{a+b+c})$ of such interference
terms, drawn in fig.\ \ref{fig:diags}.%
\footnote{
  On the technical side, see the discussion of
  ``$\Delta t{\to}0$'' UV divergences and their cancellation
  for QCD diagrams similar to our (a+b+c) here
  in sections II.A.4 and II.B.2 of ref.\ \cite{seq}, or the
  earlier discussion for a different set of QCD diagrams in
  ref.\ \cite{2brem}.
  Also, our language is somewhat loose above.  When we refer to
  the sum $2\Re(\mbox{a+b+c})$ we actually mean the difference between
  (i) that sum and (ii) what one would have gotten by instead treating
  the process as two successive, {\it independent}\/
  single splittings each calculated
  using {\it leading}-order formulas for single spitting rates.
  In our notation here and elsewhere,
  we call this difference ``$\Delta\,d\Gamma/dx\,dy$.''  Physically,
  $\Delta\,d\Gamma/dx\,dy$ corresponds to the {\it correction}
  to double emission due to overlap effects.
  See sections I.A and II.A of ref.\ \cite{seq}.
}
This is a general issue
with individual {\it time-ordered}\/ (as opposed to Feynman) diagrams.
For the same reason, the UV divergence of an individual
virtual-correction diagram
like diagram (h) is different from merely the UV divergence one
associates with a photon self-energy, because diagram (h) has a
time-order constraint that the emission in the conjugate amplitude
occur {\it during}\/ the electron pair fluctuation of the photon.
Because the sum of virtual diagrams $2\Re(\mbox{h+i+j})$
of fig.\ \ref{fig:diagsVIRT}
is related by back-end transformation to the sum of diagrams
$2\Re(\mbox{a+b+c})$, and because UV divergences cancel in the sum
$2\Re(\mbox{a+b+c})$,
as they must, this means that all the UV divergences of the
sum $2\Re(\mbox{h+i+j})$ must cancel as well, including UV divergences
associated with the photon self-energy.  That leaves diagram
(k) as the only uncanceled UV
divergence among the first line of fig.\ \ref{fig:diagsVIRT}
(and it turns out to be the only uncanceled UV divergence among
all the diagrams).  Reassuringly, we will find that the
time ordering represented by diagram (k), by itself,
indeed gives {\it exactly} the right amount of divergence
to produce, in our calculation, the known
renormalization of $\alphaqed$.


\subsection{Front-end transformations}
\label {sec:frontend}

\begin {figure}[tp]
\begin {center}
  \includegraphics[scale=0.45]{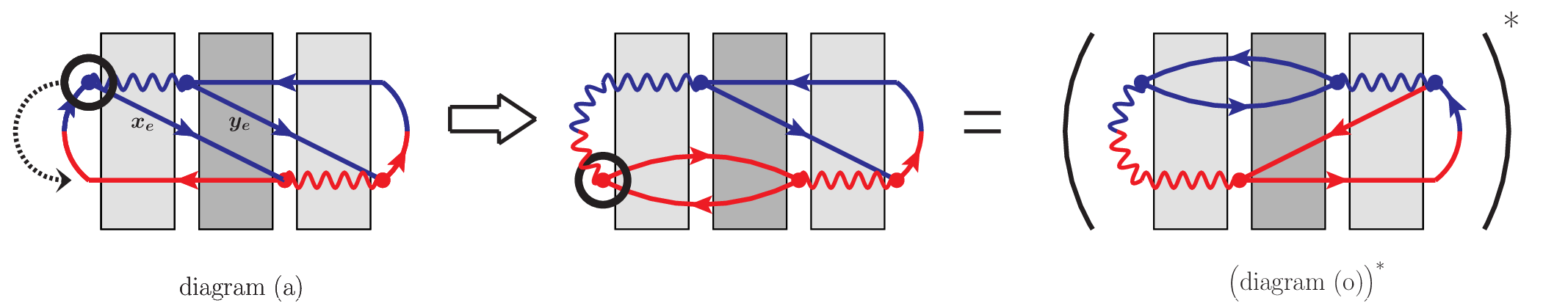}
  \caption{
     \label{fig:frontend}
     Like fig.\ \ref{fig:backend} but for a ``front-end'' transformation.
  }
\end {center}
\end {figure}

There is a somewhat related relation between pairs of
diagrams where we instead take the {\it earliest}-time
vertex in a diagram and slide it around the
{\it left}\/ end of the diagram to move it from the amplitude to
the conjugate amplitude or vice versa.  We will refer to this as
a front-end transformation, an example of which is shown
in fig.\ \ref{fig:frontend}.  We'll discuss how to implement
a front-end transformation in a moment, but first note its utility.
Using a front-end transformation,
along with complex conjugation as necessary, the $\gamma \to e\bar e$
virtual correction diagrams of
figs.\ \ref{fig:diagsVIRT2}(o,p,q,s,u) can be related to
the real double bremsstrahlung diagrams of
figs.\ \ref{fig:diags}(a,b,c,f,g), respectively.
Using {\it both}\/ a front-end and a back-end transformation on a single
diagram, figs. \ref{fig:diagsVIRT}(m) and \ref{fig:diagsVIRT2}(t) can
be related to figs.\ \ref{fig:diags}(e,f), respectively, and
fig.\ \ref{fig:diagsVIRT2}(r) can be related to fig.\ \ref{fig:diagsVIRT}(k).
In consequence, the only virtual diagram calculation that we will need to
do from scratch is the boxed diagram of fig.\ \ref{fig:diagsVIRT}(k):
an (in-medium) photon self-energy loop in the ``middle'' of a
single splitting $e \to \gamma e$.

\begin {figure}[tp]
\begin {center}
  \includegraphics[scale=0.5]{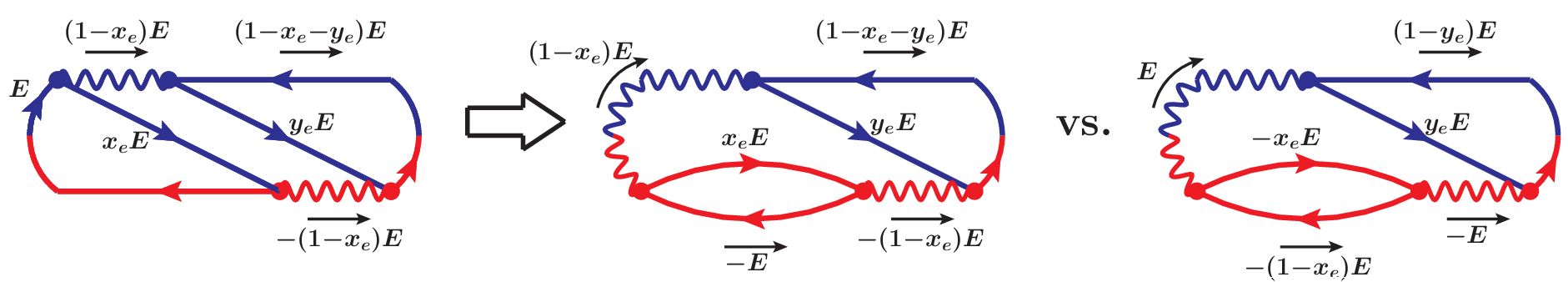}
  \caption{
     \label{fig:frontend2}
     The first diagram shows the same contribution $e \to e\bar e e$ as
     in fig.\ \ref{fig:frontend}, but here we have labeled all longitudinal
     momenta.  The second diagram shows what would
     happen if we slid the initial vertex around as in the front-end
     transformation of fig.\ \ref{fig:frontend} {\it without} changing
     the labeling of longitudinal momenta.  For comparison, the last
     diagram shows the actual labeling of longitudinal momenta for
     this $\gamma \to e\bar e$ process with an initial photon of
     energy $E$.  In all of these diagrams, momenta are taken to
     flow in the direction of fermion-line arrows unless indicated otherwise by
     a small black arrow.  This generally corresponds to labeling the
     momenta as flowing from left to right in the interference diagram.
  }
\end {center}
\end {figure}

Front-end transformations are more complicated to implement than
back-end ones.  If we were to simply slide the vertex
around as in fig.\ \ref{fig:frontend} while keeping the labeling of
all longitudinal momenta the same, we would get the transformation
shown by the first two diagrams of fig.\ \ref{fig:frontend2}.  There
are two problems with the middle diagram in fig.\ \ref{fig:frontend2},
which is supposed to represent a virtual correction to
$\gamma \to e \bar e$.
The first problem is that our convention is always to let $E$
refer to the energy of the original high-energy particle in any
process, which in this case would be the photon.  The second is that
$\xe$ is positive in the diagram of fig.\ \ref{fig:frontend2},
which is then inconsistent with the fact that the longitudinal
momentum $\xe E$ shown in the second diagram should be {\it negative},
given our convention that longitudinal momenta of
conjugate-amplitude particles are negative in our interference diagrams.
This second problem can
also be visualized by comparing the earlier figs.\ \ref{fig:backend}
and \ref{fig:frontend} for back-end and front-end transformations.
In the back-end case of
fig.\ \ref{fig:backend}, both the original diagram and the transformed
diagram had exactly the same particles evolving in each of the three
shaded regions.  In the front-end case given by the first two
diagrams of fig.\ \ref{fig:frontend}, the last shaded region is the
same, but an ``amplitude'' particle (blue line) has been switched to
a ``conjugate amplitude'' particle (red line) in each of the first
two shaded areas.  That means that the evolution in those shaded
areas is not exactly the same for those two diagrams, and so their
relation is not as simple as it was for a back-end transformation.

There is a simple way to overcome both obstacles, which is to
make the change of variables
\begin {equation}
   (\xe,\ye,E) \to
   \Bigl(\frac{-\xe}{1-\xe}\,,\,\frac{\ye}{1-\xe}\,,\,(1{-}\xe)E\Bigr)
\label {eq:frontendxyE}
\end {equation}
when making a front-end transformation like fig.\ \ref{fig:frontend}.
This converts the momenta of
the middle diagram of fig.\ \ref{fig:frontend2} into those of the last
diagram.  It also negates the value of $\xe$.  As long as one writes
formulas for evolution in diagrams in a way that is general enough
to correctly handle the cases of both positive or negative longitudinal
momentum fractions, similar to ref.\ \cite{2brem} (AI1), then all will be well.
The transformation (\ref{eq:frontendxyE}) is its own inverse,
and so works just as well transforming $\gamma \to e\bar e$ back
again to $e \to e\bar e e$.  In terms of the diagram labels in
figs.\ \ref{fig:diags} and \ref{fig:diagsVIRT2},
\begin {equation}
   2\Re
   \left[ \frac{d\Gamma}{d\ye} \right]_{\rm (o,p,q,s,u)}
   = - 2\Nf\Re \int_0^1 d\xe \>
   \left\{
     \left[ \frac{d\Gamma}{d\xe\,d\ye} \right]_{\rm (a,b,c,f,g)}
     \mbox{with substitutions (\ref{eq:frontendxyE})}
   \right\}
\label {eq:frontend1}
\end {equation}
is the front-end analog of (\ref{eq:backend}).%
\footnote{
  \label {foot:frontnorm}
  The change of $E$ to $(1{-}\xe)E$ in (\ref{eq:frontendxyE}) might
  have also led to some power of $1{-}\xe$ appearing as an
  overall normalization factor in (\ref{eq:frontend1}).
  However, for the front-end transformation of
  $d\Gamma/d\xe\,d\ye$ here,
  there turns out to be no such factor in four space-time dimensions.
  (See appendix \ref{app:frontenddimreg} for a more general discussion).
}
Here we have taken
$2\Re[\cdots]$ of all the diagrams, which we must do anyway at the end
of the calculation,
and which here obviates the need to specify exactly which diagrams need to
be complex conjugated after the front-end transformation.
Note that we have written the $\gamma{\to}e\bar e$ rate as $d\Gamma/d\ye$ to
conform with our convention that the electron produced by pair production is
labeled $\ye$, as in the last diagram of fig.\ \ref{fig:frontend2}.
The relative factor $\Nf$ in (\ref{eq:frontend1}) just reflects the
fact that one of the electron flavors in $e \to e\bar e e$ is fixed
to be that of the initial electron but there is no similar
flavor constraint for NLO $\gamma \to e\bar e$.

\begin {figure}[tp]
\begin {center}
  \includegraphics[scale=0.49]{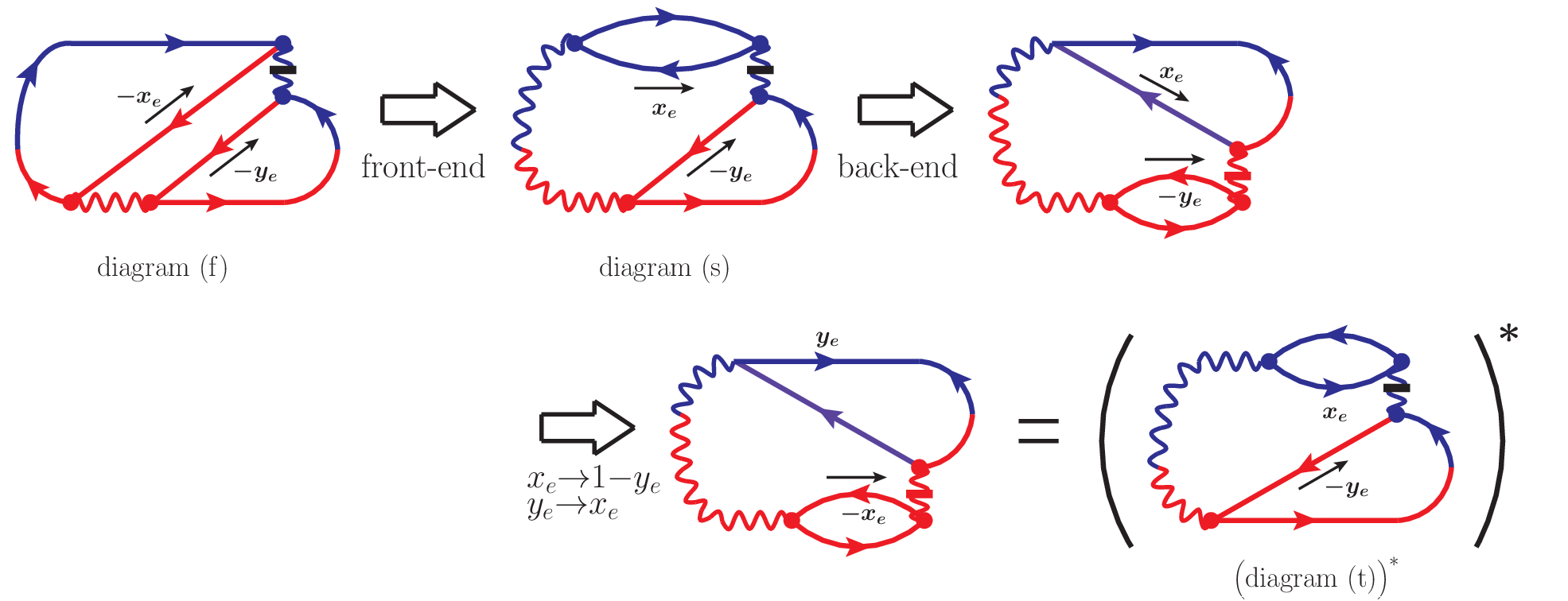}
  \caption{
     \label{fig:frontbackend1}
     The transformation of diagram (f) to diagram (t) by front-
     and back-end transformations, relabeling momentum fractions, and
     conjugation.
  }
\end {center}
\end {figure}

Performing both a front-end {\it and}\/ back-end transformation
relates
\begin {equation}
   2\Re
   \left[ \frac{d\Gamma}{d\ye} \right]_{\rm (t)}
   = + 2\Nf\Re \int_0^1 d\xe \>
   \left\{
     \left[ \frac{d\Gamma}{d\xe\,d\ye} \right]_{\rm (f)} \quad
       \vcenter{
         \hbox{with substitutions (\ref{eq:frontendxyE})}
         \hbox{followed by $(\xe,\ye)\to(1{-}\ye,\xe)$}
       }       
   \right\} .
\label{eq:frontend2}
\end {equation}
As shown in fig.\ \ref{fig:frontbackend1}, the additional step of
$(\xe,\ye)\to(1{-}\ye,\xe)$
is needed to make the labeling of the momentum fractions in diagram
(t) match up with our conventions.

\begin {figure}[tp]
\begin {center}
  \includegraphics[scale=0.5]{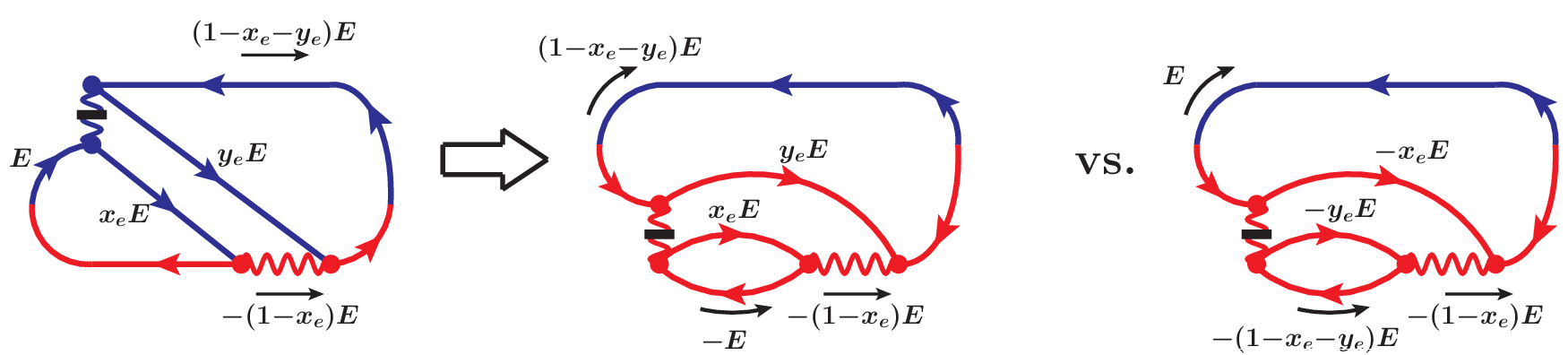}
  \caption{
     \label{fig:frontendI}
     A front-end transformation somewhat like
     fig.\ \ref{fig:frontend2} but for a diagram
     whose earliest-time vertex is instead
     an instantaneous 4-fermion interaction.
     The last diagram above is related by back-end transformation
     to the complex conjugate of 
     fig.\ \ref{fig:diagsVIRT}(m).
  }
\end {center}
\end {figure}

A different variant of the front-end transformation is
needed for interference diagrams whose earliest vertex is an
instantaneous 4-fermion interaction, such as
fig.\ \ref{fig:diags}(e).
Fig.\ \ref{fig:frontendI} shows diagrammatically the
difference between (i) ``simply
sliding the vertex around'' while keeping the labeling the same
vs.\ (ii) the actual result we want to achieve with a front-end
transformation.  We can implement the necessary relabeling of
momenta by making the change of variables
\begin {equation}
   (\xe,\ye,E) \to
   \Bigl(
     \frac{-\ye}{1{-}\xe{-}\ye} \,,\,
     \frac{-\xe}{1{-}\xe{-}\ye} \,,\,
     (1{-}\xe{-}\ye)E\Bigr) ,
\label {eq:frontendI}
\end {equation}
somewhat similar to (\ref{eq:frontendxyE}).
Performing both a front-end and back-end transformation then
relates diagrams (m) and (e) as
\begin {equation}
   2\Re
   \left[ \frac{d\Gamma}{d\xe} \right]_{\rm (m)}
   = + 2\Re \int_0^{1-\xe} d\ye \>
   \left\{
     \left[ \frac{d\Gamma}{d\xe\,d\ye} \right]_{\rm (e)}
     \mbox{with substitutions (\ref{eq:frontendI})}
   \right\} .
\label {eq:mtransform}
\end {equation}

\subsection{The fundamental virtual diagram}

\subsubsection {Overview}

\begin {figure}[tp]
\begin {center}
  \includegraphics[scale=0.7]{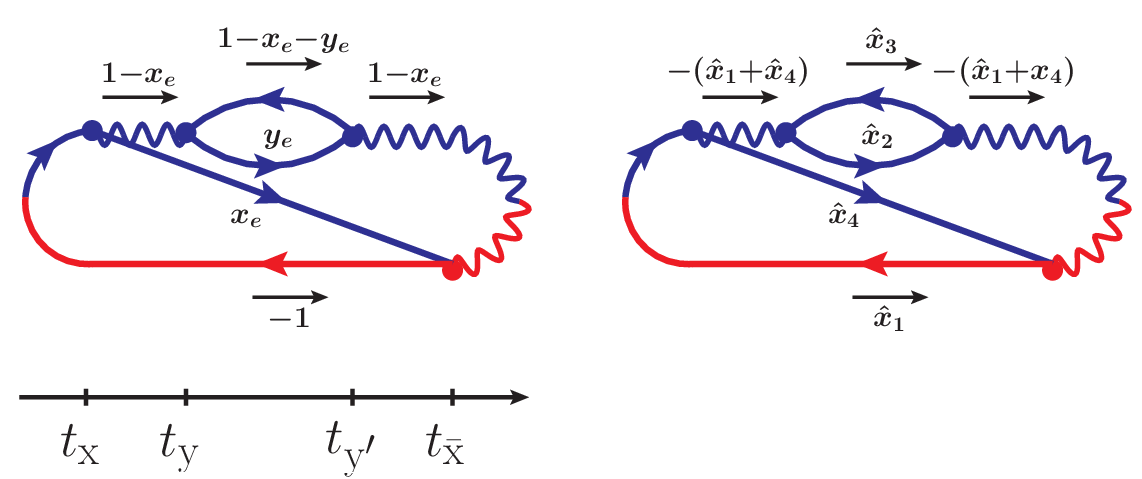}
  \caption{
     \label{fig:FundamentalVirt}
     The basic virtual diagram $xyy\bar x$
     of fig.\ \ref{fig:diagsVIRT}k, here with
     longitudinal momentum fractions labeled in terms of $\xe$ and
     $\ye$ [left] or the labels $\hat x_i$ (\ref{eq:xhat}) [right].
     Our notation for vertex times is also shown.
  }
\end {center}
\end {figure}

As previously discussed, the one virtual correction diagram which cannot
be related to the $e{\to}e\bar e e$ diagrams
is fig.\ \ref{fig:FundamentalVirt}, which is a more detailed version of
fig.\ \ref{fig:diagsVIRT}k.
Following notation similar to that
of ref.\ \cite{2brem} (AI1), we will refer to this as the $xyy\bar x$
diagram since it involves, in order, (i) emission of an $\xe$ electron in
the amplitude, (ii) emission of a $\ye$ electron in
the amplitude, (iii) re-absorption of the $\ye$ electron
in the amplitude, followed finally by (iv) emission of the $\xe$
electron in the conjugate amplitude.
This virtual correction diagram contributes
\begin {equation}
   \left[\frac{dI}{d\xe}\right]_{xyy\bar x}
   =
   \int_0^{1-\xe} d\ye \>
   \left[\frac{dI}{d\xe\,d\ye}\right]_{xyy\bar x}
\label {eq:Fund1int}
\end {equation}
to the differential emission probability for $e \to \gamma e$, where,
adapting the notation of refs.\ \cite{2brem,seq} (AI1,ACI2),
\begin {align}
   \left[\frac{dI}{d\xe\,d\ye}\right]_{xyy\bar x}
   =
   \left( \frac{E}{2\pi} \right)^2
   &
   \int_{t_\xx < t_\yx < t_{\yx'} < t_\xbx}
   \sum_{\rm pol.}
   \langle|i\,\overline{\delta H}|\B^\xbx\rangle \,
   \langle\B^\xbx,t_\xbx|{\B^{\yx'}},t_{\yx'}\rangle
\nonumber\\ &\times
   \langle{\B^{\yx'}}|{-}i\,\delta H|{\C_{41}^{\yx'}},{\C_{23}^{\yx'}}\rangle \,
   \langle\C_{41}^{\yx'},\C_{23}^{\yx'},t_{\yx'}|\C_{41}^\yx,\C_{23}^\yx,t_\yx\rangle
\nonumber\\ &\times
   \langle\C_{41}^\yx,\C_{23}^\yx|{-}i\,\delta H|\B^\yx\rangle \,
   \langle\B^\yx,t_\yx|\B^\xx,t_\xx\rangle \,
   \langle\B^\xx|{-}i\,\delta H|\rangle .
\label {eq:Fund1}
\end {align}
In broad outline,
this formula simply convolves the relevant vertex matrix elements at
different vertex times with
medium-averaged propagators between those times.%
\footnote{
  In particular, compare and contrast to AI1 (4.10) \cite{2brem}.
}
The notation $\C_{ij}$ refers to the $\B_{ij} \equiv (\b_i{-}\b_j)/(x_i{+}x_j)$
variables defined in (\ref{eq:Bdef}); we use the letter $\C$ just to
help distinguish $\B_{ij}$ variables used in the context of
4-particle (effectively 2-particle) evolution.
The indices
on $\C_{ij}$ refer to the particles involved in the 4-particle evolution
according to the convention of the
right-hand drawing in fig.\ \ref{fig:FundamentalVirt},
for which we introduce the labeling \cite{2brem}
\begin {equation}
   (\hat x_1,\hat x_2,\hat x_3,\hat x_4) \equiv (-1,\ye,1{-}\xe{-}\ye,\xe)
\label {eq:xhat}
\end {equation}
of longitudinal momentum fractions.

Initially, the evaluation of (\ref{eq:Fund1}) proceeds very similarly
to that of other diagrams, following refs.\ \cite{2brem,seq} (AI1,ACI2).
One puts in explicit formulas for the $\delta H$ matrix elements and for
the time-evolution factors.  It is then possible to analytically
perform all the integrations implicit in (\ref{eq:Fund1}) over
the intermediate transverse positions specified by the $\B$'s and $\C$'s,
as well as all but one of the time integrations.  Due to time translation
invariance, one combination of those time integrations just gives
a factor of total time $\Time$.  After all these integrations, the
differential rate $d\Gamma \equiv dI/\Time$ corresponding to
(\ref{eq:Fund1}) may be reduced to a single remaining integral
over the time difference $\Delta t$ between the two intermediate
vertex times in the process under consideration.
Schematically,
\begin {equation}
   \left[\frac{dI}{d\xe\,d\ye}\right]_{xyy\bar x}
   = \int_0^\infty d(\Delta t) \> F(\Delta t) .
\label {eq:intFdef}
\end {equation}
For the $xyy\bar x$ process of fig.\ \ref{fig:FundamentalVirt},
$\Delta t \equiv t_{\yx'}{-}t_\yx$ is the duration of the photon self-energy loop.
The integrand $F(\Delta t)$ has a somewhat complicated formula
(schematically similar to other diagrams),
which we leave to appendix \ref{app:Fund1}.
Here, in the main text, our goal is to give a broad overview of
what is involved in dealing with the UV divergence of the
$\Delta t$ integral in (\ref{eq:intFdef}).
To be concrete: For small $\Delta t$, the integrand turns out
(see appendix \ref{app:smalldt})
to have non-integrable divergences
\begin {equation}
   F(\Delta t) \propto
   \frac{\ln(2i\Omega_\ix \,\Delta t)}{(\Delta t)^2}
   - \frac{i\Omega_\ix}{\Delta t}
   \qquad
   \mbox{(for $\Delta t \ll |\Omega_\ix|^{-1}$)} ,
\label {eq:smalldt}
\end {equation}
where the magnitude of
\begin {equation}
   \Omega_\ix = \sqrt{-\frac{i(1{-}\xe)\hat q}{2 \xe E}}
\label {eq:Omegai}
\end {equation}
can be interpreted physically as the scale of
the QED inverse formation time (\ref{eq:tQED})
for leading-order $e \to \gamma e$.

UV divergences from $\Delta t{\to}0$ in the
calculation of time-ordered interference diagrams are not restricted just
to loop diagrams like fig.\ \ref{fig:FundamentalVirt}.
Individually, the real double splitting diagrams of
fig.\ \ref{fig:diags}(a--c) for $e \to e\bar e e$
also have $\Delta t{\to}0$ divergences
when written in the form $\int d(\Delta t)\>F(\Delta t)$, analogous
to the detailed discussion of the QCD case in refs.\ \cite{2brem,seq,dimreg}
(AI1--ACI3).
Those divergences cancel between the three diagrams of
fig.\ \ref{fig:diags}(a--c), also similar to refs.\ \cite{2brem,seq,dimreg}.
Consistently evaluating the remaining finite part is a little
tricky and was accomplished \cite{dimreg} by computing the individual
diagrams using dimensional regularization.

Because the virtual diagrams of fig.\ \ref{fig:diagsVIRT}(h--j)
are related to fig.\ \ref{fig:diags}(a--c) by a simple back-end
transformation (\ref{eq:backend}), the divergences of (h--j) will
also cancel.  Additionally, we find that diagrams involving
instantaneous longitudinal photon interactions are all individually
finite.  That means that the only uncanceled UV divergence in our
calculations of $e \to e\bar e e$ and $e \to \gamma e$
is the one for the $xyy\bar x$ diagram shown in
both figs.\ \ref{fig:diags}(k) and \ref{fig:FundamentalVirt}.
[The divergence of the
$\gamma \to e\bar e$ diagram of fig.\ \ref{fig:diagsVIRT2}(r) is
similarly uncanceled, but, as already noted, that diagram is related
to the other by front- and back-end transformations.]

Unfortunately, our previous method \cite{dimreg} for calculating
diagrams in dimensional regularization breaks down for the
$xyy\bar x$ diagram, for reasons that will be explained later.
We will leave the (we think interesting) details to appendices.
Here we convey the gist of the method.


\subsubsection {Dimensional Regularization: Strategy}

In this paper, the symbol $d \equiv d_\perp$ will represent
the number
\begin {equation}
     d=2-\eps
\end {equation}
of {\it transverse} dimensions, not the total number $4{-}\eps$ of
space-time dimensions.  It might reduce the opportunity for
confusion if we always wrote our $d$ as $d_\perp$, but we drop the
subscript to save space in what will be some
complicated equations.

We do not know how to do the entire integral (\ref{eq:intFdef}) for
arbitrary $d$.  Fortunately, we do not have to.  The only divergence
comes from $\Delta t\to 0$, and so we only need to regulate the
small-$\Delta t$ portion of the integral.  Let $a$ be an arbitrarily
small but finite time scale, and split (\ref{eq:intFdef}) up into
\begin {equation}
   \int_0^\infty d(\Delta t) \> F(\Delta t)
   =
   \int_0^a d(\Delta t) \> F(\Delta t)
   +
   \int_a^\infty d(\Delta t) \> F(\Delta t) .
\end {equation}
We only need dimensional regularization for the first term on the
right-hand side.  Specifically,
\begin {equation}
   \int_0^\infty d(\Delta t) \> F_d(\Delta t)
   =
   \int_0^a d(\Delta t) \> F_d(\Delta t)
   +
   \int_a^\infty d(\Delta t) \> F_2(\Delta t)
   + O(\eps) ,
\label {eq:Fsplit0}
\end {equation}
where $F_d(\Delta t)$ is the integrand for $d$ transverse dimensions.
Roughly speaking, our goal is to use small-$\Delta t$ approximations
to do the $\Delta t < a$ integral analytically, and then use
numerical integration to do the $\Delta t > a$ integral.
But there is a problem: As we take $a \to 0$ (in order to make
our calculation of the first integral arbitrarily accurate),
the second integral $\int_a^\infty F_2(\Delta t)$ will blow up, which
is undesirable for numerical integration.  We can isolate this
problem by rewriting the right-hand side of (\ref{eq:Fsplit0}) as
\begin {equation}
   \int_0^a d(\Delta t) \> F_d(\Delta t)
   +
   \int_a^\infty d(\Delta t) \> {\cal D}_2(\Delta t)
   +
   \int_a^\infty d(\Delta t) \>
      \bigl[ F_2(\Delta t) - {\cal D}_2(\Delta t) \bigr]
   + O(\eps) ,
\label {eq:Fsplit1}
\end {equation}
where ${\cal D}_2(\Delta t)$ is any convenient function
that
\begin {itemize}
\item
  matches the divergence of $F_2(\Delta t)$ as
  $\Delta t \to 0$
  [which is proportional to (\ref{eq:smalldt})];
\item
  falls off fast enough as $\Delta t \to \infty$ so that
  $\int_a^\infty d(\Delta t) \> {\cal D}_2(\Delta t)$ will
  converge for non-zero $a$;
\item
  is simple enough that $\int_a^\infty d(\Delta t) \> {\cal D}_2(\Delta t)$
  can be performed analytically.
\end {itemize}
The last integral in (\ref{eq:Fsplit1}) is then convergent by itself
in the $a \to 0$ limit, and so we may replace (\ref{eq:Fsplit1}) by
\begin {equation}
   \lim_{\mbox{\small``$\scriptstyle{a\to 0}$''}} \Biggl[
   \int_0^a d(\Delta t) \> F_d(\Delta t)
   +
   \int_a^\infty d(\Delta t) \> {\cal D}_2(\Delta t)
   \Biggr]
   +
   \int_0^\infty d(\Delta t) \>
      \bigl[ F_2(\Delta t) - {\cal D}_2(\Delta t) \bigr] 
   + O(\eps) .
\label {eq:Fsplit}
\end {equation}

Our strategy will be to evaluate the first two integrals in
(\ref{eq:Fsplit}) analytically in the limit of small $a$
(and verify cancellation of the $a$ dependence) and to evaluate
the (convergent) third integral numerically.
The choice of ${\cal D}_2(\Delta t)$ is not unique, but the final
answer for (\ref{eq:Fsplit})
will be independent of the details of that choice.
The particular choice we will find convenient is
\begin {equation}
   {\cal D}_2(\Delta t) \propto
   \frac{\ln(2i\Omega_\ix \,\Delta t)}{(\Delta t)^2}
   - i\Omega_\ix^3\,\Delta t\,\csc^2(\Omega_\ix\, \Delta t) \,,
\label {eq:calD2prop}
\end {equation}
which reduces to (\ref{eq:smalldt}) for small $\Delta t$ but also
falls off quickly enough as $\Delta t \to \infty$ [because
$\csc(\Omega\,\Delta t)$ falls exponentially for
the $\Omega \propto \sqrt{-i}$ of (\ref{eq:Omegai})].
But there is no difficulty if the reader thinks some alternative choice
would have been simpler or more natural.

We placed scare quotes around the ``$a{\to}0$'' limit in
(\ref{eq:Fsplit}) because we should clarify how it coordinates with
the $\eps \to 0$ limit of dimensional regularization.  The earlier
split (\ref{eq:Fsplit0}) of the $\Delta t$ integration is invalid
unless we can set $d{=}2$ for $\Delta t \ge a$, up to corrections that
disappear as $\eps \to 0$.
One feature of dimensional
regularization is that it converts integer power-law divergences into
non-integer power-law divergences.
For example, $\int d(\Delta t)/\Delta t$ might become
proportional to $\int d(\Delta t)/(\Delta t)^{1-\eps}$
or $\int d(\Delta t)/(\Delta t)^{1-2\eps}$.
(We will see details later.)
When integrated over
$\int_a^\infty d(\Delta t)$, the results will correspondingly have
non-integer power-law dependence on the lower cut-off $a$, and we
will encounter factors like $(a\Omega)^\eps$, where $\Omega$ is the relevant
frequency scale of the problem.  We need to be able to expand such a
factor in small $\eps$ as $1 + \eps\ln(a\Omega) + \cdots$ in order to
recover the $d{=}2$ result for the $\Delta t \ge a$ integral, which
means that $\eps\ln(a\Omega)$ will need to be small.  The ``$a{\to}0$''
limit in (\ref{eq:Fsplit}) is therefore notational short-hand for
expanding expressions assuming that the small $a$ has been chosen
relative to the (eventually) arbitrarily small $\eps$ such that
\begin {equation}
   \eps \ll \frac{1}{|\ln(a\Omega)|} \to 0 .
\label {eq:smalladef}
\end {equation}

Before moving on to discuss the application of these techniques to the
$xyy\bar x$ diagram of fig.\ \ref{fig:FundamentalVirt}, we should
clarify how we plan to handle photon and fermion helicities in
dimensional regularization.  We will implement the Conventional
Dimensional Regularization (CDR) scheme
with $\MSbar$ renormalization.
These are,
for instance, the conventions used by the Particle Data Group
\cite{PDG} and the community at large to quote, compare, and take
world-averages of determinations of $\alphas(M_{\rm Z})$.
For a
nicely brief summary of different flavors of ``dimensional regularization''
used by some authors,
see, for example, section II.A of ref.\ \cite{HLaP}.%
\footnote{
  For the quantities computed in this paper,
  there will be no difference between using
  what ref.\ \cite{HLaP} identifies as CDR and HV dimensional regularization,
  once we write final formulas for differential rates in terms of MS-bar
  renormalized $\alpha$ and set $\epsilon=0$.
  In particular, for the quantities we calculate,
  we will not have any IR/collinear divergences that
  need to be canceled between real and virtual emission diagrams, and
  so it does not matter if we treat real and virtual particles on
  exactly the same footing in $d{=}2-\eps$ dimensions.
  In part this is thanks to the fact that medium effects cut off infrared
  and collinear singularities for splittings contained within the medium.
}


\subsubsection {Dimensional Regularization: Application to $xyy\bar x$}
\label{sec:applyDR}

The $\delta H$ matrix element in (\ref{eq:Fund1}), which are the vertex
factors of figs.\ \ref{fig:VertexEtoGE} and \ref{fig:VertexEtoGEmore}
and appendix \ref{app:VertexRules}, are all proportional to
the transverse momentum $\P$ characterizing the splitting, which is
equal to the $\P_{ij}$ of any pair of the particles directly
involved in the splitting (or merging).  Below, take this to be the $\P_{ij}$
of the two daughters.  The proportionality of $\delta H$ matrix elements
to $\P$ holds in any dimension $d$.
The proportionality constant, which involves DGLAP splitting amplitudes
and consideration of particle helicities, depends on $d$, but
for the moment we won't keep track of those details.
In transverse-position space, the factors of $\P$ become factors
of $\pm i\grad\delta^{(d)}(\B)$.  Plugging this into the starting
formula (\ref{eq:Fund1}) for the $xyy\bar x$ diagram, and using
the $\delta$ functions to perform the implicit integrations
over the $\B$ variables, gives
\begin {align}
   \left[\frac{d\Gamma}{d\xe\,d\ye}\right]_{xyy\bar x}
   \propto
   \int_{\rm times} &
   \grad_{\B^\xbx}
   \langle\B^\xbx,t_\xbx|{\B^\yx}',t_{\yx'}\rangle \Bigr|_{\B^\xbx=0}
\nonumber\\ &\times
   \grad_{{\C_{23}^\yx}'} \grad_{\C_{23}^\yx}
   \langle\C_{41}^{\yx'},\C_{23}^{\yx'},t_{\yx'}|\C_{41}^\yx,\C_{23}^\yx,t_\yx\rangle
   \Bigr|_{
      \C_{23}^{\yx'}=0=\C_{23}^\yx; ~ \C_{41}^{\yx'}=\B^{\yx'};
      ~ \C_{41}^\yx=\B^\yx
   }
\nonumber\\ &\times
   \grad_{\B^\xx} \langle\B^\yx,t_\yx|\B^\xx,t_\xx\rangle \Bigr|_{\B^\xx=0} ,
\label {eq:Fund2}
\end {align}
where the various gradients $\grad$ are contracted in ways that depend on
the helicity-dependent DGLAP splitting functions
not explicitly shown here.  The integral over the relative vertex
times in the $xyy\bar x$ time-ordered diagram of
fig.\ \ref{fig:FundamentalVirt} can be written as
\begin {equation}
   \int_{\rm times} \cdots
   =
   \int_0^\infty d(\Delta t)
   \int_0^\infty d(t_\yx{-}t_\xx)
   \int_0^\infty d(t_\xbx{-}t_{\yx'})
   \cdots .
\end {equation}
In the multiple scattering ($\hat q$) approximation, the 3-particle
(effectively 1-particle) propagators
$\langle \B',t' | \B,t \rangle$ are equivalent to those of a single
$d$-dimensional harmonic oscillator in non-relativistic quantum
mechanics, whose mass $M$ and (complex) frequency $\Omega$ are related to the
longitudinal momentum fractions of the three particles.  In the
particular case of the $xyy\bar x$ diagram, the momentum fractions
are the same for both regions of 3-particle evolution and give
the $\Omega_\ix$ of (\ref{eq:Omegai}) and
\begin {equation}
   M_\ix = \xe(1-\xe)E .
\label {eq:Mi}
\end {equation}
The formula for a harmonic oscillator propagator,
\begin {equation}
   \left( \frac{M\Omega\csc(\Omega \, \Delta t)}{2\pi i} \right)^{d/2}
      \exp\Bigl(
        \tfrac{i}2 M\Omega
        \bigl[ (\B^2+\B'^2) \cot(\Omega \Delta t)
              - 2\B\cdot\B' \csc(\Omega \Delta t) \bigr]
      \Bigr) ,
\label {eq:dprop}
\end {equation}
is simple enough that the
integrals over the initial and final times in (\ref{eq:Fund2})
can be carried out explicitly \cite{dimreg}, giving%
\footnote{
  \label{foot:absval1}
  The absolute value signs on $|M_\ix|$ in (\ref{eq:Fund3}) may at
  first look redundant, since $M_\ix$ given by (\ref{eq:Mi})
  is positive for physical values ($0<\xe<1$) of $\xe$.
  However, we would like to be able to make front-end transformations
  (\ref{eq:frontendxyE}) of our results, and that transformation
  replaces the original $\xe$ by negative values (for physical values
  of the {\it new} $\xe$) and so replaces the $M_\ix$ of (\ref{eq:Mi})
  by something negative.  AI1 section V.A \cite{2brem} discusses
  how allowing for negative values of $M_\ix$ requires introducing
  absolute value signs in certain places, and the generalization to
  $d$ dimensions is discussed in ACI3 section IV.A \cite{dimreg}.
}
\begin {align}
   \left[\frac{d\Gamma}{d\xe\,d\ye}\right]_{xyy\bar x}
   \propto
   \int_0^\infty & d(\Delta t) \>
   \int_{{\B^\yx}',\B^\yx}
   \frac{ {\B^\yx}' \B^\yx }{ [({B^\yx}')^2 (B^\yx)^2]^{d/4} }
     K_{d/4}\bigl(\tfrac12 |M_\ix| \Omega_\ix ({B^\yx}')^2\bigr)
     K_{d/4}\bigl(\tfrac12 |M_\ix| \Omega_\ix (B^\yx)^2\bigr)
\nonumber\\ &\times
   \grad_{{\C_{23}^\yx}'} \grad_{\C_{23}^\yx}
   \langle{\C_{41}^\yx}',{\C_{23}^\yx}',t_\yx'|\C_{41}^\yx,\C_{23}^\yx,t_\yx\rangle
   \Bigr|_{
      {\C_{23}^\yx}'=0=\C_{23}^\yx; ~ {\C_{41}^\yx}'={\B^\yx}';
      ~ \C_{41}^\yx=\B^\yx
    } ,
\label {eq:Fund3}
\end {align}
where now it is the two $\B$s and two $\grad$s that are contracted in
ways that depend on the helicity-dependent DGLAP splitting functions.
The 4-particle (effectively 2-particle) propagator
$\langle\C_{41}',\C_{23}',t'|\C_{41},\C_{23},t\rangle$ is also
a harmonic oscillator propagator, but this time for a coupled set
of two harmonic oscillators.  A number of terms are generated when
one takes derivatives $\grad_{\C_{23}'} \grad_{\C_{23}}$ of this
propagator as in (\ref{eq:Fund3}), but only one of them leads
to divergences of the $\Delta t$ integral when $\eps=0$, and that
is then the only term we need to treat with dimensional
regularization.  Details are given in appendix \ref{app:Fund1}.
The small-$\Delta t$ behavior of that term, which is all we need
to regulate to implement (\ref{eq:Fsplit}), is specifically%
\begin {align}
   \left[\frac{d\Gamma}{d\xe\,d\ye}\right]_{xyy\bar x}^{(\Delta t < a)}
   \propto
   P_{e\to e}^{(d)}(\xe&) \,
   P_{\gamma\to e}^{(d)}\bigl( \tfrac{\ye}{1-\xe} \bigr)
   \int_0^a \frac{d(\Delta t)}{(\Delta t)^{d+1}} \>
\nonumber\\ \times &
   \int_{{\B^\yx}',\B^\yx}
   \frac{ {\B^\yx}'\cdot\B^\yx }{ [({B^\yx}')^2 (B^\yx)^2]^{d/4} }
     K_{d/4}\bigl(\tfrac12 |M_\ix| \Omega_\ix (B^\yx)^2\bigr)
     K_{d/4}\bigl(\tfrac12 |M_\ix| \Omega_\ix ({B^\yx}')^2\bigr)
\nonumber\\ &\times
   \exp\Bigl[
     - \tfrac12
     \calX_\yx (B^\yx)^2
     -
     \tfrac12
     \calX_{\yx'} ({B^\yx}')^2
     +
     \calX_{\yx\yx'} \B^\yx \cdot {\B^\yx}'
   \Bigr] ,
\label {eq:FundDiv}
\end {align}
where the exponential factor is analogous to the exponential factor
in the single harmonic oscillator propagator (\ref{eq:dprop})
and corresponds to the exponential of the double harmonic
oscillator problem after setting $\C_{23}' = 0 = \C_{23}$.
The small-$\Delta t$ expansions of the coefficients
$\calX$ in the exponent turn out to be
\begin {subequations}
\label {eq:Xexpansion}
\begin {align}
   \calX_\yx = \calX_{\yx'}
   &= - \frac{i M_\ix}{\Delta t} + \frac{i M_\ix \Omega_\ix^2 \Delta t}{3}
      + O\bigl(M \Omega^4 (\Delta t)^3\bigr) ,
\\
   \calX_{\yx\yx'}
   &= - \frac{i M_\ix}{\Delta t} - \frac{i M_\ix \Omega_\ix^2 \Delta t}{6}
      + O\bigl(M \Omega^4 (\Delta t)^3\bigr) .
\end {align}
\end {subequations}
[The fact that the expansions to this order can be written in terms
of $\Omega_\ix$ (\ref{eq:Omegai}), without reference to the actual
eigenfrequencies $\Omega_\pm$ of the double harmonic oscillator
problem, is non-trivial.  See appendix \ref{app:OmegaSqr}.]

In (\ref{eq:FundDiv}), we also now show explicitly the particular
combination of DGLAP splitting functions that appear, which for
this divergent term happen to combine into a simple product of the
($d{=}2{-}\eps$ dimensional) spin-{\it averaged} splitting functions
$P_{e\to e}^{(d)}$ and $P_{\gamma\to e}^{(d)}$
for $e \to \gamma e$ and $\gamma \to e\bar e$.

The technical problem we face to carry out dimensional
regularization is how to do the integrals in (\ref{eq:FundDiv})
with (\ref{eq:Xexpansion}) in the small-$a$ limit (\ref{eq:smalladef}).
For similar integrals in ref.\ \cite{dimreg} (ACI3), we were able to
argue that the exponential factor in the integrand limited the
range of the $B$s enough in the small-$\Delta t$ limit that one
could make small-argument expansions of the Bessel functions
$K_{\rm d/4}(\tfrac12 |M_\ix| \Omega_\ix B^2)$.  Unfortunately, that
is not the case here.  If the exponent in (\ref{eq:FundDiv})
were evaluated with (\ref{eq:Xexpansion})
at leading order in small $\Delta t$, the
exponential would be
\begin {equation}
   \exp\Bigl[
     \frac{i M_\ix}{2\,\Delta t} ({\B^\yx}'-\B^\yx)^2
   \Bigr] .
\label {eq:LOexp}
\end {equation}
This oscillating exponential will suppress contributions to the
integral unless $|{\B^\yx}' - \B^\yx | \lesssim \Delta t/M_\ix$, and so
\begin {equation}
  | {\B^\yx}' - \B^\yx | \lesssim \frac{a}{M_\ix}
\end {equation}
since $\Delta t \le a$ in (\ref{eq:FundDiv}).
But (\ref{eq:LOexp}) places no constraint on how large ${\B^\yx}' \simeq \B^\yx$
can be.  That is cut off instead by the exponential fall-off of
the Bessel functions
in (\ref{eq:FundDiv}) for large arguments.
[The sub-leading terms of (\ref{eq:Xexpansion}) are not large enough
to cut off the integrand sooner.]
The Bessel functions suppress contributions unless
\begin {equation}
  B \lesssim |M_\ix \Omega_\ix|^{-1/2} .
\label {eq:Bbound}
\end {equation}
The integrand remains unsuppressed
when the arguments of the Bessel function are of order 1, and so
small-argument expansions of the Bessel functions are not applicable.
Having to deal with the Bessel functions as they are makes the
integral harder to do.

In appendix \ref{app:bbI}, we show how to evaluate the integral
\begin {align}
   \bbI &\equiv
   \frac{i M (|M|\Omega)^{d/2}}{\pi} 
   \int_0^a \frac{d(\Delta t)}{(\Delta t)^{d+1}}
   \int_{\B,\B'}
   \frac{\B\cdot\B'}{(B^2)^{d/4} ({B'}^2)^{d/4}}
     K_{d/4}\bigl(\tfrac12 |M| \Omega B^2\bigr)
     K_{d/4}\bigl(\tfrac12 |M| \Omega {B'}^2\bigr)
\nonumber\\ & \hspace{10em} \times
   \exp\Bigl[
     - \tfrac12
     \calX_\yx B^2
     -
     \tfrac12
     \calX_{\yx'} {B'}^2
     +
     \calX_{\yx\yx'} \B \cdot \B'
   \Bigr] ,
\label {eq:bbIdef}
\end {align}
which is proportional to the one shown in (\ref{eq:FundDiv}),
to obtain
\begin {equation}
  \bbI =
  2\pi^2 (i \barOmega)^{d-1} \Bigl[
    - \Bigl( \frac{2}{\eps} - \gammaE + \ln(4\pi) \Bigr)
    - \frac{\ln(2 i \barOmega a)+1}{i \barOmega a}
    - \ln(i\barOmega a)
    + 3\ln(2\pi)
  \Bigr] + O(a) + O(\eps) ,
\label {eq:bbIresult}
\end {equation}
where $\gammaE$ is Euler's constant and%
\footnote{
  The notation $\barOmega$ defined by (\ref{eq:barOmega}) is unrelated
  to the notation $\bar\Omega_\fx$ in AI1 section VI.B \cite{2brem}:
  There are only so many accent marks, and we've been forced to recycle.
}
\begin {equation}
  \barOmega \equiv \Omega \sgn(M) .
  \label {eq:barOmega}
\end {equation}
The $a$ dependence will be canceled, as it must,
when one adds together all the terms of (\ref{eq:Fsplit})
for the calculation of $xyy\bar x$.
When we later combine
the $xyy\bar x$ diagram with leading-order $e \to \gamma e$,
the $1/\epsilon$ pole will be absorbed,
as it must,
by the known renormalization of $\alphaqed$ in QED.

The rest of the calculation of $xyy\bar x$ is mostly
just a matter of combining the above with
calculations and formulas that are closely analogous to those for
diagrams analyzed in our previous work
\cite{2brem,seq,dimreg}.
We will write our decomposition (\ref{eq:Fsplit}) of the calculation
as
\begin {equation}
  2\Re \left[ \frac{d\Gamma}{d\xe} \right]_{xyy\bar x}
  =
  \lim_{\mbox{\small``$\scriptstyle{a\to 0}$''}} \Biggl\{
    2\Re \left[ \frac{d\Gamma}{d\xe} \right]_{xyy\bar x}^{(\Delta t < a)}
    +
    2\Re \left[ \frac{d\Gamma}{d\xe} \right]_{xyy\bar x}^{({\cal D}_2)}
  \Biggr\}
  +
  2\Re \left[ \frac{d\Gamma}{d\xe} \right]_{xyy\bar x}^{({\rm subtracted})}
  ,
\label {eq:xyyxSplit}
\end {equation}
where $[ d\Gamma/d\xe ]^{(\rm subtracted)}$ represents the
$\int d(\Delta t) \> \bigl[ F_2(\Delta t) - {\cal D}_2(\Delta t) \bigr]$
term in (\ref{eq:Fsplit}).
Details are given in appendix \ref{app:Fund1}.


\subsubsection {Renormalization}

\begin {figure}[tp]
\begin {center}
  \includegraphics[scale=0.7]{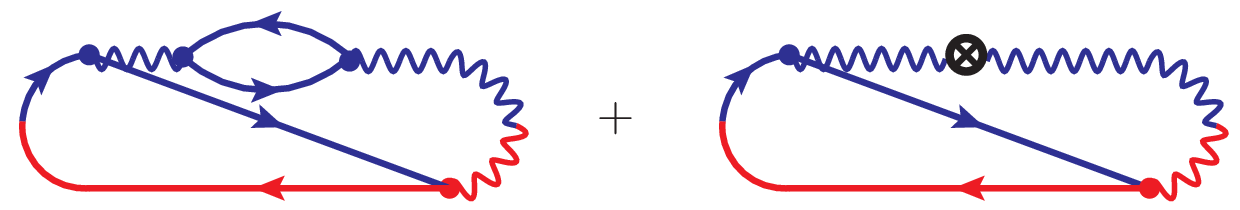}
  \caption{
     \label{fig:counterterm}
     Renormalization of the $xyy\bar x$ diagram by adding the
     corresponding (time-ordered) counter-term diagram.
  }
\end {center}
\end {figure}

There are two equivalent ways to think about renormalization and the ultimate
cancellation of the $xyy\bar x$ UV divergence appearing in
(\ref{eq:bbIresult}).  One is renormalized perturbation theory:
Add counter-term diagrams
such as fig.\ \ref{fig:counterterm}, so that the combination
\begin {equation}
   \left[ \frac{d\Gamma}{d\xe} \right]_{\xx\yx\yx\bar\xx}
   +
   \left[ \frac{d\Gamma}{d\xe}
          \right]_{\substack{\rm counterterm\\ \rm diagram}}
\end {equation}
is finite.  Note that in the large-$\Nf$ limit of QED, there is no electron
wave function renormalization and no vertex renormalization, and
in QED the
photon wave function renormalization is equivalent to charge
renormalization.

\begin {figure}[tp]
\begin {center}
  \includegraphics[scale=0.7]{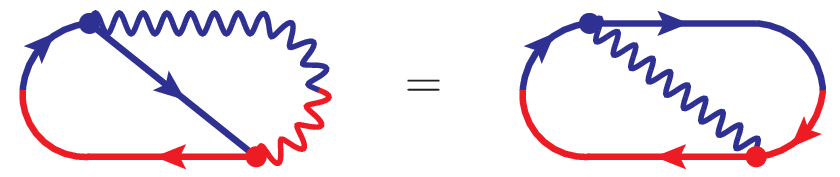}
  \caption{
     \label{fig:diagLO}
     (left)
     Leading-order time-ordered diagram $x\bar x$ for $e \to \gamma e$.  This
     diagram should be added to its complex conjugate by taking
     $2\Re[\cdots]$.
     (right) A completely equivalent way to draw this diagram.
  }
\end {center}
\end {figure}

The other viewpoint, which we find more straightforward
to implement,%
\footnote{
  The second viewpoint avoids having to sort out
  a few possible sources of confusion.  For example,
  when computing amplitudes in quantum field theory, one multiplies
  external legs by $Z_{\rm field}^{1/2}$ instead of $Z_{\rm field}$, but
  it is the latter that comes from the divergences of self-energy
  loops such as the photon self-energy loop in the $xyy\bar x$
  diagram.  The seeming difference can be resolved by realizing that the
  leading-order $x\bar x$ diagram represents a {\it rate} rather than an
  amplitude, and so should be multiplied by
  $|Z_{\rm field}^{1/2}|^2 = Z_{\rm field}$, which generates the
  counter-term diagram in fig.\ \ref{fig:counterterm}.  But one
  could also worry about why we have not included other time-orderings
  of that counter-term diagram, such as ones where the counter-term
  occurs {\it later} than both emission vertices.
  Those contributions turn
  out to cancel between diagrams with the counter-term in the amplitude and
  diagrams with the counter-term in the conjugate amplitude,
  because of the different signs in the corresponding evolution
  operators $\exp(\mp i\,\delta H\, t)$.
  The second viewpoint on renormalization allows us to bypass all
  of these considerations.
}
is to imagine
that the contribution to $d\Gamma/d\xe$ from all $e{\to}\gamma e$
diagrams, including the leading-order process represented by
the $x\bar x$ diagram of
fig.\ \ref{fig:diagLO} (plus its complex
conjugate), are initially computed using the bare coupling instead
of the renormalized coupling.  Afterwards, we convert to the
$\MSbar$-renormalized coupling using
the known relation
\begin {equation}
   \alpha_{\rm bare} =
   \alpha_{\rm ren}
     + \frac{\Nf \alpha_{\rm ren}^2}{3\pi}
       \Bigl( \frac{2}{\eps} - \gammaE + \ln(4\pi) \Bigr)
     + O(\alpha^3)
   \qquad
   \mbox{(QED).}
\label {eq:renorm}
\end {equation}
When expressed in terms of renormalized $\alpha$, the $1/\eps$
divergences will then cancel in the combination
\begin {equation}
   \left[ \frac{d\Gamma}{d\xe} \right]_{\xx\xbx}
   +
   \left[ \frac{d\Gamma}{d\xe} \right]_{\xx\yx\yx\bar\xx}
\end {equation}
through order $\alpha^2$.
Since the leading order $[ d\Gamma/d\xe ]_{\xx\bar\xx}^{\rm bare}$
is proportional to $\alpha_{\rm bare}$, (\ref{eq:renorm}) gives
\begin {equation}
   \left[ \frac{d\Gamma}{d\xe} \right]_{\xx\bar\xx}^{(\rm bare)}
   =
   \left[ \frac{d\Gamma}{d\xe} \right]_{\xx\bar\xx}^{(\rm ren)}
   +
   \frac{\Nf \alpha_{\rm ren}}{3\pi}
   \left[ \frac{d\Gamma}{d\xe} \right]_{\xx\bar\xx}^{(\rm ren)}
       \Bigl( \frac{2}{\eps} - \gammaE + \ln(4\pi) \Bigr)
   + O(\alpha^3) .
\label {eq:ren1}
\end {equation}
We will combine the second term on the right-hand side with
$xyy\bar x$ to define
\begin {equation}
   \left[ \frac{d\Gamma}{d\xe} \right]_{\xx\yx\yx\bar\xx}^{(\rm ren)}
   =
   \left[ \frac{d\Gamma}{d\xe} \right]_{\xx\yx\yx\bar\xx}
   + \frac{\Nf \alpha_{\rm ren}}{3\pi}
   \left[ \frac{d\Gamma}{d\xe} \right]_{\xx\bar\xx}^{(\rm ren)}
       \Bigl( \frac{2}{\eps} - \gammaE + \ln(4\pi) \Bigr) ,
\label {eq:xyyxRen}
\end {equation}
which is equivalent to fig.\ \ref{fig:counterterm}.
Note that, because it is multiplied by $2/\eps$,
we will need to use a $d{=}2{-}\eps$ formula for
the leading-order $[d\Gamma/d\xe]_{\xx\bar\xx}$ above, expanded
through $O(\eps)$.
The explicit formula is given in (\ref{eq:xxLO}) of appendix
\ref{app:Fund1}.

Now combine the decomposition (\ref{eq:xyyxSplit}) of $xyy\bar x$
into calculable divergent and finite parts with the renormalization
(\ref{eq:xyyxRen}), and choose  ${\cal D}_2(\Delta t)$
according to (\ref{eq:calD2prop}).
We find (see appendix \ref{app:xyyxRen})
\begin {equation}
  \left[ \frac{d\Gamma}{d\xe} \right]_{xyy\bar x}^{(\rm ren)}
  =
  -
  \frac{\Nf\alphaqed}{3\pi}
  \left[ \frac{d\Gamma}{d\xe} \right]_{x\bar x} 
  \biggl(
    \ln\Bigl( \frac{\mu^2}{(1{-}\xe)E\Omega_\ix} \Bigr)
    + \gammaE
    - 2\ln2
    + \tfrac53
  \biggr)
  +
  \left[ \frac{d\Gamma}{d\xe} \right]_{xyy\bar x}^{(\rm subtracted)}
,
\label {eq:xyyxRen2}
\end {equation}
where $\mu$ is the renormalization scale.
$[d\Gamma/d\xe]^{(\rm subtracted)}$ is as defined earlier
in (\ref{eq:xyyxSplit}), with
the result given by (\ref{eq:virtxyyxsub}) of appendix \ref{app:Fund1}.

An aspect of handling $xyy\bar x$ divergences worth mentioning here,
that was unnecessary in our previous
application of dimensional regularization to {\it real}\/ double splitting
$g \to ggg$ in ref.\ \cite{dimreg} (ACI3), is that we find we need
the $d=2{-}\eps$ result for a DGLAP splitting
function---specifically $P_{\gamma \to e\bar e}^{(d)}$, which is the additional
splitting function
factor in the $xyy\bar x$ divergence
(\ref{eq:FundDiv}) compared to the leading-order
$x\bar x$ diagram of fig.\ \ref{fig:diagLO}.


\subsection{Relation to the \boldmath$\gamma \to e \bar e$ virtual diagram (r)}

As mentioned earlier in section \ref{sec:frontend},
the diagram of fig.\ \ref{fig:diagsVIRT2}r
for NLO $\gamma \to e\bar e$
is related to the diagram of fig.\ \ref{fig:diagsVIRT}k just
discussed by a combination of front-end and back-end transformations.
Specifically, similar to (\ref{eq:frontend2}),
\begin {equation}
   2\Re
   \left[ \frac{d\Gamma}{d\ye} \right]_{\rm (r)}
   = + 2\Nf\Re \int_0^1 d\xe \>
   \left\{
     \left[ \frac{d\Gamma}{d\xe\,d\ye} \right]_{\rm (k)} \quad
       \vcenter{
         \hbox{with substitutions (\ref{eq:frontendxyE})}
         \hbox{followed by $\xe{\leftrightarrow}\ye$}
       }       
   \right\} ,
\label{eq:rfromk}
\end {equation}
where $[d\Gamma/d\xe\,d\ye]_{\rm (k)}$ represents the $\ye$-integrand
that integrates to (\ref{eq:xyyxRen2}), which may be interpreted as
\begin {multline}
  \left[ \frac{d\Gamma}{d\xe\,d\ye} \right]_{xyy\bar x}^{(\rm ren)}
  =
  -
  \frac{\Nf\alphaqed}{3\pi}
  \left[ \frac{d\Gamma}{d\xe} \right]_{x\bar x} 
  \biggl(
    \ln\Bigl( \frac{\mu^2}{(1{-}\xe)E\bar\Omega_\ix} \Bigr)
    + \gammaE
    - 2\ln2
    + \tfrac53
  \biggr) \, \delta(\ye)
\\
  +
  \left[ \frac{d\Gamma}{d\xe\,d\ye} \right]_{xyy\bar x}^{(\rm subtracted)}
\label {eq:xyyxRen3}
\end {multline}
with the understanding that $\int_0^1 d\ye \> \delta(\ye) = 1$.
Above, the use of $\bar\Omega_\ix \equiv \Omega_\ix \sgn(M_\ix)$ is necessary
to generalize the $xyy\bar x$ result to work with front-end
transformations, as explained in appendix \ref{app:barOmega}.

Since we combine leading and next-to-leading order results when
we renormalize as in (\ref{eq:ren1}), we note here that
the relation between 
leading-order results for $\gamma \to e\bar e$ and
$e \to \gamma e$ is%
\footnote{
  The normalization factor $(1{-}\ye)$ in (\ref{eq:frontbackLO}) is the
  type of normalization mentioned in footnote
  \ref{foot:frontnorm}.
  For the overall sign to work in (\ref{eq:frontbackLO}), one must use
  the generalized formula (\ref{eq:xxLO2}) for $[d\Gamma/d\xe]_{x\bar x}$ that
  works for any sign of $M_\ix$.
}
\begin {equation}
  2\Re \left[ \frac{d\Gamma}{d\ye} \right]_{y\bar y}^{\gamma \to e\bar e}
  =
  2 \Nf (1{-}\ye) \Re
  \left\{
    \left[ \frac{d\Gamma}{d\xe} \right]_{x\bar x}^{e \to \gamma e} \quad
       \vcenter{
         \hbox{with substitutions (\ref{eq:frontendxyE})}
         \hbox{followed by $\xe{\leftrightarrow}\ye$}
       }       
   \right\} .
\label {eq:frontbackLO}
\end {equation}
This can be understood as an application of the same transformation
law (\ref{eq:rfromk}) if one interprets
$[d\Gamma/d\xe\,d\ye]_{x\bar x}^{e \to \gamma e} \equiv
 [d\Gamma/d\xe]_{x\bar x}^{e \to \gamma e} \, \delta(\ye).$


\section {Conclusion}
\label {sec:conclusion}

A complete summary of our final formulas for rates is given
in Appendix \ref{app:summary}.
The work in this paper completes the analytic calculation of the
effect of overlapping formation times in QED in the simplifying
parametric limit
$\alpha \ll \Nf \alpha \ll 1$.
Even though that is a QED limit only a
theorist could love, we think it will be interesting to
look at the size of those effects on shower development as
one changes the value
of $\Nf \alpha$ to be closer and closer to $O(1)$, as a first exploration
of how to carry out similar calculations for QCD if one pushes
$\Nc\,\alphas(Q_\perp)$ to be larger and larger to answer the
question of how small or overwhelming overlap effects are in
practice.  We leave that QED analysis,
as well as discussion of numerical results
for the formulas presented here, to forthcoming work \cite{qedNfstop}.

With these QED results in hand, it should now be possible to complete
similar calculations for large-$\Nc$ QCD.





\begin{acknowledgments}

We are quite grateful to Diana Vaman,
Yuri Kovchegov, Risto Paatelainen, Tuomas Lappi, Guillaume Beuf,
Aleksi Kurkela, John Collins, R. Keith Ellis,
and Zvi Bern for useful conversations.
We also thank Omar Eldegawy for helping us double check corrections
made to the original publication of this article.
This work was supported, in part, by the U.S. Department
of Energy under Grant No.~DE-SC0007984.

\end{acknowledgments}

\appendix

\section {Summary of Results}
\label {app:summary}

In this appendix, we gather our results for three different
quantities:
\begin {equation}
  \left[ \frac{\Delta \, d\Gamma}{d\xe\,d\ye} \right]_{e\to e\bar e e} ,
  \qquad
  \left[ \frac{d\Gamma}{d\xe} \right]_{e\to\gamma e} ,
  \qquad
  \mbox{and} \qquad
  \left[ \frac{d\Gamma}{d\ye} \right]_{\gamma\to e\bar e} .
\end {equation}
The first is the total overlap correction of fig.\ \ref{fig:diags} to
real double splitting $e \to \gamma e \to e\bar e e$.
The last two are the single splitting
rates expanded through next-to-leading order,
\begin {align}
  \left[ \frac{d\Gamma}{d\xe} \right]_{e\to\gamma e}
  &=
  \left[ \frac{d\Gamma}{d\xe} \right]_{e\to\gamma e}^{\rm LO}
  +
  \left[ \frac{\Delta\,d\Gamma}{d\xe} \right]_{e\to\gamma e}^{\rm NLO} ,
\\
  \left[ \frac{d\Gamma}{d\ye} \right]_{\gamma\to e\bar e}
  &=
  \left[ \frac{d\Gamma}{d\ye} \right]_{\gamma\to e\bar e}^{\rm LO}
  +
  \left[ \frac{\Delta\,d\Gamma}{d\ye} \right]_{\gamma\to e\bar e}^{\rm NLO} ,
\end {align}
where the NLO contributions are given by figs.\ \ref{fig:diagsVIRT}
and \ref{fig:diagsVIRT2} respectively.
Note that our convention is to use $\xe$ to refer to the final
momentum fraction of an electron after bremsstrahlung and $\ye$ to
refer to the final momentum fraction of the electron in an
$e\bar e$ pair that was pair produced.


\subsection{Leading-order single-splitting rates}
\label {app:LO}

As discussed in appendix \ref{app:qhat}, the leading-order results
for single splitting in QED go all the way back to Migdal \cite{Migdal}.


\subsubsection {\boldmath$e \to \gamma e$}

In our notation, the leading-order $e{\to}\gamma e$ rate is%
\footnote{
   \label{foot:Omega0}
   QCD versions go back to BDMPS \cite{BDMPS12,BDMPS3}
   and Zakharov \cite{Zakharov}.
   To relate the QED and QCD results in our own notation:
   (\ref{eq:LO}) is AI (1.5a) with $P_{g \to g}(x)$ replaced by
   $P_{e \to e}(\xe)$.   The formulas for $\Omega_0$ in QED and QCD
   are both special cases of the general formula
   $\Omega^2 = -i [(\hat q_1/x_1)+(\hat q_2/x_2)+(\hat q_3/x_3)]/(2 E)$
   of AI (2.33b), where $\hat q_i$ is zero for the photon.
}
\begin {equation}
   \left[ \frac{d \Gamma}{d\xe} \right]_{e\to\gamma e}^{\rm LO}
   = \frac{\alpha}{\pi} \, P_{e\to e}(\xe) \, \Re(i\Omega_0)
   = \frac{\alpha}{\pi} \, P_{e\to \gamma}(x_\gamma) \, \Re(i\Omega_0)
\label {eq:LO}
\end {equation}
with $x_\gamma = 1-\xe$,
\begin {equation}
   \Omega_0
   = \sqrt{ \frac{-i(1-\xe)\hat q}{2 \xe E} }
   = \sqrt{ \frac{-ix_\gamma\hat q}{2(1-x_\gamma)E} }
\end {equation}
and the relevant DGLAP splitting function
\begin {equation}
  P_{e\to e}(\xe) = P_{e\to \gamma}(x_\gamma)
  = \frac{1 + \xe^2}{1-\xe} = \frac{1 + (1-x_\gamma)^2}{x_\gamma} .
\label {eq:Pee}
\end {equation}

For the sake of later formulas for virtual corrections, it will later
be helpful to also express the above result in terms
of the $x\bar x$ diagram of fig.\ \ref{fig:diagLO} as
\begin {equation}
   \left[ \frac{d \Gamma}{d\xe} \right]_{e\to\gamma e}^{\rm LO}
   = 2\Re \left[ \frac{d \Gamma}{d\xe} \right]_{x\bar x} .
\end {equation}


\subsubsection {\boldmath$\gamma \to e \bar e$}

Leading-order
pair production $\gamma \to e\bar e$ is essentially the same
except that
\begin {equation}
   \left[ \frac{d \Gamma}{d\ye} \right]_{\gamma\to e\bar e}^{\rm LO}
   = \frac{\Nf\alpha}{\pi} \, P_{\gamma\to e}(\ye) \,\Re(i\Omega_0^{\gamma\to e\bar e})
\label {eq:LOpair}
\end {equation}
with%
\footnote{
   See footnote \ref{foot:Omega0}.
}
\begin {equation}
   \Omega_0^{\gamma\to e\bar e}
   = \sqrt{ \frac{-i\hat q}{2 \ye (1-\ye) E} }
\end {equation}
and
\begin {equation}
  P_{\gamma\to e}(\ye) = \ye^2 + (1-\ye)^2 .
\label {eq:Pge}
\end {equation}


\subsection{Overlap corrections to real double splitting}

We decompose our total results for the overlap correction to
real double splitting as
\begin {equation}
  \left[\frac{\Delta\,d\Gamma}{d\xe\,d\ye}\right]_{e\to e\bar e e} =
  \left[ \frac{\Delta\,d\Gamma}{d\xe\,d\ye} \right]_{\rm seq}
  + \left[ \frac{d\Gamma}{d\xe\,d\ye} \right]_{(I)}
  + \left[ \frac{d\Gamma}{d\xe\,d\ye} \right]_{(II)} ,
\label {eq:total}
\end {equation}
where the pieces are given below and ``$I$'' indicates a
contribution involving an instantaneous 4-fermion interaction.
Below we mostly just collect formulas. See appendix \ref{app:real}
for a discussion of where those formulas come from.


\subsubsection{Sequential diagrams}
\label {app:seqsummary}

The ``seq'' piece above is our result for diagrams (a--c) of
fig.\ \ref{fig:diags}, which are
of a type we refer to as ``sequential'' diagrams in earlier papers.
Adopting notation similar to ACI4 appendix D.3 \cite{4point},
our result is
\begin {equation}
  \left[ \frac{\Delta\,d\Gamma}{d\xe\>d\ye} \right]_\seq
   =
   2\Nf {\cal A}_\seqNf(\xe,\ye)
\label {eq:dGammaseqNf}
\end {equation}
with
\begin {equation}
   {\cal A}_\seqNf(\xe,\ye)
   \equiv
   {\cal A}^{\rm pole}_\seqNf(\xe,\ye)
   + \int_0^{+\infty} d(\Delta t) \>
     \Bigl[
        2 \Re \bigl( B_\seqNf(\xe,\ye,\Delta t) \bigr)
        + F_\seqNf(\xe,\ye,\Delta t)
     \bigr] ,
\end {equation}
\begin {align}
   B_\seqNf(\xe,\ye,\Delta t) &=
       C_\seqNf(\hat x_1,\hat x_2,\hat x_3,\hat x_4,
               \bar\alpha,\bar\beta,\bar\gamma,\Delta t)
\nonumber\\
   &=
       C_\seqNf({-}1,y,1{-}x{-}y,x,
               \bar\alpha,\bar\beta,\bar\gamma,\Delta t) ,
\label {eq:Bseq}
\end {align}
\begin {equation}
   C_\seq = D_\seq - \lim_{\hat q\to 0} D_\seq ,
\end {equation}
\begin {align}
   D_\seq(x_1,&x_2,x_3,x_4,\bar\alpha,\bar\beta,\bar\gamma,\Delta t) =
\nonumber\\ &
   \frac{\alphaqed^2 M_\ix M_\fx^\seq}{32\pi^4 E^2} \, 
   ({-}x_1 x_2 x_3 x_4) \,
   \frac{ \Omega_+\csc(\Omega_+\Delta t) }{ \Delta t }
\nonumber\\ &\times
   \Bigl\{
     (\bar\beta Y_\yx^\seq Y_\xbx^\seq
        + \bar\alpha \Ybar_{\yx\xbx}^{\,\seq} Y_{\yx\xbx}^\seq) I_0^\seq
     + (\bar\alpha+\bar\beta+2\bar\gamma) Z_{\yx\xbx}^\seq I_1^\seq
\nonumber\\ &\quad
     + \bigl[
         (\bar\alpha+\bar\gamma) Y_\yx^\seq Y_\xbx^\seq
         + (\bar\beta+\bar\gamma) \Ybar_{\yx\xbx}^{\,\seq} Y_{\yx\xbx}^\seq
        \bigr] I_2^\seq
\nonumber\\ &\quad
     - (\bar\alpha+\bar\beta+\bar\gamma)
       (\Ybar_{\yx\xbx}^{\,\seq} Y_\xbx^\seq I_3^\seq+ Y_\yx^\seq Y_{\yx\xbx}^\seq I_4^\seq)
   \Bigl\}
   ,
\label {eq:Dseq}
\end {align}
\begin {align}
   F_\seq(\xe,\ye,\Delta t) = &
   \frac{ \alphaqed^2 P_{e\to e}(\xe) P_{\gamma\to e}(\frac{\ye}{1-\xe}) }
        { 4\pi^2(1-\xe) }
   \Bigl[ 
      \Re\bigl(i(\Omega\sgn M)_\ix\bigr) \,
      \Re\bigl( \Delta t \, (\Omega_\fx^\seq)^2
                \csc^2(\Omega_\fx^\seq \, \Delta t) \bigr)
\nonumber\\ & \qquad
      +
      \Re\bigl(i(\Omega\sgn M)_\fx^\seq) \,
      \Re\bigl( \Delta t \, \Omega_\ix^2 \csc^2(\Omega_\ix \, \Delta t) \bigr)
   \Bigr] ,
\label {eq:FseqNf}
\end {align}
\begin {subequations}
\label {eq:Aseqpole}
\begin {multline}
   {\cal A}_\seq^{\rm pole}(\xe,\ye)
   = - \frac{\alphaqed^2 \, P_{e\to e}(\xe) \, P_{\gamma\to e}(\frac{\ye}{1-\xe})}
          {4\pi^2(1-\xe)}
   \Re\Bigl[
     i (\Omega\sgn M)_\ix \, (1+\tfrac{i\pi}{2} \sgn M_\fx^\seq)
\\
     +
     i (\Omega\sgn M)_\fx^\seq \, (1+\tfrac{i\pi}{2} \sgn M_\ix)
   \Bigr]
\end {multline}
or equivalently%
\astfootnote{
   For the origin of the exact form of eqs.\ (\ref{eq:Aseqpole}),
   see appendix A of P.~Arnold, T.~Gorda and S.~Iqbal,
   ``The LPM effect in sequential bremsstrahlung:
     analytic results for sub-leading (single) logarithms,''
   JHEP \textbf{04}, 085 (2022)
   [arXiv:2112.05161 [hep-ph]].
}
\begin {multline}
   {\cal A}_\seq^{\rm pole}(\xe,\ye) =
   - \frac{\alphaqed^2 \, P_{e\to e}(\xe) \, P_{\gamma\to e}(\frac{\ye}{1-\xe})}
          {4\pi^2(1-\xe)}
   \Re\bigl[
     i (\Omega\sgn M)_\ix + i (\Omega\sgn M)_\fx^\seq
   \bigr]
\\ \times
   \bigl(1-\tfrac{\pi}{2} \sgn M_\ix \sgn M_\fx^\seq\bigr) ,
\end {multline}
\end {subequations}
and
\begin {equation}
   M_\ix = x_1 x_4 (x_1+x_4) E ,
   \qquad
   M_\fx^\seq = x_2 x_3 (x_2+x_3) E .
\label {eq:Mifseq}
\end {equation}
The $I_n^{\rm seq}$ above represent
\begin {subequations}
\label {eq:Iseq}
\begin {align}
   I_0^\seq &=
   \frac{4\pi^2}{[X_\yx^\seq X_\xbx^\seq - (X_{\yx\xbx}^\seq)^2]} \,,
\displaybreak[0]\\
   I_1^\seq &=
   - \frac{2\pi^2}{X_{\yx\xbx}^\seq}
   \ln\left( 1 - \frac{(X_{\yx\xbx}^\seq)^2}{X_\yx^\seq X_\xbx^\seq} \right) \,,
\displaybreak[0]\\
   I_2^\seq &=
   \frac{2\pi^2}{(X_{\yx\xbx}^\seq)^2}
     \ln\left( 1 - \frac{(X_{\yx\xbx}^\seq)^2}{X_\yx^\seq X_\xbx^\seq} \right)
   + \frac{4\pi^2}{[X_\yx^\seq X_\xbx^\seq - (X_{\yx\xbx}^\seq)^2]} \,,
\displaybreak[0]\\
   I_3^\seq &=
   \frac{4\pi^2 X_{\yx\xbx}^\seq}
        {X_\xbx^\seq[X_\yx^\seq X_\xbx^\seq - (X_{\yx\xbx}^\seq)^2]} \,,
\displaybreak[0]\\
   I_4^\seq &=
   \frac{4\pi^2 X_{\yx\xbx}^\seq}
        {X_\yx^\seq[X_\yx^\seq X_\xbx^\seq - (X_{\yx\xbx}^\seq)^2]} \,.
\end {align}
\end {subequations}
Here and in (\ref{eq:Dseq}), the $(X,Y,Z)^\seq$ are defined by 
\begin {subequations}
\label {eq:XYZseq}
\begin {align}
   \begin{pmatrix} X_\yx^\seq & Y_\yx^\seq \\ Y_\yx^\seq & Z_\yx^\seq \end{pmatrix}
   &\equiv
   \begin{pmatrix} |M_\ix|\Omega_\ix & 0 \\ 0 & 0 \end{pmatrix}
     - i a_\yx^{-1\top}
     \begin{pmatrix}
        \Omega_+ \cot(\Omega_+\,\Delta t) & 0 \\
        0 & (\Delta t)^{-1}
     \end{pmatrix}
     a_\yx^{-1} ,
\\
   \begin{pmatrix} X_\xbx^\seq & Y_\xbx^\seq \\ Y_\xbx^\seq & Z_\xbx^\seq \end{pmatrix}
   &\equiv
   \begin{pmatrix} |M_\fx^\seq|\Omega_\fx^\seq & 0 \\ 0 & 0 \end{pmatrix}
     - i (a_\xbx^\seq)^{-1\top}
     \begin{pmatrix}
        \Omega_+ \cot(\Omega_+\,\Delta t) & 0 \\
        0 & (\Delta t)^{-1}
     \end{pmatrix}
     (a_\xbx^\seq)^{-1} ,
\label {eq:XYZxbseq}
\\
   \begin{pmatrix} X_{\yx\xbx}^\seq & Y_{\yx\xbx}^\seq \\
                   \Ybar_{\yx\xbx}^\seq & Z_{\yx\xbx}^\seq \end{pmatrix}
   &\equiv
     - i a_\yx^{-1\top}
     \begin{pmatrix}
        \Omega_+ \csc(\Omega_+\,\Delta t) & 0 \\
        0 & (\Delta t)^{-1}
     \end{pmatrix}
     (a_\xbx^\seq)^{-1} ,
\end {align}
\end {subequations}
where the $a$'s and $\Omega$'s will be given below.
The quantities $(\bar\alpha,\bar\beta,\bar\gamma)$ in (\ref{eq:Bseq})
represent
various combinations of helicity-dependent DGLAP splitting functions
and are
\begin {equation}
   \begin{pmatrix}
      \bar\alpha \\ \bar\beta \\ \bar\gamma
   \end{pmatrix}_{e \to e\bar e e}
   =
   \frac{1}{(1-\xe)^6}
   \left\{
     \begin{pmatrix} - \\ + \\ + \end{pmatrix}
       \frac{4}{|\xe\ye\ze|}
   + \begin{pmatrix} + \\ + \\ - \end{pmatrix}
       \left[ \Bigl(1 + \frac{1}{\xe^2}\Bigr)
              \Bigl(\frac{1}{\ye^2} + \frac{1}{\ze^2}\Bigr)
       \right]
   \right\}
   ,
\label {eq:abcNf}
\end {equation}
where
\begin {equation}
   \ze \equiv 1{-}\xe{-}\ye .
\end {equation}
The normal-mode frequencies $\Omega$ and matrices $a$ of normal mode
vectors that are needed in the above formulas are
\begin {equation}
   \Omega_\ix(x_1,x_2,x_3,x_4) =
     \sqrt{
       \frac{-i \hat q}{2E}
       \Bigl(
          \frac{1}{x_1} + \frac{1}{x_4}
       \Bigr)
     }
    \,,
\label {eq:Omegai2}
\end {equation}
\begin {equation}
   \Omega_\fx^\seq(x_1,x_2,x_3,x_4)
   =
     \sqrt{
       \frac{-i \hat q}{2E}
       \Bigl(
          \frac{1}{x_2} + \frac{1}{x_3}
       \Bigr)
     } ,
\label{eq:Omegafseq4}
\end {equation}
\begin {equation}
   \Omega_+ =
   \sqrt{
     -\frac{i \hat q}{2E}
     \Bigl( \frac{1}{x_1} + \frac{1}{x_2} + \frac{1}{x_3} + \frac{1}{x_4} \Bigr)
   }
   ,
\label{eq:OmegaP}
\end {equation}
\begin {equation}
   a_\yx =
   \Bigl[
     (-x_1 x_2 x_3 x_4)
     \bigl(\frac{1}{x_1}+\frac{1}{x_2}+\frac{1}{x_3}+\frac{1}{x_4}\bigr)
     E
    \Bigr]^{-1/2}
   \begin {pmatrix}
      \sqrt{\frac{x_2 x_3}{-x_1 x_4}} & ~~1\\[8pt]
     -\sqrt{\frac{-x_1 x_4}{x_2 x_3}} & ~~1
   \end{pmatrix}
   ,
\label {eq:ayNf}
\end {equation}
and
\begin {equation}
   a_\xbx^\seq =
   \begin{pmatrix} 0 & 1 \\ 1 & 0 \end{pmatrix} a_\yx
   .
\label {eq:abx}
\end {equation}
We have written the $\Omega$'s and $a$'s in
general terms above, rather than plugging in the
specific values of $(x_1,x_2,x_3,x_4)$ used in (\ref{eq:Bseq}), because
the more general form will later be useful for
other types of diagrams.


\subsubsection{Diagrams with one instantaneous vertex}
\label {app:Isummary}

The ``(I)'' piece of our result (\ref{eq:total}) is the contribution
from diagrams (e--g) of fig.\ \ref{fig:diags}.
Paralleling the notation of ACI4 \cite{4point} as closely as possible,
\begin {subequations}
\label {eq:dGIsummary}
\begin {equation}
   \left[ \frac{d\Gamma}{d\xe\>d\ye} \right]_{(I)}
   =
   2\Nf \, {\cal A}_I(\xe,\ye) ,
\label {eq:dGammaI}
\end {equation}
\begin {equation}
   {\cal A}_I(\xe,\ye)
   \equiv
   \int_0^{\infty} d(\Delta t) \>
        2 \Re \bigl( B_I(\xe,\ye,\Delta t) \bigr)
   ,
\end {equation}
\begin {align}
   B_I(\xe,\ye,\Delta t) &=
       \frac{ 4 \bigl|\xe\ye(1{-}\xe{-}\ye)\bigr|^{1/2} }{ (1-\xe)^2 }
       \Bigl[
         D_I(\hat x_1,\hat x_2,\hat x_3,\hat x_4,\zeta,\Delta t)
\nonumber\\ & \hspace{10em}
         + D_I(-\hat x_3,-\hat x_4,-\hat x_1,-\hat x_2,\zeta,\Delta t)
      \Bigr]
\nonumber\\
   &=
       \frac{ 4 \bigl|\xe\ye(1{-}\xe{-}\ye)\bigr|^{1/2} }{ (1-\xe)^2 }
       \Bigl[
         D_I({-}1,\ye,1{-}\xe{-}\ye,\xe,\zeta,\Delta t)
\nonumber\\ & \hspace{10em}
         + D_I(-(1{-}\xe{-}\ye),-\xe,1,-\ye,\zeta,\Delta t)
       \Bigr]
   ,
\label {eq:BI}
\end {align}
\begin {equation}
   D_I(x_1,x_2,x_3,x_4,\zeta,\Delta t) =
   - \frac{\alphaqed^2 M_\fx^\seq}{16 \pi^2 E} \,
   (-x_1 x_2 x_3 x_4)
   \zeta \,
   \frac{\Omega_+ \csc(\Omega_+\,\Delta t)}{\Delta t} \,
   \frac{Y_\xbx^\seq}{X_\xbx^\seq}
   \,,
\label {eq:DI}
\end {equation}
\end {subequations}
where
\begin {equation}
   \zeta =
   \frac{(1{+}|\xe|)\,(|\ye|{+}|\ze|)}
        {(1-\xe)^3 |\xe\ye\ze|^{3/2}} \,.
\label {eq:zeta}
\end {equation}
$M_\fx^\seq$, $\Omega_\fx^\seq$, and $\Omega_+$ are given here by
the previous general formulas (\ref{eq:Mifseq}), (\ref{eq:Omegafseq4}),
and (\ref{eq:OmegaP}), for use in (\ref{eq:XYZxbseq})
for $X_\xbx^\seq$ and $Y_\xbx^\seq$.

Later, we will need to refer separately to the contributions of the
three diagrams (e--g). The above formula
for $[d\Gamma/d\xe\,d\ye]_{(I)}$ can be decomposed as
\begin {equation}
   \left[ \frac{d\Gamma}{d\xe\>d\ye} \right]_{(I)}
   =
   2\Re \left[ \frac{d\Gamma}{d\xe\>d\ye} \right]_{\rm(e)}
   +
   2\Re \left[ \frac{d\Gamma}{d\xe\>d\ye} \right]_{\rm(f)}
   +
   2\Re \left[ \frac{d\Gamma}{d\xe\>d\ye} \right]_{\rm(g)}
\end {equation}
where
\begin {align}
   2\Re \left[ \frac{d\Gamma}{d\xe\>d\ye} \right]_{\rm(e)}
   &=
   \mbox{Eqs.\ (\ref{eq:dGIsummary}) using only the
         {\it first} $D_I$ term in (\ref{eq:BI});}
\label {eq:dGe}
\\
   2\Re \left[ \frac{d\Gamma}{d\xe\>d\ye} \right]_{\rm(f)}
   &=
   \mbox{Eqs.\ (\ref{eq:dGIsummary}) using only the
         {\it second} $D_I$ term in (\ref{eq:BI});}
\label {eq:dGf}
\\
   2\Re \left[ \frac{d\Gamma}{d\xe\>d\ye} \right]_{\rm(g)}
   &=
   0.
\label {eq:dGg}
\end {align}


\subsubsection{Diagrams with two instantaneous vertices}
\label {app:IIsummary}

The ``(II)'' piece of our result (\ref{eq:total}) is the contribution
from diagram (d) of fig.\ \ref{fig:diags}, which gives
\begin {equation}
   \left[\frac{d\Gamma}{d\xe\,d\ye}\right]_{(II)}
   =
   2\Re \left[\frac{d\Gamma}{d\xe\,d\ye}\right]_{\rm(d)}
   = \frac{ 4\Nf\alphaqed^2 }{ \pi^2 }  \,
   \frac{\xe\ye\ze}{(1-\xe)^4}
      \Re(i \Omega_+) \ln 2
  .
\label {eq:II}
\end {equation}
This completes the set of formulas needed to numerically evaluate
$\Delta[d\Gamma/d\xe\,d\ye]_{e\to e\bar e e}$.


\subsection{NLO corrections to single splitting \boldmath$e\to\gamma e$}

We will decompose the contributions of fig.\ \ref{fig:diagsVIRT} to
single splitting $e\to \gamma e$ as
\begin {equation}
  \left[ \frac{\Delta\,d\Gamma}{d\xe} \right]_{e\to\gamma e}^{\rm NLO}
  =
  2\Re \biggl\{
    \left[ \frac{\Delta\,d\Gamma}{d\xe} \right]_{\rm(h+i+j)}
    +
    \left[ \frac{d\Gamma}{d\xe} \right]_{\rm(k)}
    +
    \left[ \frac{d\Gamma}{d\xe} \right]_{\rm(l)}
    +
    \left[ \frac{d\Gamma}{d\xe} \right]_{\rm(m)}
    +
    \left[ \frac{d\Gamma}{d\xe} \right]_{\rm(n)}
  \biggr\} .
\end {equation}
By the back-end transformation (\ref{eq:backend}),
\begin {align}
   2 \Re
   \left[ \frac{\Delta\,d\Gamma}{d\xe} \right]_{\rm (h+i+j)}
   &= - \int_0^{1-\xe} d\ye \>
        \left[\frac{\Delta\,d\Gamma}{d\xe\,d\ye}\right]_{\rm seq}
   ,
\\
   2 \Re
   \left[ \frac{d\Gamma}{d\xe} \right]_{\rm (l)} \quad
   &= - \int_0^{1-\xe} d\ye \>
         2\Re\left[\frac{d\Gamma}{d\xe\,d\ye}\right]_{\rm(e)}
   ,
\\
   2 \Re
   \left[ \frac{d\Gamma}{d\xe} \right]_{\rm (n)} \quad
   &= 0 ,
\end {align}
where the integrands on the right-hand side are specified in
(\ref{eq:dGammaseqNf}), (\ref{eq:dGe}) and implicitly (\ref{eq:dGg}).
By the combined front- and back-end transformation (\ref{eq:mtransform}),
\begin {equation}
   2\Re
   \left[ \frac{d\Gamma}{d\xe} \right]_{\rm (m)}
   = + \int_0^{1-\xe} d\ye \>
   \left\{
     2\Re \left[ \frac{d\Gamma}{d\xe\,d\ye} \right]_{\rm (e)}
     \mbox{with}~
     (\xe,\ye,E) \to
     \Bigl(
      \frac{-\ye}{\ze} \,,\,
      \frac{-\xe}{\ze} \,,\,
      \ze E\Bigr)
   \right\}
\end {equation}
with $2\Re[d\Gamma/d\xe\,d\ye]_{\rm(e)}$ again given by (\ref{eq:dGe}).

Finally, diagram (k) is computed in appendices \ref{app:Fund1}
and \ref{app:bbI}, giving
\begin {multline}
  2\Re\left[ \frac{d\Gamma}{d\xe} \right]_{\rm(k)}
  =
  2\Re\biggl\{ 
  -
  \frac{\Nf\alphaqed}{3\pi}
  \left[ \frac{d\Gamma}{d\xe} \right]_{x\bar x} 
  \biggl(
    \ln\Bigl( \frac{\mu^2}{(1{-}\xe)E\Omega_\ix\sgn M_\ix} \Bigr)
    + \gammaE
    - 2\ln2
    + \tfrac53
  \biggr)
  \biggr\}
\\
  +
  \int_0^{1-\xe} d\ye \>
    2\Re\left[ \frac{d\Gamma}{d\xe\,d\ye} \right]_{xyy\bar x}^{(\rm subtracted)} ,
\label {eq:dGk}
\end {multline}
where the leading-order $x\bar x$ diagram is
\begin {equation}
  \left[ \frac{d \Gamma}{d\xe} \right]_{x\bar x}
  = \frac{\alphaqed}{2\pi} \, P_{e\to e}(\xe) \, i\Omega_\ix \sgn M_\ix .
\label {eq:xxd2}
\end {equation}
The ``subtracted'' rate above is
\begin {align}
  \left[\frac{d\Gamma}{d\xe\,d\ye}\right]_{xyy\bar x}^{\rm(subtracted)} =
  - & \frac{\Nf\alphaqed^2 M_\ix^2}{16\pi^4 E^2} \,
  ({-}\hat x_1 \hat x_2 \hat x_3 \hat x_4)
  \int_0^\infty d(\Delta t)
  \Biggl[
\nonumber\\ &
   \frac{\Omega_+\csc(\Omega_+\Delta t)}{\Delta t}
   \Bigl\{
     (\bar\beta Y_\yx^2
        + \bar\gamma \Ybar_{\yx\yx'} Y_{\yx\yx'}) I_0^\new
     + (2\bar\alpha+\bar\beta+\bar\gamma) Z_{\yx\yx'} I_1^\new 
\nonumber\\ &\quad
     + \bigl[
         (\bar\alpha+\bar\gamma) Y_\yx^2
         + (\bar\alpha+\bar\beta) \Ybar_{\yx\yx'} Y_{\yx\yx'}
        \bigr] I_2^\new
\nonumber\\ &\quad
     - (\bar\alpha+\bar\beta+\bar\gamma)
       (\Ybar_{\yx\yx'} Y_\yx I_3^\new + Y_\yx Y_{\yx\yx'} I_4^\new)
   \Bigl\}
\nonumber\\ &
   - (2\bar\alpha+\bar\beta+\bar\gamma)
     \frac{\hat x_2\hat x_3}{\hat x_1\hat x_4} \, {\cal D}_2^{(\bbI)}
  \Biggr]
\end {align}
and
\begin {equation}
   {\cal D}_2^{(\bbI)}(\Delta t) =
   2\pi^2
   \left[
     \frac{\ln(2i\Omega_\ix \,\Delta t\sgn M_\ix)}{(\Delta t)^2}
     - i\Omega_\ix^3\,\Delta t \csc^2(\Omega_\ix\, \Delta t) \sgn M_\ix
   \right] .
\label {eq:D2Isummary}
\end {equation}
Here the $I^{\rm new}_n$ are the same as the
$I^\seq_n$ of (\ref{eq:Iseq}) except that the $(X,Y,Z)^\seq$ there are
replaced by
\begin {align}
   \begin{pmatrix} X_\yx^\new & Y_\yx^\new \\ Y_\yx^\new & Z_\yx^\new \end{pmatrix}
   =
   \begin{pmatrix} X_{\yx'}^\new & Y_{\yx'}^\new
                   \\ Y_{\yx'}^\new & Z_{\yx'}^\new \end{pmatrix}
   &\equiv
   \begin{pmatrix} |M_\ix|\Omega_\ix & 0 \\ 0 & 0 \end{pmatrix}
     - i a_\yx^{-1\top}
     \begin{pmatrix}
        \Omega_+ \cot(\Omega_+\,\Delta t) & 0 \\
        0 & (\Delta t)^{-1}
     \end{pmatrix}
     a_\yx^{-1} ,
\nonumber\\
   \begin{pmatrix} X_{\yx\yx'}^\new & Y_{\yx\yx'}^\new \\[2pt]
                   \Ybar_{\yx\yx'}^\new & Z_{\yx\yx'}^\new \end{pmatrix}
   &\equiv
     - i a_\yx^{-1\top}
     \begin{pmatrix}
        \Omega_+ \csc(\Omega_+\,\Delta t) & 0 \\
        0 & (\Delta t)^{-1}
     \end{pmatrix}
     a_\yx^{-1} .
\end {align}
[See (\ref{eq:Inew}) if further clarification desired.]
The $M$'s, $\Omega$'s and $a$'s are as in section
\ref{app:seqsummary} with $(x_1,x_2,x_3,x_4)$ set
to
$
  (\hat x_1,\hat x_2,\hat x_3,\hat x_4)=
  (-1,\ye,1{-}\xe{-}\ye,\xe)
$.
The only reason that the
factors of $\sgn M_\ix$ in (\ref{eq:dGk}--\ref{eq:D2Isummary}) are necessary
is to accommodate the transformation of diagram (k) that will later
be used to evaluate diagram (r).

The specific additive constants shown in (\ref{eq:dGk})
assume that the coupling $\alphaqed$ used in the leading-order calculation
(\ref{eq:LO}) of single splitting
is MS-bar $\alphaqed(\mu)$.
To use a different renormalization scheme, one would need to convert
(\ref{eq:dGk}) accordingly, but nothing else would change.


\subsection{NLO corrections to single splitting \boldmath$\gamma\to e\bar e$}

We will decompose the contributions of fig.\ \ref{fig:diagsVIRT2} to
single splitting $\gamma\to e\bar e$ as
\begin {equation}
  \left[ \frac{\Delta\,d\Gamma}{d\ye} \right]_{\gamma\to e\bar e}^{\rm NLO}
  =
  2\Re \biggl\{
    \left[ \frac{\Delta\,d\Gamma}{d\ye} \right]_{\rm(o+p+q)}
    +
    \left[ \frac{d\Gamma}{d\ye} \right]_{\rm(r)}
    +
    \left[ \frac{d\Gamma}{d\ye} \right]_{\rm(s)}
    +
    \left[ \frac{d\Gamma}{d\ye} \right]_{\rm(t)}
    +
    \left[ \frac{d\Gamma}{d\ye} \right]_{\rm(u)}
  \biggr\} .
\end {equation}
By the front-end transformation (\ref{eq:frontend1}),
\begin {align}
   2 \Re
   \left[ \frac{\Delta\,d\Gamma}{d\ye} \right]_{\rm (o+p+q)}
   &= - \Nf \int_0^1 d\xe
   \left\{
     \left[\frac{\Delta\,d\Gamma}{d\xe\,d\ye}\right]_{\rm seq}
     \mbox{with (\ref{eq:frontend1b}) below}
   \right\}
   ,
\\
   2 \Re
   \left[ \frac{d\Gamma}{d\ye} \right]_{\rm (s)} \quad
   &= - \Nf \int_0^1 d\xe \>
   \left\{
         2\Re\left[\frac{d\Gamma}{d\xe\,d\ye}\right]_{\rm(f)}
     \mbox{with (\ref{eq:frontend1b}) below}
   \right\}
   ,
\\
   2 \Re
   \left[ \frac{d\Gamma}{d\ye} \right]_{\rm (u)} \quad
   &= 0 ,
\end {align}
where
\begin {equation}
     (\xe,\ye,E) \to
     \Bigl(\frac{-\xe}{1-\xe}\,,\,\frac{\ye}{1-\xe}\,,\,(1{-}\xe)E\Bigr)
     .
\label {eq:frontend1b}
\end {equation}
By front- and back-end transformation (\ref{eq:frontend2}),
\begin {equation}
   2\Re
   \left[ \frac{d\Gamma}{d\ye} \right]_{\rm (t)}
   = \Nf \int_0^1 d\xe \>
   \left\{
     2\Re\left[ \frac{d\Gamma}{d\xe\,d\ye} \right]_{\rm (f)}
     \mbox{with}~
   (\xe,\ye,E) \to
   \Bigl(\frac{-(1{-}\ye)}{\ye},\frac{\xe}{\ye},\ye E\Bigr)
   \right\} .
\end {equation}
Finally, the result (\ref{eq:dGk}) for diagram (k) can
be transformed using (\ref{eq:rfromk}) to%
\footnote{
  See appendix \ref{app:frontend} for technicalities on getting the first
  term in (\ref{eq:dGr}) from the transformation of (\ref{eq:dGk}).
  Also, we have not bothered to write any general $\sgn M$ factors in
  (\ref{eq:dGr}) that would allow this result to in turn be transformed
  back again to diagram (k).  Instead, here we have just specialized to
  the specific values of $\sgn M$ of diagram (r).
}
\begin {multline}
  2\Re\left[ \frac{d\Gamma}{d\ye} \right]_{\rm(r)}
  =
  2\Re\biggl\{ 
  -
  \frac{\Nf\alphaqed}{3\pi}
  \left[ \frac{d\Gamma}{d\ye} \right]_{y\bar y}^{\gamma\to e\bar e} 
  \biggl(
    \ln\Bigl( \frac{\mu^2}{E\Omega_0^{\gamma\to e\bar e}} \Bigr)
    - i\pi
    + \gammaE
    - 2\ln2
    + \tfrac53
  \biggr)
  \biggr\}
\\
  +
  \Nf \int_0^1 d\xe \>
   \left\{
     2\Re\left[ \frac{d\Gamma}{d\xe\,d\ye} \right]_{xyy\bar x}^{(\rm subtracted)}
     \mbox{with}~
   (\xe,\ye,E) \to
   \Bigl(\frac{-\ye}{1{-}\ye},\frac{\xe}{1{-}\ye},(1{-}\ye)E\Bigr)
   \right\} ,
\label {eq:dGr}
\end {multline}
where
\begin {equation}
  \left[ \frac{d \Gamma}{d\ye} \right]_{y\bar y}
  = \frac{\Nf\alphaqed}{2\pi} \, P_{\gamma\to e\bar e}(\ye)
         \, i\Omega_0^{\gamma\to e\bar e}
\end {equation}
is the amplitude for which $2\Re[\cdots]$ gives the
leading-order pair production rate (\ref{eq:LOpair}).


\section{Similarities and dissimilarities with refs.\ \cite{Beuf2,HLaP}}
\label {app:smallx}

\begin {figure}[tp]
\begin {center}
  \includegraphics[scale=0.43]{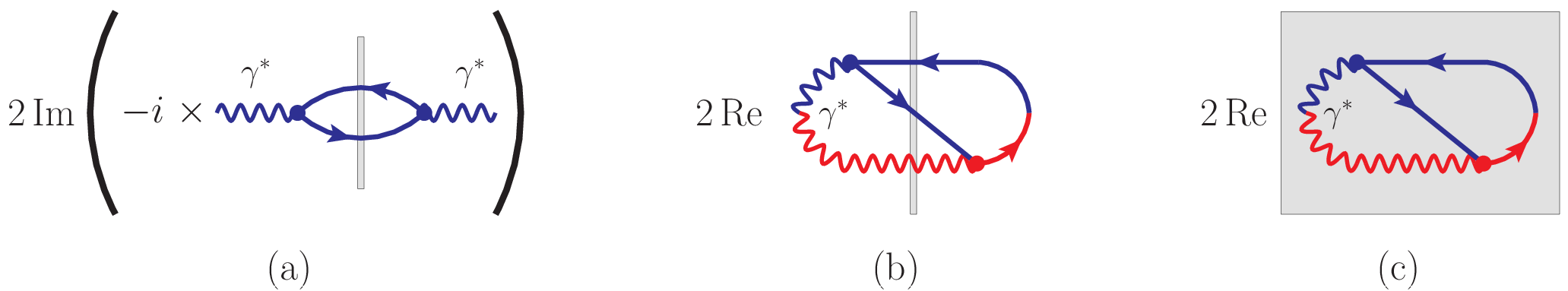}
  \caption{
     \label{fig:smallx}
     (a) The imaginary part of the leading-order (LO)
     time-ordered diagram for the virtual
     photon self-energy for an extremely thin medium, as in the 
     LO piece of the NLO results in refs.\ \cite{Beuf2,HLaP}.
     (b) The same, but drawn in the conventions \cite{2brem} of our
     fig.\ \ref{fig:diags}. (c) What the analogous process would be
     in the thick-medium limit considered in this paper.  In all cases,
     the gray area denotes the longitudinal size of the medium.
  }
\end {center}
\end {figure}

Refs.\ \cite{Beuf2,HLaP} study processes somewhat similar to ours but
in the context of next-to-leading-order DIS in the dipole
approximation appropriate to studying small-$x$ physics.  As is
standard, one can use the optical theorem to relate DIS cross-sections
to the self-energy of the virtual photon as depicted at leading order,
for example, by the time-ordered diagram in fig.\ \ref{fig:smallx}a. In
their application to small-$x$ physics, the medium is very
thin compared to the formation time, and we've depicted its extent by
the very thin gray region in fig.\ \ref{fig:smallx}a.
Fig.\ \ref{fig:smallx}b shows the same process but now drawn using the
conventions that we have used in fig.\ \ref{fig:diags}.%
\footnote{
  In case the reader is wondering how the overall sign matches up
  between fig.\ \ref{fig:smallx}a and \ref{fig:smallx}b: In our
  formalism, the red portion of the diagram represents the conjugate
  amplitude and so evolves with $e^{+i H t}$ instead of $e^{-i H t}$.
  When doing time-ordered perturbation theory, the sign difference in
  $e^{\mp i H t}$ manifests as a sign difference
  between red vertices and blue vertices in our diagram.
}
In this paper, however, our explicit calculations are
for the case where the medium is thick compared to the relevant
formation times, and so would be analogous to fig.\ \ref{fig:smallx}c
rather than fig.\ \ref{fig:smallx}b.

The photon in DIS is virtual, with a
virtuality $Q^2$ that should not be ignored.  In contrast, in our
calculations of in-medium showering, we approximate the initial
high-energy particle in figs.\ \ref{fig:diags}--\ref{fig:diagsVIRT2}
as on-shell, with negligible mass or virtuality.

Yet another difference is that refs.\ \cite{Beuf2,HLaP} write their
final answers explicitly in terms of Wilson lines running through the medium,
which should be averaged over medium fluctuations.
Because the medium is thin, these Wilson lines do not have time to
move transversely as they cross the medium.  For thick media, the
paths inside the medium do move transversely, and this dynamics is
incorporated in our treatment of the LPM effect \cite{2brem}
(based on Zakharov's picture \cite{Zakharov})
in the language of two-dimensional quantum mechanics with
a non-Hermitian potential energy.  This is an approximation that relies on
correlation lengths in the medium being small compared to both formation
times and the thickness of the medium.  Finally, our explicit calculation
in this paper
further takes this potential to be given by the $\hat q$ approximation,
appropriate for the LPM effect at high energy (but see appendix
\ref{app:qhat}).


\section{\boldmath$\hat q$ in QCD and QED}
\label {app:qhat}

In this appendix, we mention some qualitative differences between
QCD and QED concerning $\hat q$ and its logarithmic dependence on
energy at high energy.  Though the development of this paper does
not depend on the medium itself being weakly-coupled, we start by
discussing that case.


\subsection {\boldmath$\hat q$ for weakly-coupled systems}
\label {app:A1}

For weak coupling, $\hat q$ is given by
\begin {equation}
   \hat q = \int d^2 q_\perp \> q_\perp^2 \frac{d\Gamma_{\rm el}}{d^2 q_\perp},
\label {eq:qhat1}
\end {equation}
where $d\Gamma_{\rm el}/d^2 q_\perp$ is the rate of elastic scattering
from the medium for momentum transfer $\q_\perp$ perpendicular to the
direction of motion of the high-energy particle.
Coulomb interactions give
\begin {equation}
  \frac{d\Gamma_{\rm el}}{d^2 q_\perp} \propto \frac{\alpha^2}{q_\perp^4}
\label {eq:Coulomb}
\end {equation}
for $q_\perp$ large compared to the inverse electric screening length
$\xi^{-1}$ of the medium.%
\footnote{
  For a precise leading-order
  analysis of $\hat q$ in the case of ultra-relativistic plasmas,
  including screening effects with all the finicky details,
  see refs.\ \cite{LOqhat1,LOqhat2}.
}
For fixed small coupling $\alpha$,
this behavior leads to a logarithmic UV divergence in the integral
(\ref{eq:qhat1}) for $\hat q$ in both QED and QCD:
\begin {equation}
  \hat q
  \propto \int_{\sim \xi^{-1}}^\infty d(q_\perp^2) \> \frac{\alpha^2}{q_\perp^2} \,.
\label {eq:qhat2}
\end {equation}

One may then
define $\hat q(\Lambda)$ as a function of some UV cut-off
$\Lambda$ on $q_\perp$.  It turns out that the LPM calculation of
the rate for splitting processes such as hard bremsstrahlung
depends on $\hat q(Q_\perp)$ where $Q_\perp$ is the order of magnitude
of the total $q_\perp$ transferred to the high-energy particle during
a formation time.  As we'll review below, this effect (though not
in this language) goes all the way back to the results
of Migdal \cite{Migdal}.%
\footnote{
  For QCD, this observation can be found in the discussion of
  eq.\ (2.19) of ref.\ \cite{BDMPS3}.
}

That fact notwithstanding, it is interesting that
QCD $\hat q(\infty)$ as calculated by
(\ref{eq:qhat2}) is actually convergent if one accounts
for the running of the coupling as $\alpha = \alpha(q_\perp)$ in this
particular calculation.  The slow decrease of $\alpha$ with
increasing momentum due to asymptotic freedom is just enough to
then make the integral (\ref{eq:qhat2}) convergent.%
\footnote{
  See, for example, the discussion in ref.\ \cite{Peshier} or
  section 2.C of ref.\ \cite{simple}.
}
In contrast, the running of the coupling in QED
makes the divergence slightly worse since $\alphaqed(q_\perp)$ grows
with momentum.

But this apparent conclusion that $\hat q(\infty)$ is finite for QCD
is an artifact of ignoring other higher-order corrections
(besides running of the coupling) to the leading-order
analysis (\ref{eq:qhat2}), as we now review.


\subsection {Small-\boldmath$x$ logs in QCD}

In this paper, we have looked at splitting processes such as
high-energy bremsstrahlung, which are suppressed by a factor of
$\alphas(Q_\perp)$, which is small at high enough energy even if
the medium itself is strongly coupled, i.e.\ even if
$\alphas(\xi^{-1})$ is large.
Liou, Mueller, and Wu \cite{Wu0} have shown that the contribution
of high-energy gluon bremsstrahlung to $\hat q$ in QCD
is enhanced by a double logarithm, arising from emission of
nearly-collinear gluons
with longitudinal momenta between
the medium scale and the high-energy scale $E$.
This double log is related to double logs that occur in small-$x$ physics.
At large energy, the double logarithm compensates for the
small $\alphas(Q_\perp)$ and so QCD $\hat q(\infty)$ is
divergent after all, even after resummation of leading logarithms
at all orders.  Refs.\ \cite{Blaizot,Iancu,Wu} extended
this analysis to the use of $\hat q$ in the context of calculations of
high-energy splitting rates.
Ref.\ \cite{Blaizot} argues
that the effective value of $\hat q$ in splitting
calculations scales like $L^{2 \sqrt{\bar\alpha_{\rm s}}}$ for large energy $E$,
with $\bar\alpha_{\rm s} \equiv \Nc\alphas/\pi$,
and where $L$ in the present context (infinite medium) means the
formation length.  Since $L$ scales like $L \sim E^{1/2}$
(for fixed $x$ and small $\bar\alpha_{\rm s}$),
that gives $\hat q_{\rm eff} \sim E^{\sqrt{\bar\alpha_{\rm s}}}$, which diverges
as $E \to \infty$.


\subsection {The upshot for this paper}

Whether we are talking about QED or QCD, there is some sort of
logarithmic dependence of $\hat q(Q_\perp)$ on energy.
The $\hat q$ in this paper should be fixed to the one appropriate for the
energy $E$.  By making the approximation that we can describe
scattering from the medium in terms of a fixed effective value of
$\hat q$, we are ignoring one class of sub-leading corrections.
In the long term, one should be able to do a more complete calculation.
In the medium term, we will sidestep this particular issue in
forthcoming work \cite{qedNfstop} by looking
at certain characteristics of high-energy showers that are not sensitive
to the precise value of $\hat q$.


\subsection {The Coulomb logarithm in Migdal's QED result}

Since this is nominally a paper about QED (albeit large-$\Nf$ QED),
it may be helpful to relate our notation and the discussion of
Coulomb logs above to the early results by Migdal \cite{Migdal}
for high-energy QED showering off of a medium made up of
atoms, a useful summary of which can be found in the review
by Klein \cite{Klein}.
Migdal's explicit solutions to his equations rely on assuming
that Coulomb logarithms are large, and he does not try to
precisely compute the constants under the logarithms.
In the case of significant LPM suppression
(what Migdal would call $s \ll 1$),
Migdal's results for $e{\to}\gamma e$
can be rewritten in the following form at the same leading-log order:%
\footnote{
  It is easiest to take Migdal's results from Klein (72--77) \cite{Klein}
  because, among other things,
  Migdal is inconsistent about whether he works
  in units where $m_e = 1$.  In our notation, Klein's
  $E_{\rm LPM}/\xi(s)$ is $m_e^4/2\hat q_{\rm eff}(b)$, and
  Migdal and Klein's $s$ is
  $m_e^2 |\Omega|/8^{1/2}\hat q_{\rm eff}(b) \sim m_e^2/Q_\perp^2$.
  Our $x_\gamma$ is Klein's $k/E$, which he also calls $y$.
  Migdal's crisp-seeming result that $\xi(s) = 2$ for
  $s < s_1$ (what we refer to as $b \lesssim R_A$)
  is actually an approximation based on the coincidence
  that $\ln(a_Z/R_A) \approx 2\ln(a_Z/m_e^{-1})$.
}
\begin {equation}
   \frac{d \Gamma_{\rm LPM}}{d x_\gamma}
   = n \, \frac{d \sigma_{\rm LPM}}{d x_\gamma}
   \simeq \frac{\alpha}{\pi} \, P_{e\to\gamma}(x_\gamma) \, \Re(i\Omega)
   ,
\label {eq:Migdal1}
\end {equation}
with $P_{e\to\gamma}(x_\gamma)$ the DGLAP splitting function
$[1+(1-x_\gamma)^2]/x_\gamma$,
\begin {equation}
   \Omega =
   \sqrt{
      \frac{-i x_\gamma\, \hat q_{\rm eff}\bigl(|\hat q/\Omega|^{1/2}\bigr) }
           { 2(1-x_\gamma)E }
   } ,
\label{eq:OmegaMigdal}
\end {equation}
and
\begin {equation}
   \hat q_{\rm eff}(b^{-1}) = 8\pi Z^2 \alpha^2 n
      \begin {cases}
          \ln\bigl( \frac{a_Z}{b} \bigr) ,
                & R_{\rm A} \lesssim b \ll m_e^{-1} ; \\[4pt]
          \ln\bigl( \frac{a_Z}{R_A} \bigr) ,
                & b \lesssim R_A .
      \end {cases}
\label {eq:qhatMigdal}
\end {equation}
Physically, the argument of $\hat q_{\rm eff}$ in (\ref{eq:OmegaMigdal})
represents $Q_\perp \sim \sqrt{\hat q t_{\rm form}}$ (from the definition
of $\hat q$), remembering that the frequency $\Omega$
is of order $1/t_{\rm form}$ and that rough approximations are
all that are needed
for the argument of the logarithm in (\ref{eq:qhatMigdal}) for a
leading-log analysis.  The $\hat q$ in the argument of $\hat q_{\rm eff}$
can be interpreted as self-consistently $\hat q_{\rm eff}$ itself,
but it does not matter at leading-log order.
In (\ref{eq:qhatMigdal}), our notation $\hat q_{\rm eff}(b^{-1})$
is motivated by $b \sim 1/Q_\perp$
because we find it more convenient to express the right-hand side of
(\ref{eq:qhatMigdal})
in terms of transverse
distance scales.
Above, $m_e$ is the electron mass,
$R_A$ is the nuclear radius (which Migdal somewhat obscures
by approximating $R_A \simeq 0.5 Z^{1/3} \alpha / m_e$),
$a_Z$ is the length scale for screening of the nucleus's Coulomb
field by atomic electrons (for which Migdal uses the Thomas-Fermi
approximation $a_Z \simeq Z^{-1/3} a_0$), and $Z$ is atomic number.
The case
$b \gg m_e^{-1}$ not shown above would correspond to no LPM
suppression (what Migdal would call $s \gg 1$), and the
case $b \simeq m_e^{-1}$ would correspond to the transition where
the LPM effect is first turning on (which Migdal would call $s \sim 1$).

The point of writing these formulas in the above form is that
(\ref{eq:Migdal1}) is precisely the leading-order result
(\ref{eq:LO}) in our
notation.  And the leading-log calculation (\ref{eq:qhat1})
of $\hat q$ in Migdal's application is
\begin {equation}
   \hat q
   = \int d^2 q_\perp \> q_\perp^2 \frac{d\Gamma_{\rm el}}{d^2 q_\perp}
   = n \int \frac{d^2 q_\perp}{(2\pi)^2} \> \frac{Z^2 g^4}{q_\perp^4}
   = 8\pi Z^2 \alpha^2 n \int_{\sim a_Z^{-1}}^\Lambda \frac{d q_\perp}{q_\perp}
   = 8\pi Z^2 \alpha^2 n \ln\left( \frac{a_Z}{\Lambda^{-1}} \right) ,
\label {eq:qhatMigdal2}
\end {equation}
where $n$ is the density of atoms.
There are now two cases to consider.  (i) We mentioned previously
that the scale $Q_\perp$ acts as a UV cut-off on the relevant value
of $\hat q_{\rm eff}$ for splitting calculations.  Taking
$\Lambda \sim Q_\perp$, and then rewriting $Q_\perp$ as $1/b$
for the sake of expressing scales in terms of transverse distance
scales instead of transverse momentum, gives the logarithm shown
for the $R_A \lesssim b \ll m_e^{-1}$ case of (\ref{eq:qhatMigdal}).
(ii) Ignoring the sub-structure of the nucleus (whose effects are
suppressed by powers of $Z^{-1}$ except at very much higher energies%
\footnote{
  For $R_{\rm p} \lesssim b \ll R_A$, where $R_{\rm p}$ is the proton radius,
  the $q_\perp$ integrand in (\ref{eq:qhatMigdal2}) for $\Lambda = 1/b$
  is suppressed for $q_\perp \ll R_A$ because the charges inside the
  nucleus do not then contribute coherently.  (This suppression replaces
  $Z^2$ by $Z$ in the integrand by the time $q_\perp$ gets as small as
  $q_\perp \sim R_{\rm p}$.)  As a result, Migdal's
  leading-log result for what we'd call ${\hat q}_{\rm eff}(1/b)$
  does not change as $b$ drops below $R_A$.
  If one goes to energies high
  enough (and so $b$ small enough)
  to probe the substructure of the nucleons (which Migdal
  did not know about), then, as a matter of principle, eventually
  the contribution of scattering from individual quarks
  would become important at sufficiently small $b \ll R_{\rm p}$.
}%
),
there is a UV-cutoff on the nucleus's Coulomb field at
$b \sim R_A$.
The effective cut-off for splitting calculations
is therefore $\Lambda^{-1} \sim \min(Q_\perp,R_A^{-1})$
instead of just $\Lambda \sim Q_\perp$, which accounts for the other
case of (\ref{eq:qhatMigdal}).


\section{Diagrammatic vertex rules}
\label {app:VertexRules}

\begin {figure}[p]
\begin {center}
  \begin{picture}(468,535)(0,0)
  \put(0,460){
    \begin{picture}(470,53)(0,-8)
    \put(0,4){\includegraphics[scale=0.5]{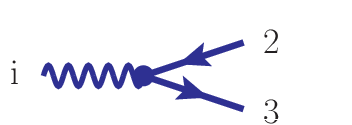}}
    \put(100,12){$
      \langle 2,3 | \delta H | \ix \rangle_{\rm rel} =
      -g \bar u_3 \slashed{\varepsilon}_\ix v_2
    $}
    \put(-1,-1){\framebox(222,32){}}
    \end{picture}
  }
  \put(0,405){
    \begin{picture}(470,45)(0,0)
    \put(-2,0){\includegraphics[scale=0.5]{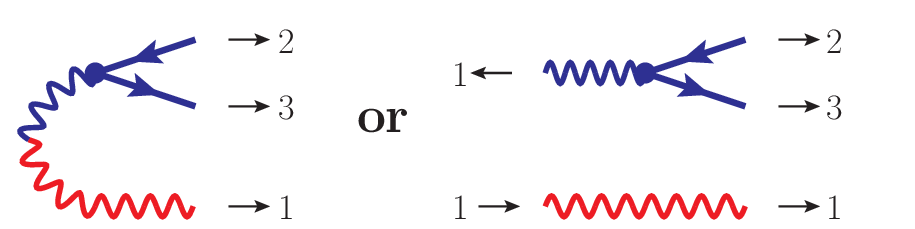}}
    \put(215,30){$
      \langle \P | {-}i\,\delta H | \rangle =
      -i \frac{\rm above}
              {\sqrt{|2 E_1| |2 E_2| |2 E_3|}}
      = \frac{i g}{2 E^{3/2}} \, \bcalP_{\gamma\to e\bar e}\cdot\P_{32}
    $}
    \put(215,5){$
      \langle \B | {-}i\,\delta H | \rangle
      = \frac{g}{2 E^{3/2}} \, \bcalP_{\gamma\to e\bar e}\cdot
        \grad \delta^{(2)}(\B_{32})
    $}
    \end{picture}
  }
  \put(0,320){
    \begin{picture}(470,53)(0,-18)
    \put(88,-8){\includegraphics[scale=0.5]{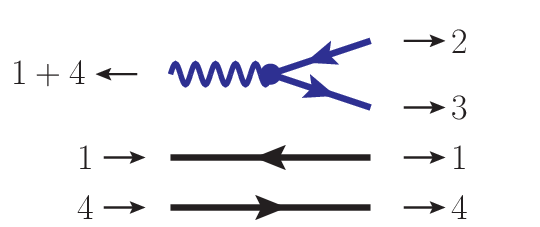}}
    \put(215,37){$
      \langle \P_{32},\P_{41} | {-}i\,\delta H | \P' \rangle
    $}
    \put(250,22){$
      = {\rm previous} \times
        |x_2+x_3|^{-1} (2\pi)^2 \delta^{(2)}(\P'_{41}{-}\P_{41})
    $}
    \put(215,3){$
      \langle \B_{32},\B_{41} | {-}i\,\delta H | \B' \rangle
    $}
    \put(250,-12){$
      = {\rm previous} \times |x_2+x_3|^{-1} \delta^{(2)}(\B'_{41}{-}\B_{41})
    $}
    \end{picture}
  }
  \put(0,245){
    \begin{picture}(470,53)(0,-8)
    \put(-2,-4){\includegraphics[scale=0.5]{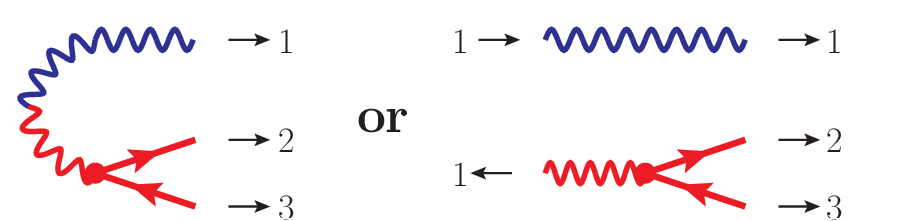}}
    \put(215,37){$
      \langle \P | {+}i\,\overline{\delta H} | \rangle
      = \langle \P | {-}i\,\delta H | \rangle^*_{(x,\p)\to(-x,-\p)}
    $}
    \put(250,22){$
      = - \frac{i g}{2 E^{3/2}} \, \bcalP_{\gamma\to e\bar e}^*\cdot\P_{32}
    $}
    \put(215,3){$
      \langle \B | {+}i\,\overline{\delta H} | \rangle
      = \langle \B | {-}i\,\delta H | \rangle^*_{(x,\b)\to(-x,\b)}
    $}
    \put(250,-12){$
      = - \frac{g}{2 E^{3/2}} \, \bcalP_{\gamma\to e\bar e}^*\cdot
        \grad \delta^{(2)}(\B_{32})
    $}
    \end{picture}
  }
  \put(0,160){
    \begin{picture}(470,53)(0,-18)
    \put(88,-8){\includegraphics[scale=0.5]{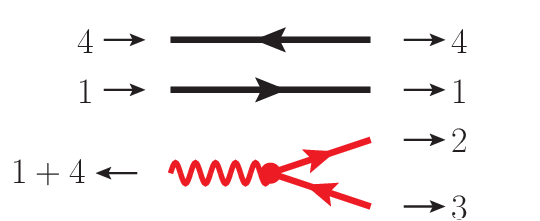}}
    \put(215,37){$
      \langle \P_{32},\P_{41} | {+}i\,\overline{\delta H} | \P' \rangle
    $}
    \put(250,22){$
      = {\rm previous} \times
        |x_2+x_3|^{-1} (2\pi)^2 \delta^{(2)}(\P'_{41}{-}\P_{41})
    $}
    \put(215,3){$
      \langle \B_{32},\B_{41} | {+}i\,\overline{\delta H} | \B' \rangle
    $}
    \put(250,-12){$
      = {\rm previous} \times |x_2+x_3|^{-1} \delta^{(2)}(\B'_{41}{-}\B_{41})
    $}
    \end{picture}
  }
  \put(0,80){
    \begin{picture}(470,53)(0,-8)
    \put(-2,-4){\includegraphics[scale=0.5]{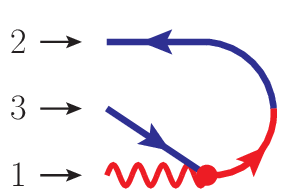}}
    \put(215,30){$
      \langle | {+}i\,\overline{\delta H} | \P \rangle =
      \langle \P | {-}i\,\delta H | \rangle^*
      = - \frac{i g}{2 E^{3/2}} \, \bcalP_{\gamma\to e\bar e}^*\cdot\P_{32}
    $}
    \put(215,9){$
      \langle | {+}i\,\overline{\delta H} | \B \rangle
      = \langle \B | {-}i\,\delta H | \rangle^*
    $}
    \put(250,-4){$
      = \frac{g}{2 E^{3/2}} \, \bcalP_{\gamma\to e\bar e}^*\cdot
        \grad \delta^{(2)}(\B_{32})
    $}
    \end{picture}
  }
  \put(0,0){
    \begin{picture}(470,53)(0,-8)
    \put(-2,5){\includegraphics[scale=0.5]{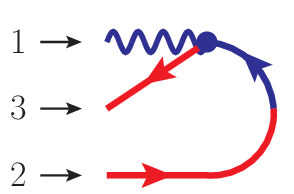}}
    \put(215,45){$
      \langle | {-}i\,\delta H | \P \rangle =
      \langle | {+}i\,\overline{\delta H} | \P \rangle^*_{(x,\p)\to(-x,-\p)}
    $}
    \put(250,30){$
      = \frac{i g}{2 E^{3/2}} \, \bcalP_{\gamma\to e\bar e}\cdot\P_{32}
    $}
    \put(215,9){$
      \langle | {-}i\,\delta H | \B \rangle
      = \langle | {+}i\,\overline{\delta H} | \B \rangle^*_{(x,\b)\to(-x,\b)}
    $}
    \put(250,-6){$
      = - \frac{g}{2 E^{3/2}} \, \bcalP_{\gamma\to e\bar e}^*\cdot
        \grad \delta^{(2)}(\B_{32})
    $}
    \end{picture}
  }
  \end{picture}
  \caption{
     \label{fig:VertexPair}
     Like figs.\ \ref{fig:VertexEtoGE} and \ref{fig:VertexEtoGEmore}
     but for pair production.  We only show elements that are directly
     used in the diagrams of figs.\ \ref{fig:diags}--\ref{fig:diagsVIRT2}.
  }
\end {center}
\end {figure}

\begin {figure}[p]
\begin {center}
  \begin{picture}(468,290)(0,0)
  \put(0,215){
    \begin{picture}(470,53)(0,-8)
    \put(0,4){\includegraphics[scale=0.5]{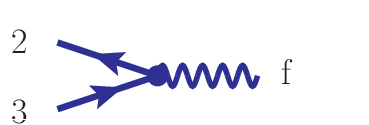}}
    \put(100,12){$
      \langle \fx | \delta H | 2,3 \rangle_{\rm rel} =
       - g \bar v_2 \slashed{\varepsilon}_\fx u_3
    $}
    \put(-1,-1){\framebox(222,32){}}
    \end{picture}
  }
  \put(0,160){
    \begin{picture}(470,45)(0,0)
    \put(-2,0){\includegraphics[scale=0.5]{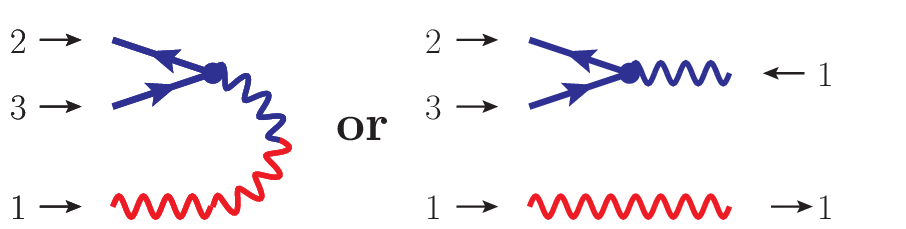}}
    \put(215,30){$
      \langle | {-}i\,\delta H | \P \rangle =
      -i \frac{\rm above}
              {\sqrt{|2 E_1| |2 E_2| |2 E_3|}}
      = \frac{i g}{2 E^{3/2}} \, \bcalP_{\gamma\to e\bar e}^*\cdot\P_{32}
    $}
    \put(215,5){$
      \langle | {-}i\,\delta H | \B \rangle
      = - \frac{g}{2 E^{3/2}} \, \bcalP_{\gamma\to e\bar e}^*\cdot
        \grad \delta^{(2)}(\B_{32})
    $}
    \end{picture}
  }
  \put(0,75){
    \begin{picture}(470,53)(0,-18)
    \put(88,-8){\includegraphics[scale=0.5]{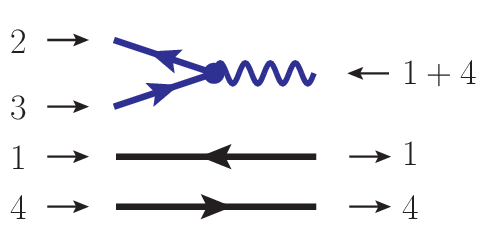}}
    \put(215,37){$
      \langle \P' | {-}i\,\delta H | \P_{32},\P_{41} \rangle
    $}
    \put(250,22){$
      = {\rm previous} \times
        |x_2+x_3|^{-1} (2\pi)^2 \delta^{(2)}(\P'_{41}{-}\P_{41})
    $}
    \put(215,3){$
      \langle \B' | {-}i\,\delta H | \B_{32},\B_{41} \rangle
    $}
    \put(250,-12){$
      = {\rm previous} \times |x_2+x_3|^{-1} \delta^{(2)}(\B'_{41}{-}\B_{41})
    $}
    \end{picture}
  }
  \put(0,0){
    \begin{picture}(470,53)(0,-8)
    \put(-2,-4){\includegraphics[scale=0.5]{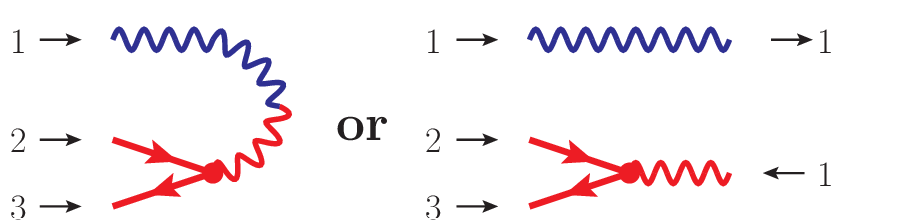}}
    \put(215,37){$
      \langle | {+}i\,\overline{\delta H} | \P \rangle
      = \langle | {-}i\,\delta H | \P \rangle^*_{(x,\p)\to(-x,-\p)}
    $}
    \put(250,22){$
      = - \frac{i g}{2 E^{3/2}} \, \bcalP_{\gamma\to e\bar e}\cdot\P_{32}
    $}
    \put(215,3){$
      \langle | {+}i\,\overline{\delta H} | \B \rangle
      = \langle | {-}i\,\delta H | \B \rangle^*_{(x,\b)\to(-x,\b)}
    $}
    \put(250,-12){$
      = \frac{g}{2 E^{3/2}} \, \bcalP_{\gamma\to e\bar e}\cdot
        \grad \delta^{(2)}(\B_{32})
    $}
    \end{picture}
  }
  \end{picture}
  \caption{
     \label{fig:VertexPairMerge}
     Like fig.\ \ref{fig:VertexPair}
     but for {\it inverse} pair production.
  }
\end {center}
\end {figure}

\begin {figure}[p]
\begin {center}
  \begin{picture}(468,255)(0,0)
  \put(0,170){
    \begin{picture}(220,30)(0,0)
    \put(12,0){\includegraphics[scale=0.5]{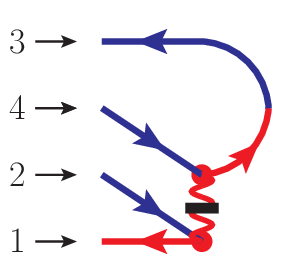}}
    \put(156,45){$
      \langle | {+}i\,\overline{\delta H} | \P_{34},\P_{12} \rangle =
      \langle \P_{34},\P_{12} | {-}i\,\delta H | \rangle^*
      = \frac{ig^2}{|x_3+x_4|^3 E^2}
    $}
    \put(156,20){$
      \langle | {+}i\,\overline{\delta H} | \B_{34},\B_{12} \rangle
      = \frac{ig^2}{|x_3+x_4|^3 E^2} \,
        \delta^{(2)}(\B_{34}) \, \delta^{(2)}(\B_{12})
    $}
    \end{picture}
  }
  \put(0,85){
    \begin{picture}(220,30)(0,0)
    \put(-2,0){\includegraphics[scale=0.5]{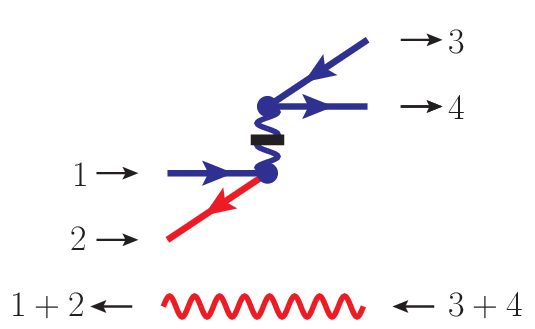}}
    \put(156,45){$
      \langle \P' | {-}i\,\delta H | \P \rangle
      =
      -i
      \frac{\rm boxed}{\sqrt{|2 E_1| |2 E_2| |2 E_3| |2 E_4|}}
      \,
      (x_3{+}x_4)^{-2}
      = \frac{-ig^2}{(x_3+x_4)^4 E^2} \,
    $}
    \put(156,20){$
      \langle \B' | {-}i\,\delta H | \B \rangle
      = \frac{-ig^2}{|x_3+x_4|^3 E^2} \,
        \delta^{(2)}(\B'_{34}) \, \delta^{(2)}(\B_{12})
    $}
    \end{picture}
  }
  \put(0,0){
    \begin{picture}(470,70)(0,0)
    \put(12,0){\includegraphics[scale=0.5]{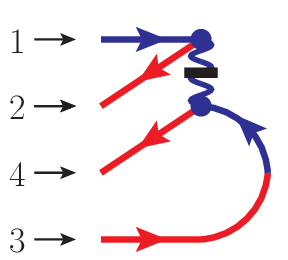}}
    \put(156,45){$
      \langle | {-}i\,\delta H | \P_{34},\P_{12} \rangle
      = \langle \P_{34},\P_{12} | {+}i\,\overline{\delta H}
          | \rangle^*_{(x,\p)\to(-x,-\p)}
      = \frac{-ig^2}{|x_3+x_4|^3 E^2}
    $}
    \put(156,20){$
      \langle | {-}i\,\delta H | \B_{34},\B_{12} \rangle
      = \frac{-ig^2}{|x_3+x_4|^3 E^2} \,
        \delta^{(2)}(\B_{34}) \, \delta^{(2)}(\B_{12})
    $}
    \end{picture}
  }
  \end{picture}
  \caption{
    \label{fig:VertexEtoEEE2}
    The other variations of fig.\ \ref{fig:VertexEtoEEE}
    that appear in figs.\ \ref{fig:diags}--\ref{fig:diagsVIRT2}.
    Above, ``boxed'' refers to the boxed LCPT rule in
    fig.\ \ref{fig:VertexEtoEEE}.
  }
\end {center}
\end {figure}

\begin {figure}[tp]
\begin {center}
  \begin{picture}(460,225)(0,0)
  \put(0,170){
    \begin{picture}(220,30)(0,-8)
    \put(0,0){\includegraphics[scale=0.5]{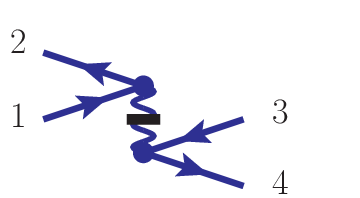}}
    \put(100,19){$
      \langle 3,4 | \delta H | 1,2 \rangle_{\rm rel} =
      -
      \frac{g^2
        (\bar v_2 \gamma^+ u_1)
        (\bar u_4 \gamma^+ v_2)
      }
      {(p_3^++p_4^+)^2}
    $}
    \put(-1,-1){\framebox(285,45){}}
    \end{picture}
  }
  \put(0,85){
    \begin{picture}(410,53)(0,-18)
    \put(0,-8){\includegraphics[scale=0.5]{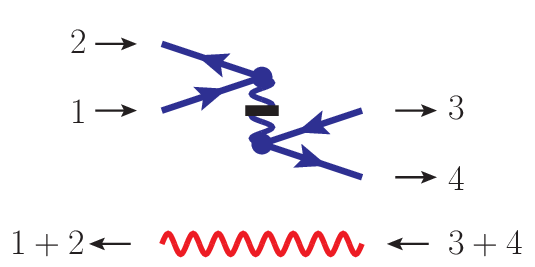}}
    \put(156,40){$
      \langle \P' | {-}i\,\delta H | \P \rangle =
      -i
      \frac{\rm above}{\sqrt{|2 E_1| |2 E_2| |2 E_3| |2 E_4|}} \,
      (x_3{+}x_4)^{-2}
      = \frac{ig^2}{(x_3+x_4)^4 E^2}
    $}
    \put(156,0){$
      \langle \B' | {-}i\,\delta H | \B \rangle
      = \frac{ig^2}{(x_3+x_4)^4 E^2} \,
        \delta^{(2)}(\B'_{34}) \, \delta^{(2)}(\B_{12})
    $}
    \end{picture}
  }
  \put(0,0){
    \begin{picture}(470,53)(0,-8)
    \put(14,-4){\includegraphics[scale=0.5]{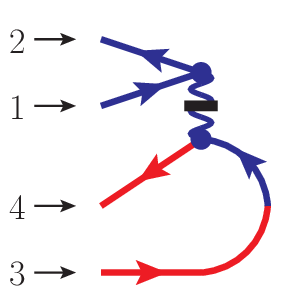}}
    \put(156,37){$
      \langle | {-}i\,\delta H | \P_{34},\P_{12} \rangle
      = \frac{-i g^2}{|x_3+x_4|^3 E^2}
    $}
    \put(156,3){$
      \langle | {-}i\,\delta H | \B_{34},\B_{12} \rangle
      = \frac{-i g^2}{|x_3+x_4|^3 E^2} \,
        \delta^{(2)}(\B_{34}) \, \delta^{(2)}(\B_{12})
    $}
    \end{picture}
  }
  \end{picture}
  \caption{
     \label {fig:VertexeeLee}
     Like figs.\ \ref{fig:VertexEtoEEE} and \ref{fig:VertexEtoEEE2}
     but for the LCPT $e\bar e \to e\bar e$
     vertex.
  }
\end {center}
\end {figure}

In the main text, figs.\ \ref{fig:VertexEtoGE}--\ref{fig:VertexEtoGEmore}
gave examples of LCPT vertex rules and the corresponding rules in our formalism.
The rest of the rules that we need for the large-$\Nf$ QED diagrams
of figs.\ \ref{fig:diags}--\ref{fig:diagsVIRT2} are shown explicitly
in figs.\ \ref{fig:VertexPair}--\ref{fig:VertexeeLee}.
In some transformations, we must negate momentum variables
$(x_n,\p_n) \to (-x_n,-\p_n)$, as indicated, to account for our convention
that momentum variables are negated for particles in conjugate-amplitudes
(red lines in the diagrams).  Transverse position variables $\b_n$
are unaffected.  When applied to the definitions (\ref{eq:Pdef})
and (\ref{eq:Bdef}) of $\P_{ij}$ and $\B_{ij}$, this transformation takes
$(\P_{ij},\B_{ij}) \to (\P_{ij},-\B_{ij})$.

The annoying part of working with the rules laid out for matrix
elements is that one must be careful about Fermi statistics in the
representation of states, e.g.\ $|e \bar e\rangle = -|\bar e
e\rangle$.  The boxed LCPT rule for $\bar e e \to \gamma$ in
fig.\ \ref{fig:VertexPairMerge}, for example, could just as well have
been written as $\langle \fx | \delta H | 3,2 \rangle_{\rm rel} = + g
\bar v_2 \slashed{\varepsilon}_\fx u_3$.  The sign would be
compensated, in a photon self-energy diagram for example, by whether
the propagator between the two vertices was $\langle 2,3;t' |
2,3;t\rangle$ or its negative $\langle 2,3;t' | 3,2;t\rangle$.  (The
minus sign in the last case would be equivalent to the usual
accounting where one says that fermion loops come with minus signs.)

We should also clarify a point about our formulas for
$e \to e\bar e e$ matrix elements, such as
$\langle 2,3,4|\delta H|\ix\rangle_{\rm rel}$ and related formulas in
figs.\ \ref{fig:VertexEtoEEE} and \ref{fig:VertexEtoEEE2}.
In these cases, we are {\it only}
giving the specific contribution associated with the
accompanying vertex diagram.  For example,
the contribution from swapping the two final-state
electrons (which in any case is sub-leading in $1/\Nf$)
is not included, and would correspond to drawing a different
vertex diagram where the ``$2$'' and ``$4$'' lines were switched.

\begin {figure}[tp]
\begin {center}
  \begin{picture}(410,170)(0,0)
    \put(25,110){\includegraphics[scale=0.7]{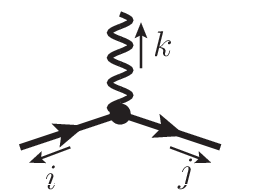}}
    \put(140,160){$
     = - \frac{g}{2 E^{3/2}} \, \bcalP_{\bar ee\gamma}(x_i,x_j,x_k) \cdot
       \grad \delta^{(2)}(\bcalB_{ji})
    $}
    \put(200,130){$
       \times
       \begin {cases}
          +1,  & \mbox{vertex in amplitude} \\
          -1 , & \mbox{vertex in conjugated amplitude}
       \end {cases}
    $}
    \put(0,0){\includegraphics[scale=0.7]{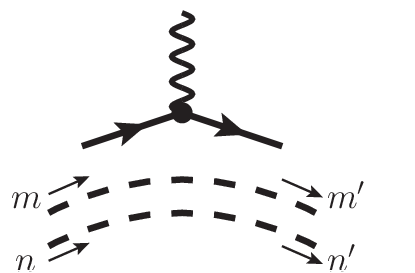}}
    \put(140,45){$
     = (\mbox{above}) \times |x_m+x_n|^{-1} \,
            \delta^{(2)}({\bcalB}_{mn}-{\bcalB}_{mn}')
    $}
  \end{picture}
  \caption{
     \label {fig:rule}
     (top)
     A general diagrammatic 3-point
     rule
     that covers all the various cases for $0{\leftrightarrow}3$
     and $2{\leftrightarrow}3$ particle transitions in the interference
     diagrams of
     figs.\ \ref{fig:diags}--\ref{fig:diagsVIRT2} and \ref{fig:diagLO},
     provided one also
     includes a minus sign for every fermion loop in the interference
     diagram. There is no arrow of time in the drawing of the figure:
     each of the three lines could be initial or final ones in the
     corresponding matrix element.
     Above,
     ${\bcalB}_{ij} \equiv (\b_i-\b_j)/(x_i+x_j)$,
     the signs of the momentum fractions $x$ are to be taken
     according to the flow of longitudinal momentum in the direction
     of the small arrows in the figure, and
     ${\bcalB}_{ij} = {\bcalB}_{jk} = {\bcalB}_{ki}$.
     (bottom) The corresponding rule for $3{\leftrightarrow}4$ particle
     transitions, where the dashed lines represent spectators that
     could be fermions or photons.
  }
\end {center}
\end {figure}

\begin {figure}[tp]
\begin {center}
  \includegraphics[scale=0.7]{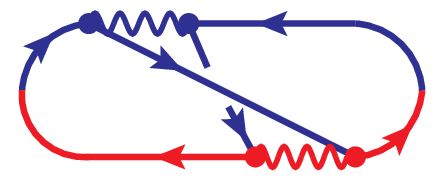}
  \caption{
     \label {fig:exchange}
     Example of a final-state electron exchange contribution (which is
     sub-leading in $1/\Nf$).
 }
\end {center}
\end {figure}

Keeping track of all the signs and different cases for matrix elements
is painstaking.  However, there is an equivalent way to formulate our
rules for diagrams that makes it easier and a little bit more like the
conventions for Feynman rules.  All the cases of
$1{\leftrightarrow}2$ splittings can be subsumed by the rule shown
in fig.\ \ref{fig:rule}, supplemented by a minus sign for
each fermion loop in the interference diagrams of
figs.\ \ref{fig:diags}--\ref{fig:diagsVIRT2}
and \ref{fig:diagLO}.
The rule of fig.\ \ref{fig:rule} sometimes differs by a sign
from the particular conventions of the $-i\,\delta H$ and
$+i\,\overline{\delta H}$
matrix elements we have written down for individual vertices,
but the supplemental minus signs for each fermion loop brings
the two different procedures into agreement and also
would produce the correct Fermi-statistics sign for
exchange diagrams such as fig.\ \ref{fig:exchange}
(which are sub-leading in $1/\Nf$).
There are different ways
one could assign signs to the vertices that would give the same
net overall sign for interference diagrams such as figs.\
\ref{fig:diagsVIRT}--\ref{fig:diagsVIRT2}; we've picked one of them.
The rule of fig.\ \ref{fig:rule} has the same form as the similar
rule given in AI1 \cite{2brem} for the 3-gluon vertex except for
details about signs.
In the rule, the factor ${\bcalP}_{\bar e e\gamma}(x_i,x_j,x_k)$ depends
implicitly on the helicities $h$ of the lines as measured in the directions
of the small arrows.  The definition of ${\bcalP}_{\bar e e\gamma}$ is
as in AI1 section IV.E, adapted here for the
QED case as
\begin {equation}
   \bcalP_{ijk}^{\bar e e\gamma} =
   \frac{{\bm e}_{(h_i+h_j+h_k)}}{ |x_i x_j x_k| }
   \sqrt{P^{\bar e e\gamma}_{h_i,h_j,h_k}(x_i,x_j,x_k)}
\end {equation}
with spin-dependent DGLAP splitting functions
\begin {subequations}
\label{eq:Pexplicit}
\begin {align}
   P^{\bar e e\gamma}_{-++}(x_i,x_j,x_k)
   \equiv P^{e \to e\gamma}_{+\to++}({-}x_i \to x_j x_k)
   = P^{\gamma \to e\bar e}_{-\to-+}({-}x_k \to x_i x_j)
      &= \frac{x_i^2}{|x_k|} \,,
\\
   P^{\bar e e\gamma}_{-+-}(x_i,x_j,x_k)
   \equiv P^{e \to e\gamma}_{+\to+-}({-}x_i \to x_j x_k)
   = P^{\gamma \to e\bar e}_{+\to-+}({-}x_k \to x_i x_j)
      &= \frac{x_j^2}{|x_k|} \,,
\\
   P^{\bar e e\gamma}_{--+} \equiv P^{e \to e\gamma}_{+\to-+} = P^{\gamma\to e\bar e}_{-\to--}
     &= 0 ,
\\
   P^{\bar e e\gamma}_{---} \equiv P^{e \to e\gamma}_{+\to--} = P^{\gamma\to e\bar e}_{+\to--}
    &= 0 ,
\end {align}
\end {subequations}
and ${\bm e}_\pm \equiv {\bm e}_x \pm i {\bm e}_y$.
The zeros above are a consequence of chirality conservation.
As in AI1 \cite{2brem}, the $P(x_\ix \to x_j x_k)$ are
defined in terms of the usual DGLAP splitting functions by
\begin {equation}
   P(x_\ix \to x_j x_k) \equiv |x_\ix| \, P(z_j,z_k)
\label {eq:Prelate}
\end {equation}
where $z \equiv x/x_\ix$ are the momentum fractions of the daughters
relative to their immediate parent.  The advantage of the
$P(x_i,x_j,x_k)$ is that they are normalized so that they are
symmetric with respect to permuting the parent with the daughters and
so are the same for $e \to \gamma e$ and $\gamma \to e\bar e$,
as indicated in (\ref{eq:Pexplicit}).
Eqs.\ (\ref{eq:Pexplicit}) only show half of the helicity cases;
the other half are given by $P_{h_1,h_2,h_3} = P_{-h_1,-h_2,-h_3}$ and
so
\begin {equation}
   \bcalP_{h_1,h_2,h_3} = \bcalP_{-h_1,-h_2,-h_3}^* .
\end {equation}

\begin {figure}[tp]
\begin {center}
  \begin{picture}(455,70)(5,0)
    \put(0,0){\includegraphics[scale=0.7]{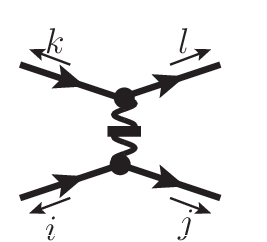}}
    \put(85,35){$
     = - \frac{i g^2}{|x_k+x_l|^3 E^2} \,
       \delta^{(2)}(\bcalB_{ji}) \, \delta^{(2)}(\bcalB_{lk})
       \times
       \begin {cases}
          +1,  & \mbox{interaction in amplitude} \\
          -1 , & \mbox{interaction in conjugated amplitude}
       \end {cases}
    $}
  \end{picture}
  \caption{
     \label {fig:ruleL}
     Similar to fig.\ \ref{fig:rule} but for the 4-fermion vertex
     associated with longitudinal photon exchange.
     The rule here is normalized to cover all the cases in this paper
     for $0 \leftrightarrow 4$
     or $2 \leftrightarrow 4$ particle transitions, which involve
     instantaneous
     $e \to e\bar e e$ in either the amplitude or conjugate amplitude.
     [We have not bothered presenting a generalized rule for $3 \to 3$
     transitions such as in figs.\ \ref{fig:diags}(g) and
     \ref{fig:diagsVIRT}(n) for $e \to e\bar e e$ and
     \ref{fig:diagsVIRT2}(u) for $e\bar e \to e\bar e$,
     since these diagrams give zero by
     symmetry arguments similar to those of ACI4 section III.B
     \cite{4point}.]
  }
\end {center}
\end {figure}

Similar rules for the instantaneous photon interactions are given
by fig.\ \ref{fig:ruleL}, which is similar in implementation
to the 4-gluon vertex rule of ACI4 fig.\ 10 \cite{4point}.

\section{Calculation of real double splitting (fig.\ \ref{fig:diags})}
\label {app:real}

The calculations of real double splitting diagrams $e \to e\bar e e$
proceed the same way as the $g \to ggg$ calculations of
refs.\ \cite{2brem,seq,4point} (AI1,ACI2,ACI4),
with schematically very similar results.

In large $\Nf$, we distinguish the final-state
electron that carries the flavor of the initial-state electron
and refer to it as $\xe$.  The electron in the pair produced from the
photon (which has a different flavor in large $\Nf$) is referred
to as $\ye$.  Because we distinguish these particles in our
large-$\Nf$ formulas, $\Delta d\Gamma/d\xe\,d\ye$ is normalized so
that, in applications, final-state integrations should be performed
as
\begin {equation}
   \int_0^1 d\xe \> d\ye \> \theta(1{-}\xe{-}\ye)
\label {eq:finalint}
\end {equation}
without any factor of $\tfrac12$ for identical final-state particles.


\subsection{Sequential diagrams}
\label{app:seq}

\subsubsection {Generic Formulas}

The calculation of the diagrams of fig.\ \ref{fig:diags}(a--c) mimics
that of the corresponding diagrams in ACI2 \cite{seq}.
Using the same notation, the results can be directly taken over,
except for some normalization factors and low-level details that
we will get to in a moment.  Adapting the results summarized in
ACI4 appendix D.3 \cite{4point}%
\footnote{
  The summary in ACI4 appendix D \cite{4point} incorporates some
  corrections \cite{dimreg} to the analysis of ``pole'' terms in
  the earlier paper ACI2 \cite{seq}.  So it is better to take formulas
  from ACI4 appendix D than directly from the original
  ACI2 analysis.
}%
,
the sum of figs.\ \ref{fig:diags}(a--c) yields
the equations (\ref{eq:dGammaseqNf}--\ref{eq:Mifseq}) given
in our summary of results.
There are a few small differences between those equations
and $g \to ggg$ formulas.
\begin {itemize}
\item
  The naive translation of $\CA^2 \alphas$ to QED is
  \begin {equation}
     \CA^2 \alphas \to \Nf \alphaqed \qquad (\mbox{sequential diagrams})
  \label {eq:alphaTranslate}
  \end {equation}
  since (i) the factor of $\CA^2$ in QCD sequential diagrams came from
  $\dA^{-1} \tr(T^a T^a T^b T^b)$, which is 1 in QED, and (ii) there are
  $\Nf$ possible flavors of the $e\bar e$ pair produced by the
  virtual photon in $e \to \gamma^* e \to e\bar e e$.
\item
  In the QED analysis,
  we do not have to worry about large-$\Nc$ ``color routings,''
  which affected the choice of how to write the
  $g \to ggg$ version of (\ref{eq:dGammaseqNf}), as described in
  ACI2 section 2.B.1 \cite{seq}.  Each QCD diagram was the sum
  of two color routings, with each routing represented by ${\cal A}$.
  If we adopt the translation (\ref{eq:alphaTranslate}), that means
  that ${\cal A}$ will represent {\it half}\/ of each QED diagram,
  which is the reason for the overall factor of $2$ in
  (\ref{eq:dGammaseqNf}).  We could have instead absorbed this factor
  of 2 into the normalizations of $D_\seq$, $F_\seq$ and
  ${\cal A}^{\rm pole}_\seq$, but it seemed more convenient to normalize the
  lower-level formulas in exactly the same way as the QCD case.
\item
  For $g \to ggg$, there are three identical particles in the final
  state, and refs.\ \cite{2brem,seq,4point} correspondingly add
  together all permutations of ${\cal A}(x,y)$ corresponding to
  permuting $x$, $y$, and $z \equiv 1{-}x{-}y$.  For $e \to e\bar e e$,
  there are two identical particles in the final state in
  ordinary $\Nf{=}1$ QED, and so one might
  reasonably think that, analogously,
  the ${\cal A}_\seqNf(\xe,\ye)$ on
  the right-hand side of (\ref{eq:dGammaseqNf})
  above should be
  ${\cal A}_\seqNf(\xe,\ye) + {\cal A}_\seqNf(\ye,\xe)$.
  If so, one would integrate over final-state particles as
  $\frac12 \int_0^1 d\xe \> d\ye \> \theta(1{-}\xe{-}\ye)$ in applications
  of $d\Gamma/d\xe\,d\ye$, where the factor of $\frac12$ would avoid
  double counting of identical final states.  In large $\Nf$,
  however, we distinguish the two electrons as explained previously.
  To allow applications the option
  of tracking the fate of the initial-flavored electron, we have chosen
  not to include the $\xe{\leftrightarrow}\ye$ permutation in
  (\ref{eq:dGammaseqNf}).  When using our large-$\Nf$
  $d\Gamma/d\xe\,d\ye$ in applications, one should correspondingly
  integrate as in (\ref{eq:finalint}),
  without any final-state factor of $\frac12$.
\end {itemize}

The quantities $I_n^\seq$ in (\ref{eq:Dseq}) are defined the same way
(\ref{eq:Iseq}) as in ACI4 appendix D.2 \cite{4point}.
The formulas for $(X,Y,Z)^\seq$ there and in (\ref{eq:Dseq}),
which are expressed
in terms
of the eigenfrequencies and eigenmodes of the problem, are also
given in ACI4 appendix D.2 \cite{4point}.  We will see
below that one of the eigenfrequencies,
$\Omega_-$, vanishes in our QED application, and so the formulas
for $(X,Y,Z)^\seq$
specialize to (\ref{eq:XYZseq}),
where
\begin {equation}
   a_\yx =
   \begin{pmatrix} C^+_{41} & C^-_{41} \\ C^+_{23} & C^-_{23} \end{pmatrix}
\label {eq:ay}
\end {equation}
is a matrix of appropriately normalized modes of the double harmonic
oscillator problem in the basis $(\C_{41},\C_{23})$ used by ACI2 \cite{seq}
at the $y$ vertex of the $xy\bar x\bar y$ diagram, and
\begin {equation}
   a_\xbx^\seq =
   \begin{pmatrix} C^+_{23} & C^-_{41} \\ C^+_{23} & C^-_{41} \end{pmatrix}
   =
   \begin{pmatrix} 0 & 1 \\ 1 & 0 \end{pmatrix} a_\yx
\end {equation}
is a permutation appropriate to the $\bar x$ vertex.


\subsubsection {QED formulas for $(\bar\alpha,\bar\beta,\bar\gamma)$}

As in ACI2 \cite{seq},
the $(\bar\alpha,\bar\beta,\bar\gamma)$ in (\ref{eq:Bseq})
are functions of $\xe$ and $\ye$ that represent various combinations of
helicity-dependent DGLAP splitting functions from the vertices of
fig.\ \ref{fig:diags}a.  The relevant QED splitting functions are
different from those of the $g \to ggg$ process in QCD.
Performing the same calculation as in ACI2 appendix E \cite{seq},
but using the QED splitting functions for $e \to \gamma e$ and
$\gamma \to e\bar e$ appropriate to fig.\ \ref{fig:diags}a, we
find the results shown in (\ref{eq:abcNf}).%
\footnote{
  \label {foot:absval2}
  Specifically, (\ref{eq:abcNf})
  is the $e \to e\bar e e$ analog of ACI2 eq.\ (E4)
  \cite{seq}.    The use of the letter $z$
  here is unrelated to the use in (\ref{eq:Prelate}).
  The absolute value signs on $|\xe\ye\ze|$
  may seem redundant here, but they are included for a reason similar
  to footnote \ref{foot:absval1}: to make sure that
  $(\bar\alpha,\bar\beta,\bar\gamma)$ behave appropriately under
  front-end transformations (\ref{eq:frontendxyE}).
  With the absolute value signs, a front-end transformation maps
  $(\bar\alpha,\bar\beta,\bar\gamma)$ into $(1-\xe)^{10}$ times
  the analogous helicity-averaged product of ${\cal P}$'s that
  one would have constructed for the last diagram of
  fig.\ \ref{fig:frontend2}.
}
Note that
\begin {equation}
   \bar\gamma = -\bar\alpha .
\end {equation}
One may check that
\begin {equation}
   \bar\alpha + \tfrac12 \bar\beta + \tfrac12 \bar\gamma
   =
   \frac{P_{e\to e}(\xe)}{\xe^2(1{-}\xe)^2} \,\,
   \frac{P_{\gamma\to e}\bigl(\frac{\ye}{1{-}\xe}\bigr)}
        {(1{-}\xe) \ye^2 (1{-}\xe{-}\ye)^2}
   \,,
\label {eq:abcPP}
\end {equation}
which is the straight-forward translation to QED of a similar relation
for $g\to ggg$.%
\footnote{
  See ACI2 eq.\ (E5) \cite{seq}.  This is a re-assuring check because it
  has to hold in order for sequential diagrams to match up with
  sequential ``Monte Carlo'' when the two splittings are far separated
  in time.  That requirement is implied by ACI2 footnote 28 \cite{seq}
  and the need for 
  ACI2 eqs.\ (C7) and (C13) to match up accordingly.
  Analogous statements must hold for QED.
}


\subsection{Frequencies $\Omega$ and eigenmodes}
\label {app:omegapm}

For all of the $e\to e\bar e e$ diagrams of fig.\ \ref{fig:diags},
we will need the relevant frequencies $\Omega$ for 3-particle and
4-particle evolution, and the eigenmodes for 4-particle evolution.


\subsubsection{3-particle evolution frequency}

Quite generally, 3-particle frequencies are given by
\begin {equation}
   \Omega(x_1,x_2,x_3) =
   \sqrt{
       \frac{-i}{2E}
       \Bigl(
          \frac{\hat q_1}{x_1} + \frac{\hat q_2}{x_2} + \frac{\hat q_3}{x_3}
       \Bigr)
    } .
\end {equation}
(See, for example, the review leading up to AI1 eq.\ (2.33b) \cite{2brem}.)
For QED, the $\hat q$ of a photon is zero, and so this formula becomes
\begin {equation}
   \Omega(x_1^{(\bar e)},x_2^{(\gamma)},x_3^{(e)}) =
   \sqrt{
       \frac{-i \hat q}{2E}
       \Bigl(
          \frac{1}{x_1^{(\bar e)}} + \frac{1}{x_3^{(e)}}
       \Bigr)
    } .
\label {eq:Omega3particle}
\end {equation}
For the initial 3-particle evolution of the
sequential diagrams of fig.\ \ref{fig:diags}(a--c) and
virtual diagrams of fig.\ \ref{fig:diagsVIRT}(h--k), this gives
\begin {equation}
   \Omega_\ix =
   \sqrt{
       \frac{-i \hat q}{2E}
       \Bigl(
          -1 + \frac{1}{\xe}
       \Bigr)
    }
    = \sqrt{-\frac{i(1{-}\xe)\hat q}{2\xe E}} \,,
\end {equation}
which is equivalent to (\ref{eq:Omegai2})%
\footnote{
   We should clarify that the electron line that is called
   ``$x_3^{(e)}$'' in the context of the
   3-particle expression (\ref{eq:Omega3particle})
   happens to be called ``$x_4$'' in the context of our
   4-particle variables $(x_1,x_2,x_3,x_4)$ used in the context
   of (\ref{eq:Omegai2}).
}
and is also the frequency we quoted for $xyy\bar x$ in (\ref{eq:Omegai}).
For the sequential diagrams, the corresponding final 3-particle evolution
has frequency
\begin {equation}
   \Omega_\fx^\seq =
   \sqrt{
       \frac{-i \hat q}{2E}
       \Bigl(
          \frac{1}{\ye} + \frac{1}{1{-}\xe{-}\ye}
       \Bigr)
    }
    = \sqrt{-\frac{i(1{-}\xe)\hat q}{2\ye(1{-}\xe{-}\ye)E}}
\label {eq:Omegafseq}
\end {equation}
for figs.\ \ref{fig:diags}(a,b) and its
complex conjugate $(\Omega_\fx^\seq)^*$ for fig.\ \ref{fig:diags}(c).
Eq.\ (\ref{eq:Omegafseq}) is equivalent to (\ref{eq:Omegafseq4}) in the case
$x_i{=}\hat x_i$ relevant to sequential diagrams, as
in (\ref{eq:Bseq}).


\subsubsection{4-particle evolution frequencies and modes}
\label {app:modes}

For 4-particle evolution, one just needs to repeat the derivation of
AI1 section V.B \cite{2brem}, which studied the medium-averaged
evolution of four high-energy gluons in (large-$\Nc$) QCD.
One can see from the diagrams of figs.\
\ref{fig:diags}--\ref{fig:diagsVIRT2} that, for large-$\Nf$ QED,
the only intermediate 4-particles states are $\bar e e \bar e e$,
and so that is the only case we address here.  (Beyond the
large-$\Nf$ limit, one would need to also consider
$\bar e \gamma e \gamma$.)  The only difference in the derivation
is that the potential for four large-$\Nc$ gluons
[AI1 eq.\ (4.19) \cite{2brem}] is replaced by the potential
\begin {equation}
  V =
  - \frac{i\hat q}{4}
  \bigl[
     b_{12}^2 + b_{23}^2 + b_{34}^2 + b_{41}^2 - b_{13}^2 - b_{24}^2
  \bigr]
\label {eq:V0}
\end {equation}
for $(\bar e,e,\bar e,e)$, where $\b_{ij} \equiv \b_i{-}\b_j$ and
where the signs in front of the terms above are minus the product of
the corresponding charges $\pm1$.  (\ref{eq:V0}) is algebraically equivalent to
\begin {equation}
  V =
  - \frac{i\hat q}{4} (\b_1 - \b_2 + \b_3 - \b_4)^2 .
\label {eq:V}
\end {equation}
Proceeding as in AI1 \cite{2brem}, one finds the normal mode
frequencies (\ref{eq:OmegaP}) for $\Omega_+$ and
\begin {equation}
   \Omega_- = 0 .
\label {eq:OmegaM}
\end {equation}
The corresponding eigenvectors $\vec C^\pm$ are given by
(\ref{eq:ayNf}) for $a_\yx$ (\ref{eq:ay})
and have been appropriately normalized so that
\begin {equation}
   \begin {pmatrix} C_{41}^j \\ C_{23}^j \end{pmatrix}^\top
   {\mathfrak M}'
   \begin {pmatrix} C_{41}^{j'} \\ C_{23}^{j'} \end{pmatrix}
   = \delta^{jj'}
\label {eq:Cnorm}
\end {equation}
with
\begin {equation}
   {\mathfrak M}'
   =
   \begin{pmatrix}
      x_4 x_1 (x_4+x_1) & \\ & x_2 x_3 (x_2+x_3)
   \end {pmatrix} E
   =
   \begin{pmatrix}
      x_1 x_4 & \\ & -x_2 x_3
   \end {pmatrix} (x_1+x_4) E .
\label {eq:frakMp}
\end {equation}
We've chosen to work in the basis $(\C_{41},\C_{23})$ here, rather
than the basis $(\C_{34},\C_{12})$ used in AI1 \cite{2brem}, in order
to match the numbering used on sequential diagrams in
ACI2 fig.\ 24 \cite{seq}.%
\footnote{
  One may convert to the basis $(C_{34},C_{12})$ by
  simple permutation of the indices.  That's because of
  charge conjugation symmetry, which means that the result
  for $(\bar e,e,\bar e,e)$ is the same as that for
  $(e,\bar e,e,\bar e)$, and then cyclically permute the indices
  of the latter to get back to $(\bar e,e,\bar e,e)$ with
  $(x_1,x_2,x_3,x_4) \to (x_2,x_3,x_4,x_1)$.
}
One may check that
\begin {equation}
  a_\yx a_\yx^\top = ({\mathfrak M}')^{-1} ,
\label {eq:aa}
\end {equation}
as implied by
the normalization condition (\ref{eq:Cnorm}).


\subsection{Diagrams with instantaneous vertices}

\begin {figure}[tp]
\begin {center}
  \includegraphics[scale=0.43]{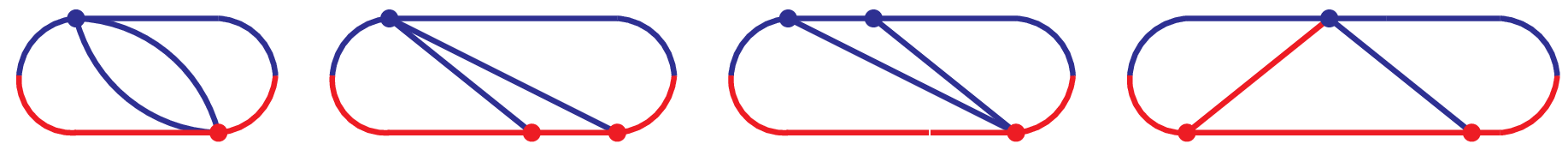}
  \caption{
     \label{fig:4point}
     QCD diagrams involving 4-gluon vertices, evaluated (for large $\Nc$)
     in ACI4 \cite{4point}.
     In this figure, the solid lines above all represent gluons.
     As usual,
     the complex conjugates of the above interference diagrams should also be
     included by taking $2\Re[\cdots]$ of the above.
  }
\end {center}
\end {figure}

The calculation of the real double-spitting diagrams involving
instantaneous vertices, shown in fig.\ \ref{fig:diags}(d-g), is
very similar to the calculation of the QCD diagrams involving
4-point gluon vertices, shown in fig.\ \ref{fig:4point},
which were computed in ACI4 \cite{4point}.

\subsubsection{The $I\bar I$ diagram}

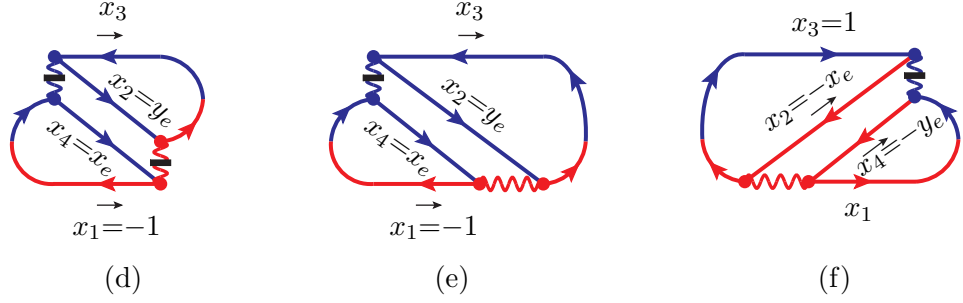
\begin {figure}[t]
\begin {center}
  \begin{picture}(370,110)(0,0)
    \put(0,0) {
      \begin{picture}(100,90)(0,-20)
        \put(0,8){\includegraphics[scale=0.5]{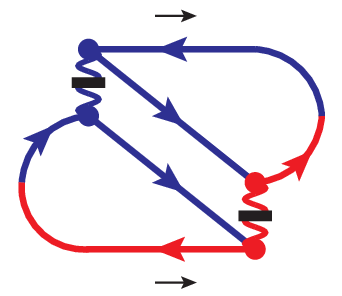}}
        \put(28,0){$x_1{=}{-}1$}
        \put(17,40){\rotatebox{325}{$x_4{=}\xe$}}
        \put(40,58){\rotatebox{325}{$x_2{=}\ye$}}
        \put(38,83){$x_3$}
        \put(40,-20){(d)}
      \end{picture}
    }
    \put(120,0) {
      \begin{picture}(100,90)(0,-20)
        \put(0,8){\includegraphics[scale=0.5]{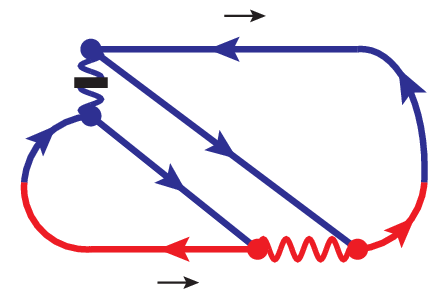}}
        \put(28,0){$x_1{=}{-}1$}
        \put(17,40){\rotatebox{325}{$x_4{=}\xe$}}
        \put(46,56){\rotatebox{325}{$x_2{=}\ye$}}
        \put(53,83){$x_3$}
        \put(45,-20){(e)}
      \end{picture}
    }
    \put(260,7){
      \begin{picture}(100,80)(0,-20)
        \put(0,8){\includegraphics[scale=0.5]{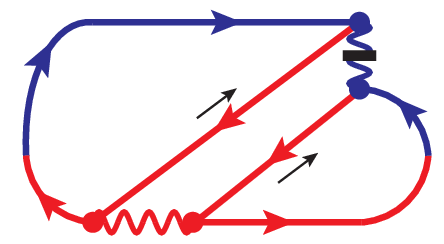}}
        \put(40,70){$x_3{=}1$}
        \put(28,32){\rotatebox{35}{$x_2{=}{-}\xe$}}
        \put(63,15){\rotatebox{35}{$x_4{=}{-}\ye$}}
        \put(60,0){$x_1$}
        \put(50,-27){(f)}
      \end{picture}
    }
  \end{picture}
  \caption{
     \label{fig:Idiags}
     Labeling of lines for the (d) $I\bar I$, (e) $I\bar x\bar y$,
     and (f) $\bar x\bar y I$ diagrams of figs.\ \ref{fig:diags}(d--f).
  }
\end {center}
\end {figure}

Consider fig.\ \ref{fig:Idiags}(d), which is a labeled version
of fig.\ \ref{fig:diags}(d).
We will call this the $I\bar I$ diagram
with ``$I$'' short for ``instantaneous 4-fermion vertex.''
Its evaluation closely parallels that of the $4\bar 4$
gluon diagram of ACI4 section III.C \cite{4point}, with the
only change being that the overall factor
\begin {align}
   S & \equiv \frac{1}{2\dA} \sum_{\mbox{\tiny$h$'s}} \sum_{\rm color} \Bigl[
      f^{a_\ix a_\xx e} f^{a_\yx a_\zx e}
      (\delta_{h_\ix,h_\yx} \delta_{h_\zx,-h_\xx}
        - \delta_{h_\ix,h_\zx} \delta_{h_\xx,-h_\yx})
\nonumber\\ & \qquad\qquad\qquad
    + f^{a_\ix a_\yx e} f^{a_\xx a_\zx e}
      (\delta_{h_\ix,h_\xx} \delta_{h_\yx,-h_\zx}
        - \delta_{h_\ix,h_\zx} \delta_{h_\xx,-h_\yx})
\nonumber\\ & \qquad\qquad\qquad
    + f^{a_\ix a_\zx e} f^{a_\xx a_\yx e}
      (\delta_{h_\ix,h_\xx} \delta_{h_\yx,-h_\zx}
        - \delta_{h_\ix,h_\yx} \delta_{h_\zx,-h_\xx})
    \Bigr]^2
    = 9 \CA^2
\end {align}
coming from the two 4-gluon vertices in ACI4 eq.\ (3.12) \cite{4point} is
replaced here by the factor
\begin {equation}
   S \equiv \frac{\Nf}{2} \sum_{\mbox{\tiny$h$'s}}
    \left[
      \frac{4|x_1 x_2 x_3 x_4|^{1/2}}{|x_1+x_4|^2} \,
      \delta_{h_\ix,h_x} \delta_{h_y,{-}h_z}
    \right]^2
   = \frac{32 \Nf |x_1 x_2 x_3 x_4|}{|x_1+x_4|^4} \,.
\end {equation}
To see this,
compare the 4-gluon vertex rule of ACI4 fig.\ 10 \cite{4point} with
the rule of fig.\ \ref{fig:ruleL} here.
From ACI4 eq.\ (3.16) \cite{4point} (times 3 to sum up the equal results
from all three color routings), the QCD gluon result was
\begin {equation}
   \left[\frac{d\Gamma}{d\xe\,d\ye}\right]_{4\bar 4}
   =
   - \frac{9\CA^2 \alphas^2}{16 \pi^2}
   \int_0^\infty d(\Delta t) \>
   \Omega_+\Omega_- \csc(\Omega_+\,\Delta t) \csc(\Omega_-\,\Delta t)
  .
\label {eq:44}
\end {equation}
The corresponding result here is then
\begin {align}
   \left[\frac{d\Gamma}{d\xe\,d\ye}\right]_{I\bar I}
   &=
   - \frac{2\Nf \alphaqed^2}{\pi^2} \,
   \frac{\xe\ye\ze}{(1-\xe)^4}
   \int_0^\infty d(\Delta t) \>
   \Omega_+\Omega_- \csc(\Omega_+\,\Delta t) \csc(\Omega_-\,\Delta t)
\nonumber\\
   &=
   - \frac{2\Nf \alphaqed^2}{\pi^2} \,
   \frac{\xe\ye\ze}{(1-\xe)^4}
   \int_0^\infty \frac{d(\Delta t)}{\Delta t} \,
   \Omega_+ \csc(\Omega_+\,\Delta t) .
\end {align}
Following our general procedure from AI1 \cite{2brem} of subtracting
out the vacuum pieces of each diagram (which must all cancel in the
final result), this is
\begin {align}
   \left[\frac{d\Gamma}{d\xe\,d\ye}\right]_{I\bar I}
   &=
   - \frac{2\Nf \alphaqed^2}{\pi^2} \,
   \frac{\xe\ye\ze}{(1-\xe)^4}
   \int_0^\infty \frac{d(\Delta t)}{\Delta t} \,
   \left[ \Omega_+ \csc(\Omega_+\,\Delta t) - \frac{1}{\Delta t} \right]
\nonumber\\
   &= \frac{ 2\Nf\alphaqed^2}{ \pi^2}  \,
   \frac{\xe\ye\ze}{(1-\xe)^4}
      \, i \Omega_+ \ln 2
  .
\end {align}
Adding this diagram to its complex conjugate gives what we labeled as
the ``$(II)$'' contribution (\ref{eq:II})
to the total answer (\ref{eq:total}).

\subsubsection{Diagrams with one instantaneous vertex}

Now consider the interference diagram of fig.\ \ref{fig:Idiags}(e).
Remember that our (large-$\Nf$)
convention is that $\xe$ is the momentum fraction
of the particular final-state electron that is connected by an
electron line to the initial electron.

There are just a few differences with the similar calculation of the
$4\bar x\bar y$ gluon process given in ACI4 section II \cite{4point}.
First, comparing fig.\ \ref{fig:Idiags}(e) here with ACI4 fig.\ 12(a),
our labeling conventions are a little different.
The translation is that
$(\hat x_1,\hat x_2,\hat x_3,\hat x_4) = (-1,y,1{-}x{-}y,x)$ in
ref.\ \cite{4point} is permuted to
$(\hat x_1,\hat x_4,\hat x_2,\hat x_3) = (-1,\xe,\ye,1{-}\xe{-}\ye)$ here.
Another is that the 4-point vertex factor
\begin {align}
      f^{a_\ix a_\xx e} & f^{a_\yx a_\zx e}
      (\delta_{h_\ix,h_\yx} \delta_{h_\zx,-h_\xx}
        - \delta_{h_\ix,h_\zx} \delta_{h_\xx,-h_\yx})
\nonumber\\ &
    + \tfrac12 f^{a_\ix a_\yx e} f^{a_\xx a_\zx e}
      (\delta_{h_\ix,h_\xx} \delta_{h_\yx,-h_\zx}
        - \delta_{h_\ix,h_\zx} \delta_{h_\xx,-h_\yx})
\end {align}
of ACI4 eq.\ (2.9) for its color routing $4\bar y \bar x_2$
should be replaced by
\begin {equation}
   \frac{4 |x_1 x_2 x_3 x_4|^{1/2}}{(x_2+x_3)^2} \,
   \delta_{h_\ix,h_\xx} \delta_{h_\yx,-h_\zx}
   =
   \frac{4 (\xe\ye\ze)^{1/2}}{(1-\xe)^2} \,
   \delta_{h_\ix,h_\xx} \delta_{h_\yx,-h_\zx}
\end {equation}
for the instantaneous vertex in our calculation here, for which there
are no color routings to consider.  (Again, compare the
4-gluon vertex rule of ACI4 fig.\ 10 \cite{4point} with
the rule of fig.\ \ref{fig:ruleL} here.) 
Also, unlike ACI4 eq.\ (2.10),
there are no color factors associated with the other vertices.
Finally, $g_{\rm s}^4$ becomes $\Nf g_{\scriptscriptstyle\rm EM}^4$.
Making these changes, and keeping track of the signs
of the $\bcalB_{ij}$ in our fig.\ \ref{fig:rule} rule, we find
that ACI4 eq.\ (2.12) translates to
\begin {align}
   \left[\frac{dI}{dx\,dy}\right]_{I\bar x\bar y}
   &=
   - \left( \frac{E}{2\pi} \right)^2
   \int_{t_\four < t_\ybx < t_\xbx}
   \sum_{h_\xx,h_\yx,h_\zx,\bar h}
   \int_{\B^\xbx}
\nonumber\\ &\times
   \frac{4 i (\xe\ye\ze)^{1/2}}{(1-\xe)^2} \, \Nf g^4 \,
   \delta_{h_\ix,h_\xx} \delta_{h_\yx,-h_\zx}
\nonumber\\ &\times
   \tfrac12 E^{-3/2} \,
   \Bigl[
     \bcalP^{\gamma\to e\bar e}_{\bar h \to h_\yx,h_\zx}(1{-}\xe \to \ye,\ze) \Bigr]^*
       \cdot \grad_{\B^\ybx}
   \langle\B^\ybx,t_\ybx|\B^\xbx,t_\xbx\rangle
   \Bigr|_{\B^\ybx=0}
\nonumber\\ &\times
   \tfrac12 E^{-3/2} |\hat x_2+\hat x_3|^{-1} \,
   \Bigl[
     \bcalP^{e\to \gamma e}_{h_\ix \to \bar h,h_\xx}(1 \to 1{-}\xe,\xe) \Bigr]^*
          \cdot \grad_{\C_{14}^\xbx}
\nonumber\\ &\qquad\qquad
  \langle\C_{23}^\xbx,\C_{14}^\xbx,t_\xbx|\C_{23}^I,\C_{14}^I,t_I\rangle
   \Bigr|_{\C_{14}^\xbx=0=\C_{14}^I=\C_{23}^I; ~ \C_{23}^\xbx=\B^\xbx}
\nonumber\\ &\times
  (2E)^{-2} |\hat x_1\hat x_2\hat x_3 \hat x_4|^{-1/2} |\hat x_2+\hat x_3|^{-1}
  .
\end {align}
The helicity sum analogous to ACI4 eqs.\ (2.14) is
\begin {equation}
   \sum_{h_\xx,h_\yx,h_\zx}
   \Bigl[
     \sum_{\bar h}
     [{\cal P}^{\gamma\to e\bar e}_{\bar h \to h_\yx,h_\zx}]^{\bar n}(1{-}\xe \to \ye,\ze)
     [{\cal P}^{e\to \gamma e}_{h_\ix \to \bar h,h_\xx}]^{\bar m}(1 \to 1{-}\xe,\xe)
   \Bigr]^*
   \delta_{h_\ix,h_\xx} \delta_{h_\yx,-h_\zx}
   |\hat x_1\hat x_2\hat x_3 \hat x_4|^{-1/2} .
\label {eq:IxyHsum}
\end {equation}
Using the QED formulas in appendix \ref{app:VertexRules} here, we find
that the initial-helicity average of (\ref{eq:IxyHsum}) above is
\begin {equation}
   \zeta(\xe,\ye) \, \delta^{\bar n\bar m}
\end {equation}
with%
\footnote{
  \label{foot:absMi}
  Similar to the discussion in footnote \ref{foot:absval2}, absolute
  value signs have been judiciously included in (\ref{eq:zeta}) so
  that $\zeta$ does the right thing under front-end transformations,
  which for the diagram at hand is implemented by (\ref{eq:frontendI}).
}
\begin {equation}
   \zeta =
   \frac{(1{+}|\xe|)\,(|\ye|{+}|\ze|)}
        {(1-\xe)^3 |\xe\ye\ze|^{3/2}} \,.
\label {eq:zeta2}
\end {equation}
Contrast to ACI4 eq.\ (2.16).
Following ACI4, the final result, analogous to ACI4 eq.\ (2.25), is then
\begin {align}
   \left[\frac{d\Gamma}{dx\,dy}\right]_{I\bar x\bar y}
   &=
   - \frac{\Nf\alphaqed^2 M_\fx^\seq}{2 \pi^2 E} \,
   \frac{(\xe\ye\ze)^{1/2}}{(1-\xe)^2} \,
   (-\hat x_1 \hat x_2 \hat x_3 \hat x_4)
   \zeta
\nonumber\\ & \qquad\qquad \times
   \int_0^\infty d(\Delta t) \>
   \Omega_+ \Omega_- \csc(\Omega_+\,\Delta t) \csc(\Omega_-\,\Delta t)
   \frac{Y_\xbx^\seq}{X_\xbx^\seq}
\nonumber\\
   &=
   - \frac{\Nf\alphaqed^2}{2\pi^2} \,
   \frac{(1{+}|\xe|)\ye\ze(|\ye|{+}|\ze|)}{(1-\xe)^4}
   \int_0^\infty \frac{d(\Delta t)}{\Delta t} \>
   \Omega_+ \csc(\Omega_+\,\Delta t)
   \frac{Y_\xbx^\seq}{X_\xbx^\seq}
   \,,
\end {align}
where $X_\xbx^\seq$ and $Y_\xbx^\seq$ are as in (\ref{eq:XYZseq}).

Following the strategy of ACI4 section III.A,
the $\bar x\bar y I$ diagram of fig.\ \ref{fig:Idiags}(f) here is
like the mirror reflection of \ref{fig:Idiags}(e).
Given how we have labeled particles,%
\footnote{
  The labeling of figs.\ \ref{fig:Idiags}(e) and (f), and so the
  specifics of the transformation (\ref{eq:xyIsub}), are slightly
  different than in ACI4 in order to maintain our convention here that
  $\xe$ always refers to the final-state electron whose electron
  line is connected to the initial-state electron.
}
this mirror reflection transformation can be achieved by the
replacement
\begin {equation}
   (x_1,x_2,x_3,x_4) \to (-\hat x_3,-\hat x_4,-\hat x_1,-\hat x_2),
\label {eq:xyIsub}
\end {equation}
including inside
of $M_\fx^\seq = x_2 x_3 (x_2+x_3) E$ and
\begin {equation}
   \Omega_\fx^\seq(x_1,x_2,x_3,x_4)
   =
     \sqrt{
       \frac{-i \hat q}{2E}
       \Bigl(
          \frac{1}{x_2} + \frac{1}{x_3}
       \Bigr)
     } .
\label{eq:Omegafseq4b}
\end {equation}
For the $I \bar x\bar y$ diagram,
$\Omega_\fx^\seq(\hat x_1,\hat x_2,\hat x_3,\hat x_4)$
is just a way to write the $\Omega_\fx^\seq$ of (\ref{eq:Omegafseq}) in
terms of the 4-particle variables $(\hat x_1,\hat x_2,\hat x_3,\hat x_4)$.
But the form (\ref{eq:Omegafseq4b}) has the advantage that, after
the substitution (\ref{eq:xyIsub}), it also
gives the correct $\Omega$ for the initial 3-particle evolution
for $\bar x\bar y I$ as depicted in fig.\ \ref{fig:Idiags}(f).

The $\bar x I \bar y$ diagram of fig.\ \ref{fig:diags}(g) vanishes
by parity symmetry for the same reasons as the $\bar y 4\bar x$ diagram
in ACI4 section III.B.

These results for diagrams with one instantaneous vertex are summarized
in section \ref{app:Isummary}.



%


\section{Explicit formulas for evaluation of
         \boldmath$xyy\bar x$ diagram}
\label {app:Fund1}

\subsection {General formula for \boldmath$d{=}2$}
\label {app:Fund1a}

From the starting formula (\ref{eq:Fund1}), following the
same steps as our earlier work \cite{2brem,seq} (AI1,ACI2) on real
double splitting rates in (large-$\Nc$) QCD,
one obtains formulas that schematically have the same form.
Namely,
\begin {align}
  \left[\frac{d\Gamma}{d\xe\,d\ye}\right]_{xyy\bar x} =
   -
   \frac{\Nf\alphaqed^2 M_\ix^2}{16\pi^4 E^2} \, & 
   ({-}\hat x_1 \hat x_2 \hat x_3 \hat x_4)
   \int_0^\infty d(\Delta t) \>
   \Omega_+\Omega_- \csc(\Omega_+\Delta t) \csc(\Omega_-\Delta t)
\nonumber\\ &\times
   \Bigl\{
     (\bar\beta Y_\yx^2
        + \bar\gamma \Ybar_{\yx\yx'} Y_{\yx\yx'}) I_0^\new
     + (2\bar\alpha+\bar\beta+\bar\gamma) Z_{\yx\yx'} I_1^\new
\nonumber\\ &\quad
     + \bigl[
         (\bar\alpha+\bar\gamma) Y_\yx^2
         + (\bar\alpha+\bar\beta) \Ybar_{\yx\yx'} Y_{\yx\yx'}
        \bigr] I_2^\new
\nonumber\\ &\quad
     - (\bar\alpha+\bar\beta+\bar\gamma)
       (\Ybar_{\yx\yx'} Y_\yx I_3^\new + Y_\yx Y_{\yx\yx'} I_4^\new)
   \Bigl\} .
\label {eq:virtxyyx}
\end {align}
[which here must eventually be integrated over $\ye$ as in (\ref{eq:Fund1int})].
In our QED application, $\Omega_- = 0$ (\ref{eq:OmegaM}), and so
\begin {equation}
   \Omega_- \csc(\Omega_-\,\Delta t)
   =
   \frac{1}{\Delta t}
\label {eq:OmCsc}
\end {equation}
above.  The $I_n^\new$ are just like the $I_n^\seq$ of (\ref{eq:Iseq}),
\begin {subequations}
\label {eq:Inew}
\begin {align}
   I_0^\new &=
   \frac{4\pi^2}{[X_\yx^\new X_{\yx'}^\new - (X_{\yx\yx'}^\new)^2]} ,
\displaybreak[0]\\
   I_1^\new &=
   - \frac{2\pi^2}{X_{\yx\yx'}^\new}
   \ln\left( 1 - \frac{(X_{\yx\yx'}^\new)^2}{X_\yx^\new X_{\yx'}^\new} \right) ,
\label {eq:I1new}
\displaybreak[0]\\
   I_2^\new &=
   \frac{2\pi^2}{(X_{\yx\yx'}^\new)^2}
     \ln\left( 1 - \frac{(X_{\yx\yx'}^\new)^2}{X_\yx^\new X_{\yx'}^\new} \right)
   + \frac{4\pi^2}{[X_\yx^\new X_{\yx'}^\new - (X_{\yx\yx'}^\new)^2]} ,
\displaybreak[0]\\
   I_3^\new &=
   \frac{4\pi^2 X_{\yx\yx'}^\new}
        {X_{\yx'}^\new[X_\yx^\new X_{\yx'}^\new - (X_{\yx\yx'}^\new)^2]} ,
\displaybreak[0]\\
   I_4^\new &=
   \frac{4\pi^2 X_{\yx\yx'}^\new}
        {X_\yx^\new[X_\yx^\new X_{\yx'}^\new - (X_{\yx\yx'}^\new)^2]} ,
\end {align}
\end {subequations}
except that the $(X,Y,Z)^\seq$ of (\ref{eq:XYZseq})
are replaced by $(X,Y,Z)^\new$ in which
$M_\fx^\seq$ is replaced by $M_\ix$
(because the final 3-particle evolution in
the $xyy\bar x$ diagram,
fig.\ \ref{fig:FundamentalVirt},
involves the same particles as the initial
stage of 3-particle evolution)
and the $a_\xbx^\seq$ are replaced by $a_{\yx'} = a_\yx$
(because the particles that merge at the end of the
interim 4-particle evolution are the same as the ones that
split at its start):
\begin {subequations}
\label {eq:XYZnew}
\begin {align}
   \begin{pmatrix} X_\yx^\new & Y_\yx^\new \\ Y_\yx^\new & Z_\yx^\new \end{pmatrix}
   =
   \begin{pmatrix} X_{\yx'}^\new & Y_{\yx'}^\new
                   \\ Y_{\yx'}^\new & Z_{\yx'}^\new \end{pmatrix}
   &\equiv
   \begin{pmatrix} |M_\ix|\Omega_\ix & 0 \\ 0 & 0 \end{pmatrix}
     - i a_\yx^{-1\top}
     \begin{pmatrix}
        \Omega_+ \cot(\Omega_+\,\Delta t) & 0 \\
        0 & (\Delta t)^{-1}
     \end{pmatrix}
     a_\yx^{-1} ,
\\
   \begin{pmatrix} X_{\yx\yx'}^\new & Y_{\yx\yx'}^\new \\[2pt]
                   \Ybar_{\yx\yx'}^\new & Z_{\yx\yx'}^\new \end{pmatrix}
   &\equiv
     - i a_\yx^{-1\top}
     \begin{pmatrix}
        \Omega_+ \csc(\Omega_+\,\Delta t) & 0 \\
        0 & (\Delta t)^{-1}
     \end{pmatrix}
     a_\yx^{-1} .
\end {align}
\end {subequations}


\subsection {Small \boldmath$\Delta t$ expansion}
\label {app:smalldt}

\subsubsection{Structure}

To study the UV divergence of $xyy\bar x$, we want to identify which
terms of the integrand in (\ref{eq:virtxyyx}) are important as
$\Delta t \to 0$.  For that, we need the small-$\Delta t$ expansion
of (\ref{eq:XYZnew}).  At leading order in $\Delta t$, the expansion
is the same for $(X,Y,Z)^\new_\yx$ and $(X,Y,Z)^\new_{\yx\yx'}$:
\begin {equation}
   \begin{pmatrix} X^\new & Y^\new
                   \\ Y^\new & Z^\new \end{pmatrix}
   \simeq
   - i \frac{a_\yx^{-1\top} a_\yx^{-1}}{\Delta t}
   =
   \frac{-i E(x_1+x_4)}{\Delta t}
   \begin{pmatrix} x_1 x_4 & 0 \\ 0 & -x_2 x_3 \end{pmatrix} ,
\label {eq:dtleading}
\end {equation}
where the second equality uses
(\ref{eq:frakMp}) and (\ref{eq:aa}).
This approximation is fine for the $Y$'s and $Z$'s in (\ref{eq:virtxyyx}),
but it is inadequate for the combination
$X_\yx^\new X_{\yx'}^\new - (X_{\yx\yx'}^\new)^2$ that appears in
(\ref{eq:Inew}), which is zero at the order of (\ref{eq:dtleading}).

Going to next order in $\Delta t$,
\begin {subequations}
\label {eq:dtNLO}
\begin {align}
   X_\yx^\new = X_{\yx'}^\new &\simeq
     \frac{-i M_\ix}{\Delta t} + |M_\ix|\Omega_\ix + O(\Delta t) ,
\label {eq:XdtNLO1}
\\
   X_{\yx\yx'}^\new &\simeq
     \frac{-i M_\ix}{\Delta t} + O(\Delta t) ,
\label {eq:XdtNLO2}
\\
   Y^\new_{\rm any} &= O(\Delta t) ,
\\
   Z^\new_{\rm any} &= \frac{i x_2 x_3(x_1+x_4)E}{\Delta t} + O(\Delta t) ,
\label {eq:ZdtNLO2}
\end {align}
which gives
\begin {equation}
   X_\yx^\new X_{\yx'}^\new - (X_{\yx\yx'}^\new)^2
   = \frac{-2i M_\ix^2 \Omega_\ix}{\Delta t} \, \sgn M_\ix
     + O\bigl((\Delta t\bigr)^0).
\label {eq:detXleading}
\end {equation}
\end {subequations}
One then finds that the $I_n^\new$ of (\ref{eq:Inew}) are all
$O(\Delta t)$.  Using these expansions in (\ref{eq:virtxyyx}) shows
that all the terms in the integrand are finite as $\Delta t\to 0$
except for the term involving $Z_{\yx\yx'} I_1$, whose dependence on
$\Delta t$ is
\begin {equation}
   \Omega_+\Omega_- \csc(\Omega_+\Delta t) \csc(\Omega_-\Delta t)
   Z_{\yx\yx'} I_1^\new
   \sim \frac{1}{(\Delta t)^2} .
\label {eq:ZI1term}
\end {equation}
This $Z_{\yx\yx'} I_1$ term contains the small-$\Delta t$ divergence
represented by (\ref{eq:FundDiv}) in the main text.  [We will
discuss the translation to (\ref{eq:FundDiv}) later.]

Because (\ref{eq:ZI1term}) blows up as $(\Delta t)^{-2}$, we also
potentially have a sub-leading divergence $(\Delta t)^{-1}$
if there are any corrections in the small-$\Delta t$ expansion that
are suppressed by only one more power of $\Delta t$.
Tracing back through
(\ref{eq:ZI1term}) and the expansions
(\ref{eq:dtNLO}), and noting that
$\ln\bigl[ 1 - X_{\yx\yx'}^2/X_\yx X_{\yx'} \bigr]
 = \ln\bigl[ (X_\yx X_{\yx'} - X_{\yx\yx'}^2) / X_\yx X_{\yx'} \bigr]$
in (\ref{eq:I1new}),
the only correction that contributes at this
order is the $O\bigl((\Delta t)^0\bigr)$ correction to
(\ref{eq:detXleading}).  Computing this correction requires the
$O(\Delta t)$ terms in (\ref{eq:XdtNLO1}) and (\ref{eq:XdtNLO2}).

There is a simplification that will allow us to work out
a useful formula for the needed corrections to
$X_\yx^\new$ and $X_{\yx'}^\new$ more generally than for the
QED application of relevance here.
So, even though $\Omega_- = 0$
for QED, let us be more general and rewrite (\ref{eq:XYZnew}) as
\begin {subequations}
\begin {align}
   \begin{pmatrix} X_\yx^\new & Y_\yx^\new \\ Y_\yx^\new & Z_\yx^\new \end{pmatrix}
   =
   \begin{pmatrix} X_{\yx'}^\new & Y_{\yx'}^\new
                   \\ Y_{\yx'}^\new & Z_{\yx'}^\new \end{pmatrix}
   &\equiv
   \begin{pmatrix} |M_\ix|\Omega_\ix & 0 \\ 0 & 0 \end{pmatrix}
     - i a_\yx^{-1\top} \uOmega \cot(\uOmega\,\Delta t) a_\yx^{-1} ,
\\
   \begin{pmatrix} X_{\yx\yx'}^\new & Y_{\yx\yx'}^\new \\[2pt]
                   \Ybar_{\yx\yx'}^\new & Z_{\yx\yx'}^\new \end{pmatrix}
   &\equiv
     - i a_\yx^{-1\top} \uOmega \csc(\uOmega\,\Delta t) a_\yx^{-1}
\end {align}
\end {subequations}
with
\begin {equation}
   \uOmega \equiv \begin{pmatrix} \Omega_+ & \\ & \Omega_- \end{pmatrix} .
\label {eq:uOmega}
\end {equation}
Noting that (\ref{eq:xhat}) and (\ref{eq:Mifseq})
give $x_1 x_4(x_1+x_4)E = M_\ix$
in (\ref{eq:dtleading}), the
small-$\Delta t$ expansions of the $X$'s are then
\begin {subequations}
\label{eq:NLOexpandXnew}
\begin {align}
   X_\yx^\new = X_{\yx'}^\new &=
   \frac{-i M_\ix}{\Delta t} + |M_\ix|\Omega_\ix
   + \frac{i}{3} \bigl[ a_\yx^{-1\top} \uOmega^2 a_\yx^{-1} \bigr]_{11}
     \Delta t
   + O\bigl( (\Delta t)^3 \bigr) ,
\\
   X_{\yx\yx'}^\new &=
   \frac{-i M_\ix}{\Delta t}
   - \frac{i}{6} \bigl[ a_\yx^{-1\top} \uOmega^2 a_\yx^{-1} \bigr]_{11}
     \Delta t
   + O\bigl( (\Delta t)^3 \bigr) .
\end {align}
\end {subequations}
Then
\begin {equation}
  X_\yx^\new X_{\yx'}^\new - (X_{\yx\yx'}^\new)^2
   = \frac{-2i M_\ix^2 \Omega_\ix}{\Delta t} \, \sgn M_\ix
     + M_\ix^2 \Omega_\ix^2
     + M_\ix \bigl[ a_\yx^{-1\top} \uOmega^2 a_\yx^{-1} \bigr]_{11}
     + O\bigl( \Delta t \bigr) .
\label {eq:NLOdetX0}
\end {equation}
The expression $\bigl[ a_\yx^{-1\top} \uOmega^2 a_\yx^{-1} \bigr]_{11}$ sounds
like a mess that depends on detailed formulas for $\Omega_\pm$ and
$a_\yx$.  Happily, it can be recast into a very simple form.


\subsubsection{Value of ${a^{-1}}^\top \Omega^2 a^{-1}$}
\label {app:OmegaSqr}

As in appendix \ref{app:modes} above,
return again to the derivation of eigenfrequencies $\Omega_\pm$
and normal modes in AI1 section V.B \cite{2brem}.
From AI1 (5.17), the relevant Lagrangian has the form
\begin {equation}
   L
   =
   \tfrac12
   \dot{\vec C}^\top {\mathfrak M} \dot{\vec C}
   - \tfrac12 
   \vec C^\top {\mathfrak K} \vec C ,
\label {eq:L2}
\end {equation}
where $\vec C$ is $(C_{41},C_{23})$ or whatever permutation you want,
and ${\mathfrak M}$ and ${\mathfrak K}$ are $2\times2$ matrices in
that basis.  The $\frac12 \vec C^\top {\mathfrak K} \vec C$ above encodes
the potential V of the double harmonic
oscillator problem, e.g.\ (\ref{eq:V}) in our case.
For our basis choice $(C_{41},C_{23})$, ${\mathfrak M}$
is given by (\ref{eq:frakMp}).

The equation of motion from (\ref{eq:L2}) is
\begin {equation}
   {\mathfrak M} \ddot{\vec C} = - {\mathfrak K} \ddot{\vec C} ,
\end {equation}
and the corresponding normal-mode eigenvalue problem is
\begin {equation}
   {\mathfrak M} \Omega_\pm^2 \vec C^\pm = {\mathfrak K} \vec C^\pm .
\end {equation}
We can rewrite both cases ($\pm$) of this equation simultaneously as
\begin {equation}
   {\mathfrak M} a \uOmega^2 = {\mathfrak K} a ,
\label {eq:omsqr1}
\end {equation}
where
\begin {equation}
   a \equiv \Bigl( ~ \vec C^+ ~\big|~~ \vec C^- ~ \Bigr)
\end {equation}
is the $2\times2$ matrix
$a_\yx$ [(\ref{eq:ay}) in this paper, or whatever permutation is relevant to
one's choice of basis for $\vec C$].

Multiplying (\ref{eq:omsqr1}) by $a^\top$ on the left,
\begin {equation}
   a^\top {\mathfrak M} a \uOmega^2 = a^\top {\mathfrak K} a .
\end {equation}
Because the normal modes are orthogonal with respect to ${\cal M}$
and then normalized (\ref{eq:Cnorm}) so that
\begin {equation}
   (\vec C^j)^\top {\mathfrak M} \vec C^{j'}
   = \delta^{jj'} ,
\end {equation}
we have $a^\top {\mathfrak M} a = \openone$ and so
\begin {equation}
   \uOmega^2 = a^\top {\mathfrak K} a .
\end {equation}
Thus,
\begin {equation}
   (a^{-1})^\top \uOmega^2 a^{-1} = {\mathfrak K} .
\end {equation}
This result is independent of details about the eigenfrequencies or
eigenmodes and which basis we pick for $\vec C$.

In the application to $xyy\bar x$, we want in particular the first element of
this matrix,
\begin {equation}
   \bigl[ (a^{-1})^\top \uOmega^2 a^{-1} \bigr]_{11} = {\mathfrak K}_{11} ,
\end {equation}
in the $(C_{41},C_{23})$ basis (numbered as in the right-hand diagram
of fig.\ \ref{fig:FundamentalVirt}).  In that context, ${\mathfrak K}_{11}$
corresponds to the (complex) spring constant of the problem if we were
to set $C_{23}$ to zero.  That would be the same as setting $\b_2 = \b_3$ and
so the same as
placing the $e$ and $\bar e$ of the photon self-energy loop on top
of each other.  In that case, the $e\bar e$ pair is, with regard
to charge, indistinguishable from the photon that
created them (i.e.\ no charge in the QED case here).
In this case, the potential in the 4-particle evolution of
fig.\ \ref{fig:FundamentalVirt} would be the same as that in the
preceding 3-particle portion.  But that means that ${\mathfrak K}_{11}$
must be identical to the spring constant for the initial 3-particle
evolution, and so
\begin {equation}
   {\mathfrak K}_{11} = M_\ix \Omega_\ix^2
\label {eq:K11}
\end {equation}
in the context of the $xyy\bar x$ diagram.  One may verify this general
relation in our QED case by starting from the potential
(\ref{eq:V}), following the method of AI1 section V.B \cite{2brem}
to construct
\begin {equation}
  {\mathfrak K} = 
  - \frac{i\hat q}{2} (x_1+x_4)^2
  \begin {pmatrix} \phantom{-}1 & -1 \\ -1 & \phantom{-}1 \end{pmatrix}
\end {equation}
in the $(C_{41},C_{23})$ basis,
and then checking (\ref{eq:K11}) using (\ref{eq:xhat}),
(\ref{eq:Omegai}), and (\ref{eq:Mi}).
A similar check works in the case of (large-$\Nc$) QCD.


\subsubsection{Putting it together}
\label {app:barOmega}

Using the results above, (\ref{eq:NLOdetX0}) becomes
\begin {equation}
  X_\yx^\new X_{\yx'}^\new - (X_{\yx\yx'}^\new)^2
   = \frac{-2i M_\ix^2 \Omega_\ix}{\Delta t} \, \sgn M_\ix
     + 2 M_\ix^2 \Omega_\ix^2
     + O\bigl( \Delta t \bigr) .
\end {equation}
Using this and our earlier expansions in (\ref{eq:virtxyyx}) gives
the unregulated ($\eps{=}0$) divergence
\begin {equation}
   \left[\frac{d\Gamma}{d\xe\,d\ye}\right]_{xyy\bar x}^{(\Delta t<a)} =
   -
   \frac{\Nf\alphaqed^2 M_\ix^2}{16\pi^4 E^2} \,
   ({-}\hat x_1 \hat x_2 \hat x_3 \hat x_4)
   (2\bar\alpha+\bar\beta+\bar\gamma) 
   \int_0^a \frac{d(\Delta t)}{(\Delta t)^2} \>
   Z_{\yx\yx'} I_1^\new
   + O(a)
\label {eq:d2Div0}
\end {equation}
with
\begin {equation}
   \int_0^a \frac{d(\Delta t)}{(\Delta t)^2} \>
   Z_{\yx\yx'} I_1^\new
   =
   \frac{\hat x_2 \hat x_3}{\hat x_1 \hat x_4} \, \bbI_{\rm unregulated}
   ,
\label {eq:ZI1vsI}
\end {equation}
\begin {equation}
   \bbI_{\rm unregulated} =
   2\pi^2
   \int_0^a d(\Delta t) \>
   \left[
         \frac{\ln(2i\barOmega_\ix \,\Delta t)}{(\Delta t)^2}
        - \frac{i\barOmega_\ix}{\Delta t}
   \right]
   + O(a) .
\label {eq:bbIunregulated}
\end {equation}
This is the origin of (\ref{eq:smalldt}) in the main text.
As we will see below,
the $\bbI_{\rm unregulated}$ above is the unregulated ($\eps{=}0$)
version of the $\bbI$ introduced in (\ref{eq:bbIdef}) of the main text.
We will see that the
$\calX$ of (\ref{eq:FundDiv}) turn out to represent the $X^\new$ of
(\ref{eq:XYZnew}) without the $|M_\ix|\Omega_\ix$ terms, i.e.
\begin {equation}
   X_\yx^\new = |M_\ix|\Omega_\ix + \calX_\yx, \qquad
   X_{\yx\yx'}^\new = \calX_{\yx\yx'} .
\end {equation}
The expansions (\ref{eq:Xexpansion}) of the $\calX$ then follow from the
expansions (\ref{eq:NLOexpandXnew}) of the $X$.
Finally, we note that the relationship (\ref{eq:abcPP}) and the
explicit values (\ref{eq:xhat}) of the $x_i$ can be
used to rewrite (\ref{eq:d2Div0}) as
\begin {equation}
   \left[\frac{d\Gamma}{d\xe\,d\ye}\right]_{xyy\bar x}^{(\Delta t<a)} =
   -
   \frac{\Nf\alphaqed^2 M_\ix^2}{8\pi^4 E^2} \,
   \frac{ P_{e\to e}(\xe) \, P_{\gamma\to e}\bigl(\frac{\ye}{1{-}\xe}\bigr) }
        { \xe \ye (1{-}\xe{-}\ye) (1{-}\xe)^3}
   \int_0^a \frac{d(\Delta t)}{(\Delta t)^2} \>
   Z_{\yx\yx'} I_1^\new
   + O(a) .
\label {eq:d2Div1}
\end {equation}
The overall factors of $P_{e\to e} P_{\gamma\to e}$ are the source
of the similar overall factors in (\ref{eq:FundDiv}), except that we
will need to generalize the derivation to $d={2-\eps}$ transverse
dimensions.


\subsection {Needed generalizations to \boldmath$d{=}2{-}\eps$}

In the case of QCD $g \to ggg$, the dimensional regularization of
diagrams was carried out in ACI3 \cite{dimreg}.  We can adapt
intermediate results from that paper if we (i) first look at its
sequential diagram result for $xy\bar x\bar y$
[ACI3 eq.\ (5.10)], (ii) convert to QED
by making the same modifications as in section \ref{app:Fund1a} of this
paper, and then (iii) adapt the results to the virtual $xyy\bar x$
diagram as in (\ref{eq:Fund1}).  But, for the reasons described in
section \ref{sec:applyDR},
we {\it also} need to backtrack in the derivation
of ACI3 and never expand the Bessel functions
$K_{d/4}(\frac12 |M| \Omega B^2)$ [ACI3 eq.\ (4.15)].
The result is the following
generalization of the divergent $Z_{\yx\yx'} I_1$ term of
(\ref{eq:virtxyyx}) to $d$ dimensions:
\begin {align}
  \left[\frac{d\Gamma}{d\xe\,d\ye}\right]_{xyy\bar x}^{(\Delta t < a)} = &
   \left( \frac{\mu^2}{E} \right)^{\eps}
   \frac{\Nf \alphaqed^2 M_\ix^2}{2^{d+2} \pi^{2d+1} i^d E^2} \,
   \Gamma^2( \tfrac12{+}\tfrac{d}{4} ) \,
   ({-}\hat x_1 \hat x_2 \hat x_3 \hat x_4)^{d/2}
   (d\bar\alpha+\bar\beta+\bar\gamma)
\nonumber\\ &\times
   \int_0^a \frac{d(\Delta t)}{(\Delta t)^d}
   \int_{{\B^\yx}',\B^\yx}
   {\B^\yx}'\cdot\B^\yx
   \left( \frac{|M_\ix|\Omega_\ix}{({B^\yx}')^2} \right)^{d/4}
     K_{d/4}\bigl(\tfrac12 |M_\ix| \Omega_\ix ({B^\yx}')^2\bigr)
\nonumber\\ &\times
   \left( \frac{|M_\ix|\Omega_\ix}{(B^\yx)^2} \right)^{d/4}
     K_{d/4}\bigl(\tfrac12 |M_\ix| \Omega_\ix (B^\yx)^2\bigr)
\nonumber\\ &\times
   Z_{\yx\yx'}
   \exp\Bigl[
     - \tfrac12
     \calX_\yx (B^\yx)^2
     -
     \tfrac12
     \calX_{\yx'} ({B^\yx}')^2
     +
     \calX_{\yx\yx'} \B^\yx \cdot {\B^\yx}'
   \Bigr]
   + O(a) ,
\label {eq:FundDivd0}
\end {align}
where the $M_\ix$, $\Omega_\ix$ and
$(\calX,Y,Z)$ are defined exactly the same as for $d{=}2$.

To make contact with the derivation in ACI3,
compare (\ref{eq:FundDivd0}) above to the $Z$ term in ACI3 (4.14)
for
the QCD $xy\bar y\bar x$ diagram.  The structure is the same.
In addition to a factor of 2 related to converting the QCD group
factors for $xy\bar y\bar x$ to QED,
and the different way $(\bar\alpha,\bar\beta,\bar\gamma)$ are contracted
here to make $d\bar\alpha{+}\bar\beta{+}\bar\gamma$ for the $Z$ term
of $xyy\bar x$,
there is one minor change in
the overall pre-factor concerning the powers of $\mu$
and $E$ in (\ref{eq:FundDivd0}) above.  $\mu$ is the renormalization
scale, which we introduce by writing the $d$-dimensional coupling
constant $g_d$, which has dimensions of $({\rm mass})^{\eps/2}$, as
\begin {equation}
  g_d = \mu^{\eps/2} g ,
\label {eq:gdim}
\end {equation}
where $g$ is the dimensionless coupling constant for $\eps{=}0$.
As a result, the dimensionless $\alphaqed^2$ in
(\ref{eq:FundDivd0}) is associated with a factor of $\mu^{2\eps}$,
as written explicitly in the pre-factor.  The overall power of
$E$ then follows from dimensional analysis if one takes the
convention, as in ACI3 \cite{dimreg}, that the $d$-dimensional
generalizations of $(\bar\alpha,\bar\beta,\bar\gamma)$ are
defined to be dimensionless.%
\footnote{
  To wit, there was a mistake in the prefactors
  in ACI3 \cite{dimreg}, which are off by an overall factor
  of $(\mu/E)^{2\eps}$.  This mistake did not matter there because
  the $1/\epsilon$ poles all cancel when one adds up all of
  the real-double splitting diagrams (because
  there should be no UV divergence in the total result if there
  are no loops in the amplitude or in the conjugate amplitude).
  In that case, the $\epsilon$ dependence of an overall factor common
  to all diagrams will not matter when we set $\eps{=}0$ at the
  end of the day.  In this paper, however, the $1/\eps$ poles do
  {\it not}\/ cancel when we sum the virtual diagrams of
  fig.\ \ref{fig:diagsVIRT}---they can't, because we know we need
  to get coupling renormalization at this order.
  ACI3's error in the overall factor comes in the
  paragraph of ACI3 appendix A concerning ACI3 eq.\ (4.3),
  which forgets that the
  $d$-dimensional coupling is dimensionful.
\label {foot:norm}
}

We may now use the $d$-dimensional generalization
\cite{dimreg}%
\footnote{
  The argument for this relation is the same as that for ACI3 (5.17)
  \cite{dimreg}.  However, fixing the overall normalization error
  described in footnote \ref{foot:norm} of the current paper modifies
  the $(1-x)^{d-1}$ in the second denominator of that equation to
  $1-x$.  Like the overall normalization, this correction does not
  affect any of ACI3's final results because of
  the cancellation of divergence there, but it is important for our
  current work.
}
\begin {equation}
   \bar\alpha + \tfrac1{d} \bar\beta + \tfrac1{d} \bar\gamma
   =
   \frac{P^{(d)}_{e\to e}(\xe)}{\xe^2(1{-}\xe)^2} \,\,
   \frac{P^{(d)}_{\gamma\to e}\bigl(\frac{\ye}{1{-}\xe}\bigr)}
        {(1{-}\xe) \ye^2 (1{-}\xe{-}\ye)^2}
\end {equation}
of (\ref{eq:abcPP}) to rewrite (\ref{eq:FundDivd0}) as
\begin {align}
  \left[\frac{d\Gamma}{d\xe\,d\ye}\right]_{xyy\bar x}^{(\Delta t < a)} = &
   - \left( \frac{\mu^2}{E} \right)^{\eps}
   \frac{d \Nf \alphaqed^2}{2^{d+2} \pi^{2d} i^d} \,
   \Gamma^2( \tfrac12{+}\tfrac{d}{4} ) \,
   \frac{ P^{(d)}_{e\to e}(\xe) \, P^{(d)}_{\gamma\to e}(\yfrake) }
        { (1-\xe) \bigl[ \xe \ye (1{-}\xe{-}\ye) \bigr]^{\eps/2} } 
   \, \bbI
   + O(a) .
\label {eq:FundDivd1}
\end {align}
where here $\bbI$ is defined by (\ref{eq:bbIdef}) with $(M,\Omega)$ set
to $(M_\ix,\Omega_\ix)$.
The explicit formulas (\ref{eq:xhat}) and (\ref{eq:Mi})
for $\hat x_i$ and $M_\ix$ have been used above,
as well as the small-$\Delta t$ expansion (\ref{eq:ZdtNLO2}) of $Z_{\yx\yx'}$.
$\yfrake \equiv \ye/(1-\xe)$ as in (\ref{eq:yfrake})
is the longitudinal momentum fraction of the virtual
pair's electron relative to the its immediate parent, the photon.
When combined with the definition of $\bbI$,
(\ref{eq:FundDivd1}) supplies the proportionality constant that
we did not show in the corresponding version
(\ref{eq:FundDiv}) in the main text.

We will see later that we do not need the $d$-dimensional version of
the DGLAP splitting function $P^{(d)}_{e\to e}(\xe)$
in (\ref{eq:FundDiv}) because it is common to $xyy\bar x$ and the
leading-order process $x\bar x$.  But we will need the $d$-dimensional
version of the other DGLAP splitting function
$P^{(d)}_{\gamma\to e}(\yfrake)$, which is%
\footnote{
  \label{foot:Pdim}
  See, for example, eq.\ (16) of ref.\ \cite{Pdim}, which one may
  verify independently.
  Our $\eps$ is their $2\eps$, and their $T_R$ is 1 in the QED
  case we consider here.  The result implicitly depends on the
  convention \cite{tHooftVeltman}
  that the trace of the Dirac identity matrix is
  simply {\it defined}\/ as $\tr(\openone_{\rm Dirac}) \equiv 4$
  in $d$ dimensions,
  which is part of Conventional Dimensional Regularization for fermions.
}
\begin {equation}
   P_{\gamma\to e}^{(d)}(z) =
     z^2 + (1{-}z)^2 - \frac{2\eps}{2{-}\eps} \, z(1{-}z) ,
\label {eq:Pdim}
\end {equation}
which reproduces the usual result in the case $\eps{=}0$.


\subsection {The subtraction \boldmath${\cal D}_2(\Delta t)$}

We now turn to the subtraction ${\cal D}_2(\Delta t)$ introduced
in (\ref{eq:Fsplit1}) and (\ref{eq:Fsplit}),
\begin {equation}
   \lim_{\mbox{\small``$\scriptstyle{a\to 0}$''}} \Biggl[
   \int_0^a d(\Delta t) \> F_d(\Delta t)
   +
   \int_a^\infty d(\Delta t) \> {\cal D}_2(\Delta t)
   \Biggr]
   +
   \int_0^\infty d(\Delta t) \>
      \bigl[ F_2(\Delta t) - {\cal D}_2(\Delta t) \bigr] 
   + O(\eps) ,
\label {eq:FsplitB}
\end {equation}
which will allow us
to (i) turn the $d{=}2$ expression (\ref{eq:virtxyyx}) for
$xyy\bar x$ into a convergent integral that can be done
numerically, corresponding to the
$\int d(\Delta t)\>[F_2(\Delta t)-{\cal D}_2(\Delta t)]$ term above,
and (ii) cancel the $a$ dependence of
the $\Delta t{<}a$ contribution (\ref{eq:FundDivd1}), as in the
first two terms above.
We will choose ${\cal D}_2(\Delta t)$ proportional to
(\ref{eq:calD2prop}).  To get the ultimate proportionality constant,
see (\ref{eq:d2Div0}--\ref{eq:bbIunregulated}), but first we
will define a ${\cal D}_2^{(\bbI)}(\Delta t)$ by choosing
a proportionality constant corresponding to
the $\bbI_{\rm unregulated}$ of (\ref{eq:bbIunregulated}):
\begin {equation}
   {\cal D}_2^{(\bbI)}(\Delta t) =
   2\pi^2
   \left[
     \frac{\ln(2i\barOmega_\ix \,\Delta t)}{(\Delta t)^2}
     - i\barOmega_\ix^3\,\Delta t\,\csc^2(\barOmega_\ix\, \Delta t)
   \right] .
\label {eq:calD2I}
\end {equation}
This is equivalent to (\ref{eq:D2Isummary}), given the definition
(\ref{eq:barOmega}) of $\bar\Omega_\ix$.


\subsubsection {Combination with $\bbI$}

One of the integrals we need in (\ref{eq:FsplitB}) is then
\begin {equation}
   \int_a^\infty d(\Delta t) \> {\cal D}_2^{(\bbI)}(\Delta t)
   = 2\pi^2
     \Bigl(
        \frac{\ln(2 i\barOmega_\ix a) + 1}{a}
        + i\barOmega_\ix \bigl[ \ln(2 i\barOmega_\ix a) - 1 \bigr]
     \Bigr) + O(a) .
\label {eq:D2int}
\end {equation}
Combining this with the result (\ref{eq:bbIresult}) for $\bbI$, we
see that the $a$ dependence cancels, leaving
\begin {equation}
   \bbI
   +
   \int_a^\infty d(\Delta t) \> {\cal D}_2(\Delta t)
   =
   2\pi^2 (i \barOmega_\ix)^{d-1} \Bigl[
    - \Bigl( \frac{2}{\eps} - \gammaE + \ln(4\pi) \Bigr)
    + 4\ln2 + 3\ln\pi - 1
  \Bigr] + O(a) + O(\eps) .
\end {equation}
Now multiply this by the prefactors shown in (\ref{eq:FundDivd1})
to convert $\bbI$ into the $\Delta t{<}a$ result for $xyy\bar x$
to get
\begin {align}
  \lim_{\mbox{\small``$\scriptstyle{a\to 0}$''}} \Biggl\{
&
    \left[ \frac{d\Gamma}{d\xe\>d\ye} \right]_{xyy\bar x}^{(\Delta t < a)}
    +
    \left[ \frac{d\Gamma}{d\xe\>d\ye} \right]_{xyy\bar x}^{({\cal D}_2)}
  \Biggr\}
\nonumber\\ &
  =
  \left( \frac{\mu^2}{E} \right)^{\eps}
  \frac{d \Nf \alphaqed^2}{2^{d+2} \pi^{2d}} \,
  \Gamma^2( \tfrac12{+}\tfrac{d}{4} ) \,
  \frac{ P^{(d)}_{e\to e}(\xe) \, P^{(d)}_{\gamma\to e}(\yfrake) }
       { (1-\xe) \bigl[ \xe \ye (1{-}\xe{-}\ye) \bigr]^{\eps/2} } 
\nonumber\\ &\qquad \times
  2\pi^2 i \barOmega_\ix^{d-1} \Bigl[
    - \Bigl( \frac{2}{\eps} - \gammaE + \ln(4\pi) \Bigr)
    + 4\ln2 + 3\ln\pi - 1
  \Bigr] + O(\eps) .
\label {eq:smalldtWithD2}
\end{align}
We'll leave it in this form for the moment.


\subsubsection {The subtracted piece}

Using the normalization of (\ref{eq:ZI1vsI}) for the relation between
$\bbI_{\rm unregulated}$ and $\int (\Delta t)^{-2} Z I_1$, the
$\int d(\Delta t)\>[F_2(\Delta t)-{\cal D}_2(\Delta t)]$
term of (\ref{eq:FsplitB}) then corresponds to
modifying (\ref{eq:virtxyyx}) and (\ref{eq:OmCsc}) to
\pagebreak[2]
\begin {align}
  \left[\frac{d\Gamma}{d\xe\,d\ye}\right]_{xyy\bar x}^{\rm(subtracted)} =
  - & \frac{\Nf\alphaqed^2 M_\ix^2}{16\pi^4 E^2} \,
  ({-}\hat x_1 \hat x_2 \hat x_3 \hat x_4)
  \int_0^\infty d(\Delta t)
  \Biggl[
\nonumber\\ &
   \frac{\Omega_+\csc(\Omega_+\Delta t)}{\Delta t}
   \Bigl\{
     (\bar\beta Y_\yx^2
        + \bar\gamma \Ybar_{\yx\yx'} Y_{\yx\yx'}) I_0^\new
     + (2\bar\alpha+\bar\beta+\bar\gamma) Z_{\yx\yx'} I_1^\new 
\nonumber\\ &\quad
     + \bigl[
         (\bar\alpha+\bar\gamma) Y_\yx^2
         + (\bar\alpha+\bar\beta) \Ybar_{\yx\yx'} Y_{\yx\yx'}
        \bigr] I_2^\new
\nonumber\\ &\quad
     - (\bar\alpha+\bar\beta+\bar\gamma)
       (\Ybar_{\yx\yx'} Y_\yx I_3^\new + Y_\yx Y_{\yx\yx'} I_4^\new)
   \Bigl\}
\nonumber\\ &
   - (2\bar\alpha+\bar\beta+\bar\gamma)
     \frac{\hat x_2\hat x_3}{\hat x_1\hat x_4} \, {\cal D}_2^{(\bbI)}
  \Biggr]
.
\label {eq:virtxyyxsub}
\end {align}
As designed, this is an integral that can be done numerically.


\subsection {Integration over \boldmath$\ye$}

For the virtual diagram, we need to integrate over $\ye$ as in
(\ref{eq:Fund1int}).  For the subtracted piece (\ref{eq:virtxyyxsub}),
this is another integral we will do numerically:
\begin {equation}
   \left[\frac{dI}{d\xe}\right]_{xyy\bar x}^{\rm(subtracted)}
   =
   \int_0^{1-\xe} d\ye \>
   \left[\frac{dI}{d\xe\,d\ye}\right]_{xyy\bar x}^{\rm(subtracted)} .
\end {equation}
For the other piece (\ref{eq:smalldtWithD2}), we will do the $\ye$ integral
analytically.  We start by changing integration variable from
$\ye$ to $\yfrake \equiv \ye/(1-\xe)$:
\begin {align}
  \lim_{\mbox{\small``$\scriptstyle{a\to 0}$''}} \Biggl\{
&
    \left[ \frac{d\Gamma}{d\xe} \right]_{xyy\bar x}^{(\Delta t < a)}
    +
    \left[ \frac{d\Gamma}{d\xe} \right]_{xyy\bar x}^{({\cal D}_2)}
  \Biggr\}
\nonumber\\ &
  =
  \left( \frac{\mu^2}{E} \right)^{\eps}
  \frac{d \Nf \alphaqed^2}{2^{d+2} \pi^{2d}} \,
  \Gamma^2( \tfrac12{+}\tfrac{d}{4} ) \,
  \frac{ P^{(d)}_{e\to e}(\xe) }
       { \xe^{\eps/2}(1-\xe)^\eps } 
  \int_0^1 d\yfrake
    \frac{ P^{(d)}_{\gamma\to e}(\yfrake) }
         { \bigl[ \yfrake(1-\yfrake) \bigr]^{\eps/2} }
\nonumber\\ &\qquad \times
  2\pi^2 i \barOmega_\ix^{d-1} \Bigl[
    - \Bigl( \frac{2}{\eps} - \gammaE + \ln(4\pi) \Bigr)
    + 4\ln2 + 3\ln\pi - 1
  \Bigr]
  + O(\eps) .
\label{eq:divagogo}
\end{align}
$\Omega_\ix$ does not depend on $\ye$, and the $\yfrake$ integral can
be done using the formula $(\ref{eq:Pdim})$ for $P^{(d)}_{\gamma\to e}$,
giving
\begin {equation}
   \int_0^1 d\yfrake
   \frac{ P^{(d)}_{\gamma\to e}(\yfrake) }
        { \bigl[ \yfrake(1-\yfrake) \bigr]^{\eps/2} }
   = \frac{4 \, \Gamma \bigl( \frac{4-\eps}{2} \bigr) }
          { \Gamma(4{-}\eps) }
   = \tfrac23 + \tfrac59 \eps + O(\eps^2) .
\end {equation}


\subsection {Renormalization}
\label {app:xyyxRen}

To carry out renormalization as in (\ref{eq:xyyxRen}), we need
the leading-order result $[d\Gamma/d\xe]_{x\bar x}$ in $d$ transverse
dimensions.
That can be taken from the similar result in ACI3 \cite{dimreg} for
$g \to gg$
except that $\CA \alphas\, P_{g\to gg}(x)$ in QCD is replaced by
$\alphaqed\, P_{e\to e}(\xe)$ here:%
\footnote{
  Specifically, ACI3 eqs.\ (3.1), (3.2) and (3.7) \cite{dimreg} give
  $2\Re[d\Gamma/dx]_{x\bar x}$.  We've used the more general
  $\barOmega_\ix$ (\ref{eq:barOmega}) instead of $\Omega_\ix$ to fit how
  we've written results for $xyy\bar x$, which we want to have the
  option of front- and back-end transforming via
  transforming via (\ref{eq:rfromk}).
}
\begin {equation}
  \left[ \frac{d\Gamma}{d\xe} \right]_{x\bar x} =
  - \frac{\mu^\eps\alphaqed d}{8\pi} \, P_{e\to e}^{(d)}(\xe) \,
  \Beta(\tfrac12{+}\tfrac{d}{4},-\tfrac{d}{4}) \,
  \Bigl( \frac{2\pi}{M_\ix\barOmega_\ix} \Bigr)^{\eps/2}
  i \barOmega_\ix ,
\label {eq:xxLO}
\end {equation}
where $\Beta(x,y) \equiv \Gamma(x)\,\Gamma(y)/\Gamma(x{+}y)$ is the
Euler Beta function and the factor of $\mu^\eps$ associated with
$\alphaqed$ comes from (\ref{eq:gdim}).

Using (\ref{eq:xyyxSplit}) and (\ref{eq:divagogo}--\ref{eq:xxLO}),
and expanding as necessary in $\eps$, we can rewrite
\begin {equation}
  \left[ \frac{d\Gamma}{d\xe} \right]_{xyy\bar x} =
  -
  \frac{\Nf\alphaqed}{3\pi}
  \left[ \frac{d\Gamma}{d\xe} \right]_{x\bar x} 
  \biggl(
    \frac{2}{\eps}
    + \ln\Bigl( \frac{\pi\mu^2}{(1{-}\xe)E\barOmega_\ix} \Bigr)
    + \tfrac53
  \biggr)
  +
  \left[ \frac{d\Gamma}{d\xe} \right]_{xyy\bar x}^{\rm(subtracted)} .
\label {eq:lastdiv}
\end {equation}
Renormalizing using (\ref{eq:xyyxRen}) then leaves us with
\begin {multline}
  \left[ \frac{d\Gamma}{d\xe} \right]_{xyy\bar x}^{(\rm ren)}
  =
  -
  \frac{\Nf\alphaqed}{3\pi}
  \left[ \frac{d\Gamma}{d\xe} \right]_{x\bar x} 
  \biggl(
    \ln\Bigl( \frac{\mu^2}{(1{-}\xe)E\barOmega_\ix} \Bigr)
    + \gammaE
    - 2\ln2
    + \tfrac53
  \biggr)
\\
  +
  \int_0^{1-\xe} d\ye
  \left[ \frac{d\Gamma}{d\xe\,d\ye} \right]_{xyy\bar x}^{(\rm subtracted)}
  ,
\label {eq:xyyxRen2general}
\end {multline}
in which $[d\Gamma/d\xe\,d\ye]_{xyy\bar x}^{\rm(subtracted)}$ is given by
(\ref{eq:virtxyyxsub}).
Eq.\ (\ref{eq:xyyxRen2general})
is the result (\ref{eq:xyyxRen2}) quoted in the main text, but
generalized here to handle either sign of $M_\ix$ and so handle
front-end transformations such as in (\ref{eq:rfromk}).
Since the renormalized
result (\ref{eq:xyyxRen2}) is finite, we may use the
$d{=}2$ version (\ref{eq:xxd2}) of (\ref{eq:xxLO}) there, which is
\begin {equation}
  \left[ \frac{d\Gamma}{d\xe} \right]_{x\bar x} =
  \frac{\alphaqed}{2\pi} \, P_{e\to e}(\xe) \,
  i \barOmega_\ix
  \qquad
  \mbox{for $d{=}2$}.
\label {eq:xxLO2}
\end {equation}
So, as previously promised, we never
need the $d$-dimensional version of the structure function
factor $P_{e\to e}^{(d)}(\xe)$ that was common to the leading-order
$x\bar x$ and the virtual correction $xyy\bar x$.


\subsection {\boldmath$\hat q$ and dimensional regularization}

Throughout this paper, we have used $\hat q$ as an independent parameter
describing interactions with the medium, which could either be calculated
theoretically (in certain limiting cases) or used as a phenomenological
parameter.  One might worry whether or not
the use of dimensional regularization requires knowing the $O(\eps)$
{\it corrections} to the 3+1 dimensional value of $\hat q$, given that
various of our formulas along the way have involved $1/\epsilon$ divergences
multiplying
expressions that depend on $\hat q$ through complex frequencies $\Omega$.
Fortunately, this is not an issue.
Imagine that we used the full $d$-dimensional value of
$\hat q$ (whatever it is) everywhere in our intermediate calculations.
Since our final, renormalized results are finite expressions where
the $\eps\to 0$ limit is taken, the
$\hat q$ in those expressions can then be replaced in the last step
by 3+1 dimensional $\hat q$, just as we did for
$P_{e\to e}(\xe)$ above.%
\footnote{
  We are assuming here that $\hat q$ has been defined in a physically relevant
  way so that it
  itself is not infinite in 3+1 dimensions.  See appendix \ref{app:qhat}.
}


\section{The integral \boldmath$\bbI$}
\label {app:bbI}

In this appendix, we derive the result (\ref{eq:bbIresult}) for the
$\bbI$ integral defined by (\ref{eq:bbIdef}).


\subsection {Perturbative treatment of
             \boldmath$O(M\Omega^2 \, \Delta t \, B^2)$
             corrections to the exponent}

The definition (\ref{eq:bbIdef}) of $\bbI$ is
\begin {align}
   \bbI &\equiv
   \frac{i M (|M|\Omega)^{d/2}}{\pi} 
   \int_0^a \frac{d(\Delta t)}{(\Delta t)^{d+1}}
   \int_{\B,\B'}
   \frac{\B\cdot\B'}{(B^2)^{d/4} ({B'}^2)^{d/4}}
     K_{d/4}\bigl(\tfrac12 |M| \Omega B^2\bigr)
     K_{d/4}\bigl(\tfrac12 |M| \Omega {B'}^2\bigr)
\nonumber\\ & \hspace{10em} \times
   \exp\Bigl[
     - \tfrac12
     \calX_\yx B^2
     -
     \tfrac12
     \calX_{\yx'} {B'}^2
     +
     \calX_{\yx\yx'} \B \cdot \B'
   \Bigr] ,
\label {eq:bbIdef2}
\end {align}
where the $\calX$ have the small-$\Delta t$ expansions
(\ref{eq:Xexpansion})
\begin {subequations}
\label {eq:Xexpansion2}
\begin {align}
   \calX_\yx = \calX_{\yx'}
   &= - \frac{i M}{\Delta t} + \frac{i M \Omega^2 \Delta t}{3}
      + \cdots ,
\\
   \calX_{\yx\yx'}
   &= - \frac{i M}{\Delta t} - \frac{i M \Omega^2 \Delta t}{6}
      + \cdots .
\end {align}
\end {subequations}
Plugging these expansions into the exponential of
(\ref{eq:bbIdef2}) gives
\begin {equation}
   \exp\Bigl[
     - \tfrac12 \calX_\yx B^2 - \tfrac12 \calX_{\yx'} {B'}^2
     + \calX_{\yx\yx'} \B \cdot \B'
   \Bigr]
   =
   \exp\Bigl[
     \frac{i M}{2\Delta t} (\B-\B')^2
   \Bigr]
   \exp\Bigl[
     O(M \Omega^2 \, \Delta t \, \{B^2,{B'}^2\})
   \Bigr] ,
\label {eq:expform2}
\end {equation}
where the last exponential factor represents the corrections from the
second terms in the expansions (\ref{eq:Xexpansion2}) and
$O(\cdots)$ means ``of order.''
Because of the bound (\ref{eq:Bbound}) on the sizes of
both $B$ and $B'$ that contribute to $\bbI$, the exponent
of this second exponential is small:
\begin {equation}
   |M \Omega^2 \, \Delta t \, B^2|
   \lesssim |\Omega \, \Delta t|
   \le |\Omega a| \ll 1 .
\end {equation}
We will need this correction, but we can treat it as perturbative,
rewriting (\ref{eq:expform2}) as
\begin {equation}
   \exp\Bigl[
     - \tfrac12 \calX_\yx B^2 - \tfrac12 \calX_{\yx'} {B'}^2
     + X_{\yx\yx'} \B \cdot \B'
   \Bigr]
   =
   \exp\Bigl[
     \frac{i M}{2\Delta t} (\B-\B')^2
   \Bigr]
   \Bigl[
     1 +
     O(M \Omega^2 \, \Delta t \, \{B^2,{B'}^2\})
   \Bigr] .
\end {equation}
Formally, it will be convenient to implement this expansion by introducing a
redundant parameter $\xi{=}1$ into (\ref{eq:Xexpansion2}),
\begin {subequations}
\label {eq:Xexpansionxi}
\begin {align}
   \calX_\yx = \calX_{\yx'}
   &= - \frac{i M}{\Delta t} + \xi \, \frac{i M \Omega^2 \Delta t}{3}
      + O\bigl(M \Omega^4 (\Delta t)^3\bigr) ,
\\
   \calX_{\yx\yx'}
   &= - \frac{i M}{\Delta t} - \xi \, \frac{i M \Omega^2 \Delta t}{6}
      + O\bigl(M \Omega^4 (\Delta t)^3\bigr) ,
\end {align}
\end {subequations}
and then think of $\bbI$ as a function $\bbI(\xi)$ of $\xi$.
The above discussion then translates to
\begin {equation}
   \bbI = \bigl[ \bbI(\xi) \bigr]_{\xi=1}
   = \Bigl[
        \bbI(0) + \xi\,\bbI'(0) + \frac{\xi^2}{2!}\,\bbI''(0) + \cdots
     \Bigr]_{\xi=1}
   = \bbI(0) + \bbI'(0) + O(a) .
\label {eq:bbIexpandXi}
\end {equation}
For the same reason that the
unregulated ($\eps{=}0$) discussion of appendix \ref{app:smalldt}
did not require the yet-higher order terms
not explicitly shown in (\ref{eq:Xexpansionxi}),
we will not need them here either.
They give vanishing contribution to $\bbI$ for
$a\to 0$.


\subsection {Units}

In order to simplify the presentation of calculations,
in this appendix we will start by assuming $M>0$ and
evaluate $\bbI$ in units where
\begin {equation}
  M=1 \quad \mbox{and} \quad \Omega=1 .
\end {equation}
At the end, we will be able to put $M$ and the complex-valued
$\Omega$ back into the answer using
the scaling properties of the definition (\ref{eq:bbIdef2}) of $\bbI$
and then analytic continuation of $\Omega$ back to complex values.
Then we will generalize the result to also cover the case
$M < 0$.


\subsection {Representation of Bessel Functions}

Integrating complicated things involving Bessel functions is hard,
which motivates us to
replace each Bessel function with the integral representation%
\footnote{
  See, e.g., Gradshteyn \& Ryzhik (8.432.3) \cite{GR}
  for a more precise statement on range of validity.
}
\begin {equation}
  z^{-\nu} K_\nu(z) =
  \frac{\pi^{1/2}}{2^\nu\Gamma(\frac12+\nu)}
  \int_1^\infty dp \> (p^2-1)^{\nu-\frac12} e^{-p z}
  \qquad
  (z > 0).
\end {equation}
So (\ref{eq:bbIdef2}) for $\bbI$ becomes
\begin {multline}
   \bbI =
   \frac{i\, 2^{-d}}{\Gamma^2(\frac{d+2}{4})}
   \int_1^\infty dp \, dq \> (p^2-1)^{-\eps/4} (q^2-1)^{-\eps/4}
\\ \times
   \int_0^a \frac{d(\Delta t)}{(\Delta t)^{d+1}}
   \int_{\B,\B'}
   \B\cdot\B'
   \exp\Bigl[
     - \tfrac12 (\calX_\yx+p) B^2 - \tfrac12 (\calX_{\yx'}+q) {B'}^2
     + X_{\yx\yx'} \B \cdot \B'
   \Bigr] .
\label {eq:bbI2}
\end {multline}


\subsection {Doing the $\B$ integrals}

To do the $\B$ integrals above, it is convenient to take advantage
of the form of the exponential to replace
\begin {equation}
   \B\cdot\B' \exp[\cdots] \to
   \frac{\partial}{\partial \calX_{\yx\yx'}} \, \exp[\cdots] .
\label {eq:PrefactorTrick}
\end {equation}
The $\B$ integral of the exponential gives
\begin {align}
   \int_{\B,\B'}
   \exp\Bigl[
     - \tfrac12 (\calX_\yx+p) B^2 - \tfrac12 & (\calX_{\yx'}+q) {B'}^2
     + \calX_{\yx\yx'} \B \cdot \B'
   \Bigr]
\nonumber\\
   &=
   (2\pi)^d
   \left[ \det
      \begin{pmatrix}
        \calX_\yx+p & -\calX_{\yx\yx'} \\
        -\calX_{\yx\yx'} & \calX_{\yx'}+q
      \end{pmatrix}
   \right]^{-d/2}
\nonumber\\
   &=
   \frac{(2\pi)^d}
        {\bigl[
          (\calX_\yx \calX_{\yx'} - \calX_{\yx\yx'}^2)
          + \calX_{\yx'} p + \calX_\yx q + pq
         \bigr]^{d/2}}
   \,.
\end {align}
Now use (\ref{eq:PrefactorTrick}), giving
\begin {multline}
   \int_{\B,\B'}
   \B\cdot\B'
   \exp\Bigl[
     - \tfrac12 (\calX_\yx+p) B^2 - \tfrac12 (\calX_{\yx'}+q) {B'}^2
     + \calX_{\yx\yx'} \B \cdot \B'
   \Bigr]
\\
   =
   \frac{d (2\pi)^d \calX_{\yx\yx'}}
        {\bigl[
          (\calX_\yx \calX_{\yx'} - \calX_{\yx\yx'}^2)
          + \calX_{\yx'} p + \calX_\yx q + pq
         \bigr]^{\frac{d}{2} + 1}}
   \,.
\end {multline}
Now specialize to our case by using (\ref{eq:Xexpansionxi}) to get
\begin {multline}
   \int_{\B,\B'}
   \B\cdot\B'
   \exp\Bigl[
     - \tfrac12 (\calX_\yx+p) B^2 - \tfrac12 (\calX_{\yx'}+q) {B'}^2
     + \calX_{\yx\yx'} \B \cdot \B'
   \Bigr]
\\
   =
   \frac{-i d (2\pi)^d}
        {\Delta t \, \Bigl[
          - \frac{i(p+q)}{\Delta t}
          + pq + \xi
         \Bigr]^{\frac{d}{2} + 1}}
    \,
    \bigl[ 1 + O\bigl( (\Delta t)^2 \bigr) \bigr]
   \,.
\label {eq:I1}
\end {multline}
As discussed earlier,
the relative $O\bigl( (\Delta t)^2 \bigr)$ corrections can be
dropped because they will lead to a convergent $\Delta t$ integral
in $d{=}2$ that vanishes
as $a \to 0$.

Using (\ref{eq:I1}) in (\ref{eq:bbI2}) gives
\begin {multline}
   \bbI(\xi) =
   \frac{d \pi^d}{\Gamma^2(\frac{d+2}{4})}
   \int_1^\infty dp \, dq \> (p^2-1)^{-\eps/4} (q^2-1)^{-\eps/4}
\\ \times
   \int_0^a \frac{d(\Delta t)}{(\Delta t)^{(d+2)/2}}
   \bigl[ -i(p+q) + (pq + \xi)\Delta t \bigr]^{-(d+2)/2}
   + O(a)
   .
\label {eq:bbI3}
\end {multline}
We find it convenient to change integration variables to
\begin {equation}
  u = \frac{1}{p} \,,
  \qquad
  v = \frac{1}{q} \,,
\end {equation}
to get%
\footnote{
   One check of (\ref{eq:bbI4}) is to set $d{=}2$ and $\xi=1$,
   do the $(u,v)$ integrals,
   and verify that the result reproduces (\ref{eq:bbIunregulated}).
}
\begin {multline}
   \bbI(\xi) =
   \frac{d \pi^d}{\Gamma^2(\frac{d+2}{4})}
   \int_0^1 du \, dv \> (1-u^2)^{-\eps/4} (1-v^2)^{-\eps/4}
\\ \times
   \int_0^a \frac{d(\Delta t)}{(\Delta t)^{(d+2)/2}}
   \bigl[ -i(u+v) + (1 + \xi u v)\Delta t \bigr]^{-(d+2)/2}
   + O(a)
   .
\label {eq:bbI4}
\end {multline}


\subsection{Doing the \boldmath$\Delta t$ Integral}

To perform the $\Delta t$ integrals, we use%
\footnote{
  The integral (\ref{eq:dtHyperInt}) can be derived by expanding the
  integrand in a Taylor series in $\alpha$ and integrating term by
  term.
}
\begin {equation}
  \int_0^a \frac{d(\Delta t)}{(\Delta t)^r (\beta+\alpha\,\Delta t)^s}
  =
  \frac{ a^{1-r} \beta^{-s} }{ (1-r) } \;
  F\bigl( 1{-}r, s; 2{-}r; {-}\tfrac{\alpha}{\beta} a \bigr)
  ,
\label {eq:dtHyperInt}
\end {equation}
where $F = {}_2F_1$ is the hypergeometric function.
In our application (\ref{eq:bbI4}),
\begin {multline}
   \int_0^a \frac{d(\Delta t)}{(\Delta t)^{(d+2)/2}}
   \bigl[ -i(u+v) + (1 + \xi u v)\Delta t \bigr]^{-(d+2)/2}
\\
   =
   -\frac{2}{d a^{d/2} \beta^{(d+2)/2}} \;
   F\bigl( {-}\tfrac{d}{2}, 1+\tfrac{d}{2}; 1-\tfrac{d}{2};
           {-}\tfrac{\alpha}{\beta} a \bigr)
\label {eq:dtint1}
\end {multline}
with
\begin {equation}
  \alpha \equiv 1 + \xi u v ,
  \qquad
  \beta \equiv -i(u+v)
\end {equation}
(not to be confused with any other use of the letters
$\alpha$ or $\beta$ in this paper, but there are only so many
letters in the alphabet).
Use the hypergeometric transformation
\begin {multline}
  F(a,b;c;z) =
  \frac{\Gamma(c)\,\Gamma(b-a)}{\Gamma(b)\,\Gamma(c-a)} \,
    (-z)^{-a} \,
    F\bigl(a, 1-c+a; 1-b+a; \tfrac{1}{z} \bigr)
\\
  +
  \frac{\Gamma(c)\,\Gamma(a-b)}{\Gamma(a)\,\Gamma(c-b)} \,
    (-z)^{-b} \,
    F\bigl(b, 1-c+b; 1-a+b; \tfrac{1}{z} \bigr) ,
\end {multline}
together with $F(a,0;c;z) = 1$ [which follows from the series expansion
that defines the hypergeometric function],
to rewrite (\ref{eq:dtint1}) as
\begin {align}
   \int_0^a & \frac{d(\Delta t)}{(\Delta t)^{(d+2)/2}}
   \bigl[ -i(u+v) + (1 + \xi u v)\Delta t \bigr]^{-(d+2)/2}
\nonumber\\
   &=
   - \frac{2}{d} \biggl[
     \frac{ \Gamma(1-\frac{d}{2}) \, \Gamma(1+d) }{ \Gamma(1+\frac{d}{2}) }
       \, \frac{\alpha^{d/2}}{\beta^{1+d}}
     +
     \frac{ \Gamma(1-\frac{d}{2}) \, \Gamma(-1-d) }
            { \Gamma(-\frac{d}{2}) \, \Gamma(-d) }
       \,
       \frac{ F\bigl( 1+\tfrac{d}{2}, 1+d; 2+d;
                 {-}\tfrac{\beta}{\alpha a} \bigr) }
            { a^{1+d} \alpha^{1+\frac{d}{2}} }
   \biggr]
\nonumber\\
   &=
     - \frac{ 2 \, \Gamma(1-\frac{d}{2}) \, \Gamma(1+d) }
          { d \, \Gamma(1+\frac{d}{2}) }
       \, \frac{\alpha^{d/2}}{\beta^{1+d}}
     -
       \frac{ F\bigl( 1+\tfrac{d}{2}, 1+d; 2+d;
                 {-}\tfrac{\beta}{\alpha a} \bigr) }
            { (1+d) a^{1+d} \alpha^{1+\frac{d}{2}} }
   \,.
\label {eq:dtint2}
\end {align}
The advantage of this rewriting is that the second term on the right-hand
side of (\ref{eq:dtint2}) is finite if we set $d=2$.
That is, the $1/\epsilon$ divergence arising from the $\Delta t$
integration is isolated in the first term.  Moreover, we will
see that the integration of the finite second term over $(u,v)$
is also finite, so we do not need to keep dimensional regularization
in order to do those $(u,v)$ integrals: we can just set $d{=}2$ there and
be done with it.  All together, using (\ref{eq:dtint2}) in the expression
(\ref{eq:bbI4}) for $\bbI$ gives
\begin {equation}
   \bbI(\xi) = \bbI_{\rm div}(\xi) + \bbI_{\rm regular}(\xi) + O(a) + O(\eps)
\label {eq:bbIdivreg}
\end {equation}
with (dimensionally regularized) divergent piece
\begin {align}
  \bbI_{\rm div}(\xi) &=
   -\frac{d \pi^d}{\Gamma^2(\frac{d+2}{4})}
   \int_0^1 du \, dv \> (1-u^2)^{-\eps/4} (1-v^2)^{-\eps/4} \,
     \frac{ 2 \, \Gamma(1-\frac{d}{2}) \, \Gamma(1+d) }
          { d \, \Gamma(1+\frac{d}{2}) }
       \, \frac{\alpha^{d/2}}{\beta^{1+d}}
\nonumber\\
  &=
   - \frac{ 2 \pi^d \, \Gamma(1-\frac{d}{2}) \, \Gamma(1+d) }
          { \Gamma(1+\frac{d}{2}) \, \Gamma^2(\frac{d+2}{4}) }
   \int_0^1 du \, dv \> (1-u^2)^{-\eps/4} (1-v^2)^{-\eps/4} \,
       \, \frac{(1+\xi u v)^{d/2}}{[-i(u+v)]^{1+d}}
\label {eq:Idiv1}
\end {align}
and regular, finite piece
\begin {equation}
  \bbI_{\rm regular}(\xi) =
   - 2 \pi^2
   \int_0^1 du \, dv \>
       \frac{ F\bigl( 2,3;4; {-}\tfrac{\beta}{\alpha a} \bigr) }
            { 3 a^3 \alpha^2 }
   =
   - \frac{2\pi^2}{3a^3}
   \int_0^1 du \, dv \>
       \frac{ F\bigl( 2,3;4; \tfrac{i(u+v)}{(1+\xi u v) a} \bigr) }
            { (1+\xi u v)^2 }
   .
\label {eq:Ireg1}
\end {equation}
The hypergeometric function above is given in terms of elementary
functions as
\begin {equation}
   F(2,3;4;z) =
     \frac{6\,[\ln(1-z)+z+\frac12 z^2]}{z^3} + \frac{3}{1-z}
\end {equation}
and falls like $3/z^2$ for large $z$.


\subsection{Evaluating \boldmath$\bbI_{\rm regular}$}

We have not investigated whether the $(u,v)$ integrals (\ref{eq:Ireg1})
for $\bbI_{\rm regular}(\xi)$ can be performed by brute force,
but the integrals
simplify if we use the trick of expanding in $\xi$ as in
(\ref{eq:bbIexpandXi}):
\begin {equation}
   \bbI = \bbI(0) + \bbI'(0) + O(a) .
\label {eq:bbIexpandXi2}
\end {equation}
One can use symbolic integration software to do the integrals
corresponding to $\bbI_{\rm regular}(0)$ and $\bbI'_{\rm regular}(0)$.
Alternatively, one can rewrite $du \, dv = \tfrac12 \, d(u{+}v)\,d(u{-}v)$
with appropriate limits of integration and do the relatively simple
$u{-}v$ integration by hand, followed by the $u{+}v$ integration.
By either method, the results, when expanded in $a$, are
\begin {align}
   \bbI_{\rm regular}(0) &= 
   -\frac{2\pi^2}{a} \bigl[ \ln(2 i a) + 1 \bigr]
   - i \pi^2 \bigl[ 3\ln(i a) + \ln2 - 3 \bigr]
   + O(a) ,
\\
   \bbI'_{\rm regular}(0) &= 
   i \pi^2 \bigl[ \ln(i a) + 3\ln2 \bigr]
   + O(a) ,
\end {align}
so that
\begin {equation}
   \bbI_{\rm regular}(0) + \bbI'_{\rm regular}(0) = 
   -\frac{2\pi^2}{a} \bigl[ \ln(2 i a) + 1 \bigr]
   - i \pi^2 \bigl[ 2\ln(\tfrac{i a}{2}) - 3 \bigr]
   + O(a) .
\label{eq:IregExpand}
\end {equation}

The ignorability of higher-order terms in the $\xi$-expansion
(\ref{eq:bbIexpandXi2}) of $\bbI$ in the $a{\to}0$ limit
will hold for the total
$\bbI = \bbI_{\rm div}+\bbI_{\rm regular}$ but turns out not to hold
separately for $\bbI_{\rm div}$ and $\bbI_{\rm regular}$.
A check we should make on our calculation is that the
$O(\xi^2,\xi^3,\cdots)$ terms indeed cancel between
$\bbI_{\rm div}$ and $\bbI_{\rm regular}$.  For this purpose, it
will be sufficient to check that
$\bbI''_{\rm div}(\xi) + \bbI''_{\rm regular}(\xi) = O(a)$ for
arbitrary $\xi \le 1$.  So, for later reference, we
note here that differentiating the integrand of
(\ref{eq:Ireg1}) twice with respect to $\xi$ gives
\begin {align}
  \bbI''_{\rm regular}(\xi) &=
  4\pi^2 i \int_0^1 du \, dv \>
  \frac{(u v)^2}{\bigl(u + v + i(1+\xi u v) a\bigr)^3 (1 + \xi u v)}
\nonumber\\
  &=
  4\pi^2 i \int_0^1 du \, dv \>
  \frac{(u v)^2}{(u+v)^3 (1+\xi u v)}
  + O(a) .
\label {eq:bbIregHigher}
\end {align}


\subsection{Evaluating \boldmath$\bbI_{\rm div}$}

Rewrite (\ref{eq:Idiv1}) as
\begin {multline}
   \bbI_{\rm div}(\xi) =
   -\frac{ 2 \pi^d \, \Gamma(1-\frac{d}{2}) \, \Gamma(1+d) }
          { \Gamma(1+\frac{d}{2}) \, \Gamma^2(\frac{d+2}{4}) }
     \; i^{d+1} \bbA(\xi)
\\
   = 4 \pi^2 i
     \Bigl(
        \frac{2}{\eps} - \gammaE - 2\ln(i\pi) - 2
        + O(\eps)
     \Bigr) \, \bbA(\xi),
\label {eq:bbIdivA}
\end {multline}
where
\begin {equation}
  \bbA(\xi) \equiv 
   \int_0^1 du \, dv \> (1-u^2)^{-\eps/4} (1-v^2)^{-\eps/4} \,
       \, \frac{(1+\xi u v)^{d/2}}{(u+v)^{1+d}}
   \,.
\label {eq:bbAdef}
\end {equation}

As before, we take the $\xi$ expansion (\ref{eq:bbIexpandXi2})
of $\bbI$ and so here of $\bbA$.  The integral
\begin {equation}
  \bbA(0) =
   \int_0^1 \frac{du \, dv}{(u+v)^{1+d}} \, (1-u^2)^{-\eps/4} (1-v^2)^{-\eps/4}
\label {eq:bbAxi0a}
\end {equation}
does not converge for $d=2{-}\eps$ near 2, and so we cannot simply
expand the integrand in powers of $\eps$.  Dimensional regularization
tells us to imagine doing the integral in dimensions $d<1$
where it converges and then analytically continuing the result
in $d$.  But we can make that job
easier if we first rewrite (\ref{eq:bbAxi0a}) as
\begin {equation}
  \bbA(0) =
   \int_0^1 \frac{du \, dv}{(u+v)^{1+d}}
   +
   \int_0^1 \frac{du \, dv}{(u+v)^{1+d}} \,
     \bigl[ (1-u^2)^{-\eps/4} (1-v^2)^{-\eps/4} - 1 \bigr] .
\end {equation}
The first integral is relatively easy, whereas the second integral converges
for $d=2{-}\eps$ near 2 and so we may expand that integrand in powers
of $\eps$.  So, using $u {\leftrightarrow} v$ symmetry of the integrand,
\begin {align}
  \bbA(0) &=
  \int_0^1 \frac{du \, dv}{(u+v)^{1+d}}
    - \frac{\eps}{2} \int_0^1 \frac{du \, dv}{(u+v)^3} \, \ln(1-u^2)
    + O(\eps^2)
\nonumber\\
  &=
  \frac{2(1-2^{-d})}{d(1-d)}
  - \frac{\eps}{2} \left( - \frac14 - \frac{\ln 2}{2} \right)
    + O(\eps^2)
\nonumber\\
  &= -\tfrac34
     \Bigl[
        1 + \Bigl( \tfrac43 - \tfrac23 \ln 2 \Bigr)\eps
        + O(\eps^2)
     \Bigr] .
\label {eq:bbAxi0}
\end {align}

The next term in the $\xi$ expansion is
\begin {equation}
   \bbA'(0) =
   \frac{d}{2}
   \int_0^1 du \, dv \> \frac{u v}{(u+v)^{1+d}} \, (1-u^2)^{-\eps/4} (1-v^2)^{-\eps/4}
   .
\end {equation}
This integral is convergent for $d$ near 2, and so we may expand the integrand
in powers of $\eps$.  Again using $u \leftrightarrow v$ symmetry,
\begin {align}
   \bbA'(0) &=
   \frac{d}{2} \Biggl[
     \int_0^1 du \, dv \> \frac{u v}{(u+v)^3}
     - \frac{\eps}{2} \int_0^1 du \, dv \> \frac{u v}{(u+v)^3} \, \ln(1-u^2)
\nonumber\\ & \hspace{14em}
     + \eps \int_0^1 du \, dv \> \frac{u v}{(u+v)^3} \, \ln(u+v)
   \Biggr]
\nonumber\\
   &=
   \tfrac{d}{2} \Bigl[
     \tfrac14
     - \tfrac{\eps}{2} \Bigl( \tfrac14 - \tfrac12\ln 2 \Bigr)
     + \eps \Bigl( \tfrac38 - \tfrac34\ln 2 \Bigr)
   \Bigr]
\nonumber\\
   &=
   \tfrac14
   \Bigl[
     1 + \Bigl( \tfrac12 - 2\ln2 \Bigr) \eps
        + O(\eps^2)
   \Bigr] .
\end {align}
Combining this with (\ref{eq:bbAxi0}) gives
\begin {equation}
  \bbA(0) + \bbA'(0) =
   -\tfrac12
   \Bigl[
     1 + \tfrac{7}{4} \eps
        + O(\eps^2)
   \Bigr]
\end {equation}
and thence, from (\ref{eq:bbIdivA}),
\begin {equation}
  \bbI_{\rm div}(0) + \bbI'_{\rm div}(0) =
  2\pi^2 i \Bigl[
    - \frac{2}{\eps} + \gammaE + 2\ln(i \pi) - \frac32
    + O(\eps)
  \Bigr] .
\label {eq:IdivExpand}
\end {equation}

Let's also pause to check the cancellation of higher-order terms
$O(\xi^2,\xi^3,\cdots)$ in the $\xi$ expansion.  Differentiating
(\ref{eq:Idiv1}) twice with respect to $\xi$ gives
\begin {equation}
   \bbI''_{\rm div}(\xi) =
   - \frac{ 2 \pi^d \, \Gamma(1-\frac{d}{2}) \, \Gamma(1+d) }
          { \Gamma(1+\frac{d}{2}) \, \Gamma^2(\frac{d+2}{4}) }
   \int_0^1 du \, dv \> (1-u^2)^{-\eps/4} (1-v^2)^{-\eps/4} \,
       \frac{ \frac{d}{2} (\frac{d}{2}{-}1)(1{+}\xi u v)^{\frac{d}{2}-2} (uv)^2 }
            { [-i(u+v)]^{1+d} }
   \,.
\end {equation}
The integral is convergent for $d$ near 2, and expanding in $\eps$
gives
\begin {equation}
   \bbI''_{\rm div}(\xi) =
  - 4\pi^2 i \int_0^1 du \, dv \>
  \frac{(u v)^2}{(u+v)^3 (1+\xi u v)}
  + O(\eps) .
\end {equation}
As promised, this indeed cancels the corresponding behavior
(\ref{eq:bbIregHigher}) up to $O(a)$ corrections, confirming the
expansion $\bbI = \bbI(0) + \bbI'(0) + O(a)$ for the total
$\bbI = \bbI_{\rm div} + \bbI_{\rm regular}$.


\subsection{Final result for $\bbI$ for \boldmath$M > 0$}

Combining (\ref{eq:bbIexpandXi}), (\ref{eq:bbIdivreg}),
(\ref{eq:IregExpand}), and (\ref{eq:IdivExpand}), we get the
result for the integral $\bbI$ defined by (\ref{eq:bbIdef}):
\begin {equation}
  \bbI =
  2\pi^2 i \Bigl[
    - \Bigl( \frac{2}{\eps} - \gammaE + \ln(4\pi) \Bigr)
    - \frac{\ln(2 i a)+1}{i a}
    - \ln a
    + 3\ln(2\pi) + \tfrac{i\pi}{2}
  \Bigr]
\label {eq:bbIresult1}
\end {equation}
in units where $M=1$ and $\Omega=1$.
We have isolated the combination $\frac{2}{\eps} - \gammaE + \ln(4\pi)$
above because that is the combination that appears in $\MSbar$
renormalization (\ref{eq:renorm}).

To restore $M$ and $\Omega$, note that we could scale out
all the $M$ and $\Omega$ from the original integral (\ref{eq:bbIdef2})
combined with (\ref{eq:Xexpansion2}) by rescaling integration variables as
\begin {equation}
  \B \to \frac{\B}{(M\Omega)^{1/2}} \,,
  \quad
  \Delta t \to \frac{\Delta t}{\Omega} \,.
\end {equation}
This rescaling brings (\ref{eq:bbIdef2}) into a dimensionless form times
(i) the $M(M\Omega)^{d/2}$ already explicit in (\ref{eq:bbIdef2}),
(ii) a factor of $\Omega^d$ from the $d(\Delta t)/(\Delta t)^{d+1}$,
and (iii) a factor of $(M \Omega)^{-(d+2)/2}$ from the
$d^dB \, d^dB' \> \B\cdot\B' / (B^2)^{d/4} ({B'}^2)^{d/4}$.
These combine into an overall factor of $\Omega^{d-1}$.
We can therefore go backward and restore the original units
to (\ref{eq:bbIresult1}) by (a) replacing
$a$ (which is a cut-off on $\Delta t$) by $\Omega a$
and (b) multiplying by $\Omega^{d-1}$ overall, giving
\begin {equation}
  \bbI =
  2\pi^2 i \Omega^{d-1} \Bigl[
    - \Bigl( \frac{2}{\eps} - \gammaE + \ln(4\pi) \Bigr)
    - \frac{\ln(2 i \Omega a)+1}{i \Omega a}
    - \ln(\Omega a)
    + 3\ln(2\pi) + \tfrac{i\pi}{2}
  \Bigr] ,
\end {equation}
which can also be written as
\begin {equation}
  \bbI =
  2\pi^2 (i \Omega)^{d-1} \Bigl[
    - \Bigl( \frac{2}{\eps} - \gammaE + \ln(4\pi) \Bigr)
    - \frac{\ln(2 i \Omega a)+1}{i \Omega a}
    - \ln(i\Omega a)
    + 3\ln(2\pi)
  \Bigr] ,
\label {eq:bbIresult2}
\end {equation}


\subsection{Generalization to include \boldmath$M < 0$}

Let $\bbI(M,\Omega)$ represent $\bbI$ as a function of $M$ and $\Omega$.
From the definition of $\bbI$ by (\ref{eq:bbIdef2}) and (\ref{eq:Xexpansion2}),
one can see that
\begin {equation}
   \bbI(-M,\Omega) = [\bbI(M,\Omega^*)]^* .
\label {eq:bbInegateM}
\end {equation}
This relation tells us how to get the result for negative $M$ from the result
(\ref{eq:bbIresult2}) for positive $M$.  We can implement the relation
by rewriting (\ref{eq:bbIresult2}) as
\begin {equation}
  \bbI =
  2\pi^2 (i \barOmega)^{d-1} \Bigl[
    - \Bigl( \frac{2}{\eps} - \gammaE + \ln(4\pi) \Bigr)
    - \frac{\ln(2 i \barOmega a)+1}{i \barOmega a}
    - \ln(i\barOmega a)
    + 3\ln(2\pi)
  \Bigr]
\label {eq:bbIbar}
\end {equation}
with $\barOmega \equiv \Omega \sgn M$ as in (\ref{eq:barOmega}).
This is the result that was quoted in (\ref{eq:bbIresult}).


\section{Technical points concerning front-end transformations}
\label {app:frontend}

\subsection{Branch cuts for the transformation (k)\boldmath$\to$(r)}

The result given by eqs.\ (\ref{eq:dGk}) and (\ref{eq:xxd2}) for
diagram (k) contains a term
\begin {equation}
  -
  \frac{\Nf\alphaqed^2}{6\pi^2}
    \, P_{e\to e}(\xe) \,
  2\Re\biggl\{ 
    i\Omega_\ix \sgn(M_\ix)
    \ln\Bigl( \frac{\mu^2}{(1{-}\xe)E\Omega_\ix\sgn M_\ix} \Bigr)
  \biggr\} \,.
\label {eq:kLogTerm}
\end {equation}
Under a combined front-end and back-end trasnformation, this term
transforms the same way as (\ref{eq:frontbackLO}).  [See the comment
after (\ref{eq:frontbackLO}) to understand the relation to
(\ref{eq:rfromk}).]  Under this transformation,
$\Omega_\ix \to (\Omega_0^{\gamma\to e\bar e})^*$ and $\sgn(M_\ix) \to -1$.
(\ref{eq:kLogTerm}) transforms to
\begin {multline}
  -
  \frac{\Nf\alphaqed^2}{6\pi^2}
    \, P_{\gamma\to e}(\ye) \,
  2\Re\biggl\{ 
    -i (\Omega_0^{\gamma\to e\bar e})^*
    \ln\Bigl( \frac{\mu^2}{-E(\Omega_0^{\gamma\to e\bar e})^*} \Bigr)
  \biggr\}
\\
  =
  -
  \frac{\Nf\alphaqed^2}{6\pi^2}
    \, P_{\gamma\to e}(\ye) \,
  2\Re\biggl\{ 
    i \Omega_0^{\gamma\to e\bar e}
    \ln\Bigl( \frac{\mu^2}{-E\Omega_0^{\gamma\to e\bar e}} \Bigr)
  \biggr\} \,.
\label {eq:rLogTerm}
\end {multline}
This result may be confusing depending on whether one thinks that
the minus sign inside the logarithm represents $e^{i\pi}$ or $e^{-i\pi}$.
In general, we try to write expressions so that the relevant branch cut
for logarithms is along the negative real axis.  Since
$\Omega_0^{\gamma \to e \bar e} \propto \sqrt{-i}$ (which should be interpreted
as $e^{-i\pi/4}$ in LPM formulas), that means that (\ref{eq:rLogTerm}) above
is the same as
\begin {equation}
  -
  \frac{\Nf\alphaqed^2}{6\pi^2}
    \, P_{\gamma\to e}(\ye) \,
  2\Re\biggl\{ 
    i \Omega_0^{\gamma\to e\bar e}
    \left[
      \ln\Bigl( \frac{\mu^2}{E\Omega_0^{\gamma\to e\bar e}} \Bigr)
      - i\pi
    \right]
  \biggr\} \,,
\label {eq:rLogTerm2}
\end {equation}
which is how we have written it in (\ref{eq:dGr}).

To understand why this is the correct interpretation of the logarithm
in (\ref{eq:rLogTerm}), and the origin of why there is a $-i\pi$ term in
(\ref{eq:rLogTerm2}) for diagram (r) that is not in (\ref{eq:kLogTerm})
for diagram (k), we need to step back to an earlier stage of the
derivation.  The important difference revolves around the overall factor
of $1/i^d$ in (\ref{eq:FundDivd0}) for the $xyy\bar x$ (k) diagram.
This factor can be traced back to the small-$\Delta t$ expression
for the 4-particle (effectively 2-particle)
propagator in ACI3 (4.9) \cite{dimreg}, which
can be viewed as the product of a free propagator for $\C_{34}$ times
a free propagator for $\C_{12}$,
\begin {equation}
  \langle\C_{34},\C_{12},\Delta t|\C'_{34},\C'_{12},0\rangle
  \simeq
\\
  \left(\frac{{\mathfrak M}_{34}}{2\pi i \, \Delta t}\right)^{d/2}
  e^{i|\C_{34}{-}\C_{34}'|^2/2\,\Delta t}
  \times
  \left(\frac{{\mathfrak M}_{12}}{2\pi i \, \Delta t}\right)^{d/2}
  e^{i|\C_{12}{-}\C_{12}'|^2/2\,\Delta t} ,
\label {eq:CCCCsmall}
\end {equation}
with
\begin {subequations}
\label {eq:Ms}
\begin {align}
  {\mathfrak M}_{34} &= x_3 x_4 (x_3{+}x_4)E ,
\\
  {\mathfrak M}_{12} &= x_1 x_2 (x_1{+}x_2)E = -x_1 x_2 (x_3{+}x_4) E
\end {align}
\end {subequations}
For the $xyy\bar x$ diagram,
$(x_1,x_2,x_3,x_4)=(-1,\ye,1{-}\xe{-}\ye,\xe)$, which means that both
of the masses $({\mathfrak M}_{34},{\mathfrak M}_{12})$ are positive.
The total phase associated with the exponential prefactors in
(\ref{eq:CCCCsmall}) is then
\begin {equation}
   \left( \frac{1}{i} \right)^{d/2} \times \left( \frac{1}{i} \right)^{d/2}
   = \left( \frac{1}{i} \right)^d
   \qquad \mbox{for diagram (k).}
\label {eq:kphase}
\end {equation}
This is the origin of the overall $i^{-d}$ in (\ref{eq:FundDivd0}).
The reason $({\mathfrak M}_{34},{\mathfrak M}_{12})$ given by
(\ref{eq:Ms}) were positive was that for the 4-particle portion of
the evolution in fig.\ \ref{fig:diagsVIRT}k there
is only one line which is in the conjugate amplitude ($x_1<0$) and
the other three are all in the amplitude ($x_2,x_3,x_4 > 0$).
In contrast, for the 4-particle portion of the evolution in
fig.\ \ref{fig:diagsVIRT2}r, there are two lines in the conjugate
amplitude (two of the $x_i < 0$) and two in the amplitude
(the other two $x_i > 0$), and so one of
$({\mathfrak M}_{34},{\mathfrak M}_{12})$ will be positive and one will
be negative.  The analog of (\ref{eq:kphase}) is then
\begin {equation}
   \left( \frac{1}{i} \right)^{d/2} \times \left( -\frac{1}{i} \right)^{d/2}
   = 1
   \qquad \mbox{for diagram (r).}
\label {eq:rphase}
\end {equation}
One can consider (\ref{eq:rphase}) as the product of an ordinary
free propagator times the conjugate of an ordinary free propagator,
for which the overall phases cancel.

In consequence, if we had done a direct calculation of (r) [instead of using
our front-end trick], there would have been a relative factor of $i^d$
between our calculations of (k) and our calculations of (r).  When multiplied
by the $2/\eps$ divergence of those calculations, like in (\ref{eq:lastdiv}),
this leads to a relative additive term $\pm i\pi$ between the finite terms in
(\ref{eq:lastdiv}) and their analog for (r).
So, ultimately, the origin of the
$-i\pi$ difference between the
form of (\ref{eq:kLogTerm}) and (\ref{eq:rLogTerm2}) is that in
fig.\ \ref{fig:diagsVIRT}k the earliest splitting and the photon
self-energy loop both happen in the amplitude whereas in
fig.\ \ref{fig:diagsVIRT2}r, one happens in the amplitude and the
other in the conjugate amplitude.


\subsection {Front-end transformations in \boldmath$d{=}2{-}\eps$ dimensions}
\label{app:frontenddimreg}

The front-end transformation rules given in the main text were for
$d{=}2$.  On a diagram by diagram basis, there are some subtleties
conerning overall normalization when $d{=}2{-}\eps$, which we now discuss,
but they do not affect any of the results in this paper.


\subsubsection {Transformation of leading-order rates}

\begin {figure}[tp]
\begin {center}
  \includegraphics[scale=0.55]{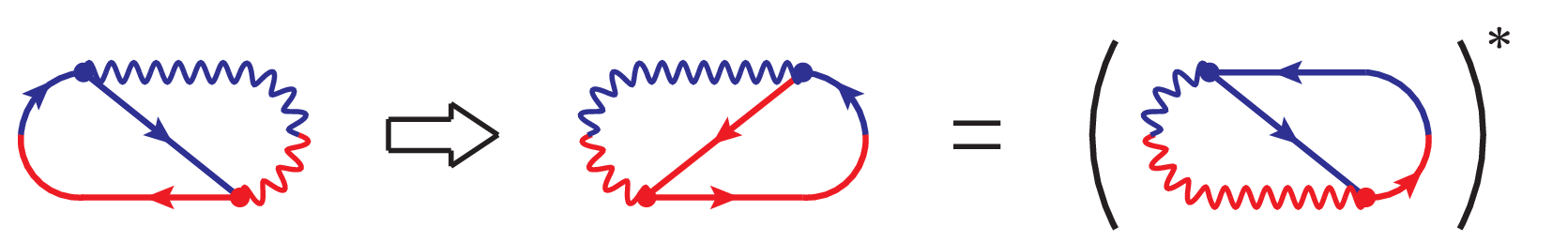}
  \caption{
     \label{fig:LOtransform}
     The conversion of the leading-order diagram $x\bar x$ for $e \to \gamma e$
     via front-end and back-end transformation into the conjugate of the
     leading-order diagram $y\bar y$ for $\gamma \to e\bar e$.
  }
\end {center}
\end {figure}

The easiest way to find the necessary modifications is to consider
the transformation (\ref{eq:frontbackLO}) that related leading-order
rates for $e\to\gamma e$ and $\gamma\to e \bar e$.
In $d=2{-}\eps$ dimensions, the $x\bar x$ diagram for
$e\to\gamma e$ is given by (\ref{eq:xxLO}) as
\begin {equation}
  \left[ \frac{d\Gamma}{d\xe} \right]_{x\bar x}^{e\to\gamma e} =
  - \frac{\mu^\eps\alphaqed d}{8\pi} \, P_{e\to e}^{(d)}(\xe) \,
  \Beta(\tfrac12{+}\tfrac{d}{4},-\tfrac{d}{4}) \,
  \Bigl( \frac{2\pi}{M_\ix\barOmega_\ix} \Bigr)^{\eps/2}
  i \barOmega_\ix
\label {eq:LOdimxx}
\end {equation}
with $\barOmega_\ix = \Omega_\ix \sgn(M_\ix)$ and $M_\ix = \xe(1{-}\xe)E$.
If we take a front-end and back-end transformation, as shown in
fig.\ \ref{fig:LOtransform}, we should get the
conjugate of the analogous $\gamma \to e\bar e$ result, which is
\begin {equation}
  \left[ \frac{d\Gamma}{d\ye} \right]_{y\bar y}^{\gamma \to e\bar e} =
  - \frac{\Nf\mu^\eps\alphaqed d}{8\pi} \, P_{\gamma\to e}^{(d)}(\ye) \,
  \Beta(\tfrac12{+}\tfrac{d}{4},-\tfrac{d}{4}) \,
  \Bigl( \frac{2\pi}{M_{\gamma\to e\bar e}\barOmega_{\gamma\to e\bar e}} \Bigr)^{\eps/2}
  i \barOmega_{\gamma\to e\bar e}
\label {eq:LOdimyy}
\end {equation}
with $M_{\gamma\to e\bar e} = \ye(1{-}\ye)E$.  In addition to the
$d$-dimensional splitting function of (\ref{eq:Pdim}), we also
need%
\footnote{
  See, for example, eq.\ (14) of ref.\ \cite{Pdim} with similar
  conversions as in our earlier footnote \ref{foot:Pdim}.
}
\begin {equation}
   P_{e\to e}^{(d)}(z) =
     \frac{1+z^2}{1-z} - \frac{\eps}{2} (1{-}z) .
\label {eq:Peedim}
\end {equation}
One can check that the transformation (\ref{eq:frontbackLO}) does
not do the job to relate (\ref{eq:LOdimxx}) to (\ref{eq:LOdimyy}).
To get it to work, one must modify the overall power of $1{-}\ye$
in (\ref{eq:frontbackLO}) to $(1{-}\ye)^{1-\eps}$.
One also needs to account for the fact that the rate
$e \to \gamma e$ is {\it averaged}\/ over electron helicities
and summed over photon helicities, whereas the rate for
$\gamma \to e\bar e$ is the opposite.  In dimensional regularization,
the number ${\cal N}_\gamma$ of initial photon helicities is $2-\eps$,
but the number of initial electron helicites ${\cal N}_e$
is fixed by the convention
$\tr(\openone_{\rm Dirac}){\equiv}4$ to be exactly $2$.
We need to account for this difference
by including a factor ${\cal N}_e/{\cal N}_\gamma$ in
our transformation.  One can check already at the level of the DGLAP
splitting functions (\ref{eq:Pdim}) and (\ref{eq:Peedim}) that
\begin {equation}   
   P_{\gamma\to e}^{(d)}(z)
   = (1{-}z) \frac{{\cal N}_e}{{\cal N}_\gamma}
       P_{e\to e}^{(d)}\bigl( \tfrac{-z}{1{-}z} \bigr)
   .
\end {equation}
Using that, one finds that the $d{=}2{-}\eps$ generalization of
(\ref{eq:frontbackLO}) is
\begin {equation}
  2\Re \left[ \frac{d\Gamma}{d\ye} \right]_{y\bar y}^{\gamma \to e\bar e}
  =
  2 \Nf (1{-}\ye)^{1-\eps} \, \frac{{\cal N}_e}{{\cal N}_\gamma} \Re
  \left\{
    \left[ \frac{d\Gamma}{d\xe} \right]_{x\bar x}^{e \to \gamma e}
     \mbox{with}~
     (\xe,E) \to
     \Bigl(\frac{-\ye}{1{-}\ye},(1{-}\ye)E\Bigr)
   \right\} .
\label {eq:frontbackLOdim}
\end {equation}

\subsubsection {Transformation of NLO rates}

To get the transformation law for NLO rates, we can just use
the obsevation after (\ref{eq:frontbackLO}) that one can interpret
$[d\Gamma/d\xe\,d\ye]_{x\bar x}^{e \to \gamma e} \equiv
 [d\Gamma/d\xe]_{x\bar x}^{e \to \gamma e} \, \delta(\ye)$
in the leading-order rate formula.
The $d{=}2{-}\eps$ analog of (\ref{eq:rfromk}) will then be
\begin {multline}
   2\Re
   \left[ \frac{d\Gamma}{d\ye} \right]_{\rm (r)}
   = 2\Nf (1{-}\ye)^{-\eps} \, \frac{{\cal N}_e}{{\cal N}_\gamma}
       \Re \int_0^1 d\xe \>
   \Biggl\{
     \left[ \frac{d\Gamma}{d\xe\,d\ye} \right]_{\rm (k)}
     \mbox{with}~
\\
     (\xe,\ye,E) \to
     \Bigl(\frac{-\ye}{1{-}\ye},\frac{\xe}{1{-}\ye},(1{-}\ye)E\Bigr)
   \Biggr\} .
\label{eq:rfromkdim}
\end {multline}
However, we do not need this in this paper.  Because the additional
factor of $(1-\ye)^{-\eps} {\cal N}_e/{\cal N}_\gamma$ we needed to add
to the $d{=}2$ transformation laws was the same for both the leading-order
and NLO rates, it can be factored out when transformationing 
the final renormalized rate (\ref{eq:xyyxRen2general}).  Since the
renormalized rate is finite, we can then set $\eps{=}0$, and so the
additional factor is just 1 and makes no difference there.  We would only
need the additional factor if we wanted to individually transform
divergent diagrams.


\subsubsection {A check: the QCD $x y \bar y \bar x$
                and $x \bar y y \bar x$ diagrams}

\begin {figure}[tp]
\begin {center}
  \begin{picture}(430,65)(0,0)
  \put(0,10){\includegraphics[scale=0.5]{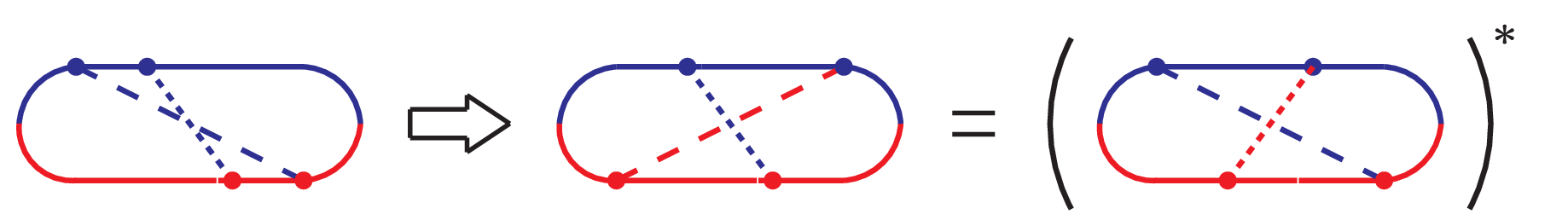}}
  \put(43,0){$xy\bar y\bar x$}
  \put(342,0){$(x\bar y y\bar x)^*$}
  \end{picture}
  \caption{
     \label{fig:frontbackcheck}
     The conversion of the QCD diagram $xy\bar y\bar x$ for $g \to ggg$
     via front-end and back-end transformation into the conjugate of the
     diagram $x\bar y y\bar x$ for $g \to ggg$.  Long and short dashed
     lines represent final-state gluons with longitudinal momentum
     fractions $x$ and $y$ respectively.
  }
\end {center}
\end {figure}

Part of the reason we went through this discussion is so that we
could make an explicit check of front-end transformations using
previous results \cite{2brem,dimreg} for QCD diagrams.
In particular, a front-end plus
back-end trasnformation should relate the two $g \to ggg$ interference
diagrams depicted in fig.\ \ref{fig:frontbackcheck}.  These diagrams
are each UV divergent (particular time-orderings of tree-level
diagrams can be UV divergent even though the sum of all time-orderings
is not), and so one must use the $d{=}2{-}\eps$ version of the
transformation.  Here, the initial particles are always gluons, so
that transformation law is simply
\begin {equation}
   2\Re
   \left[ \frac{\Delta\,d\Gamma}{dx\,dy} \right]_{x\bar y y\bar x}
   = 2(1{-}x)^{-\eps} \Re
   \Biggl\{
     \left[ \frac{\Delta\,d\Gamma}{dx\,dy} \right]_{xy\bar y\bar x}
     \mbox{with}~
     (x,y,E) \to
     \Bigl(\frac{-x}{1{-}x},\frac{y}{1{-}x},(1{-}x)E\Bigr)
   \Biggr\} .
\label {eq:frontbackcheck}
\end {equation}
Using formulas for $xy\bar y\bar x$ and $x\bar y y\bar x$ from
AI1 \cite{2brem} and ACI3 \cite{dimreg},
we have checked that the above transformation indeed works.%
\footnote{
  Specifically, $xy\bar y\bar x$ consists of (i)
  a divergent ``pole contribution'' given by ACI3 (4.36) and (4.37)
  plus (ii) the non-pole piece given by AI1 (5.45) integrated just
  over $\Delta t > a$.
  The corresponding contributions to $x\bar y y\bar x$ are then
  obtained from the transformations described in AI1 section VI.A.
  To get the overall sign in (\ref{eq:frontbackcheck}) above, one
  must add absolute value signs to the various longitudinal momentum
  fractions in the formulas AI1 (4.39) for $(\alpha,\beta,\gamma)$,
  just like we had to add absolute value signs to the
  $(\bar\alpha,\bar\beta,\bar\gamma)$ of this paper in order to
  be able to implement front-end transformations (see footnote
  \ref{foot:absval2}).  Since it is only $x$ (and not $y$ or $1{-}x$
  or $1{-}y$ or $1{-}x{-}y$) that negates under the transformation
  in (\ref{eq:frontbackcheck}), it is sufficient to just replace
  the {\it factors} of $x$ and $x^3$ by $|x|$ and $|x|^3$ in AI (4.39).
}

\end {document}